TECHNISCHE UNIVERSITÄT DARMSTADT

**Fachbereich Physik – Institut für Angewandte Physik**
**Arbeitsgruppe *Theoretische Quantenphysik***

**DIPLOMARBEIT**

**zum Thema**

# Quantenkryptographie mit kontinuierlichen Variablen

**Diskussion verschiedener Realisierungen mit Qudits**

Verfasser:
Ulrich Seyfarth

Betreuer:
Prof. Dr. Gernot Alber

# Vorwort

Informationen sind ein wichtiger Bestandteil unseres gesellschaftlichen Zusammenlebens. Jeder Mensch sollte die Möglichkeit haben, Informationen vor anderen zu schützen, um sich frei entfalten zu können. Wir leben in der heutigen Zeit in einer sogenannten Informationsgesellschaft. Viele Informationen sind digital gespeichert und damit nicht mehr physikalisch voneinander getrennt, da fast jedes digitale Gerät, welches Informationen speichern kann, auch einen Zugang zum weltweiten Netz, dem Internet hat.

Schon die Geschichte hat gezeigt, dass das Verschlüsseln von Informationen ein häufig verwendetes Mittel ist, um eben diese Informationen vor anderen zu verbergen, sei es aus dem Grund, einen eigenen Vorteil zu bewahren oder einen anderen Menschen psychisch nicht verletzen zu wollen. Trotzdem kann man als Purist auch danach fragen, ob es überhaupt sinnvoll ist, die Möglichkeit des Verschlüsselns zu bieten, weil auch dies wiederum schnell zu Konflikten führen kann.

Allerdings haben nicht nur z. B. ein Staat oder eine Firma das Interesse, Informationen vor den politischen Gegnern oder der Konkurrenz zu schützen, sondern auch der Privatmann möchte in Zeiten der Videoüberwachung und der Vorratsdatenspeicherung seine intimsten, digital festgehaltenen Informationen sicher schützen können.
In der Geschichte wurden viele Verfahren entwickelt, um Daten zu verschlüsseln, jedoch konnten die meisten immer wieder kompromittiert werden.

Die einzig bekannte Möglichkeit, Informationen wirklich sicher zu verschlüsseln ist das One-Time-Pad. Hier wird ein Text mithilfe einer ebensolangen Zeichenfolge symmetrisch verschlüsselt, wobei der Schlüssel nur ein einziges Mal Anwendung findet. Das Problem ist also vom sicheren Weitergeben der Nachricht auf ein sicheres Weitergeben des Schlüssels verlagert worden, da man das Verschlüsselungsverfahren öffentlich zugänglich macht.

Das Erzeugen und Verteilen eines sicheren Schlüssels an zwei Parteien, im Weiteren mit Alice und Bob bezeichnet, ist nun die Herausforderung, die man mithilfe der Quantenmechanik lösen möchte.

In der Theorie hat man schon vor einigen Jahren Verfahren entwickelt, die eben dieses Erzeugen und Verteilen erlauben, jedoch ist deren experimentelle Umsetzung sehr schwierig. Es wird heute auch immer wichtiger, ein sehr schnelles Verfahren zu entwickeln, da die Informationsmengen immer weiter steigen und der Schlüssel ebensolang wie die Nachricht sein muss.

Für die experimentelle Umsetzbarkeit und einen zeitnahen Schlüsselaustausch hat man in jüngster Vergangenheit neue Ansätze verfolgt.

In dieser Arbeit werden Wege aufgezeigt und diskutiert, wie man das Austauschen eines sicheren Schlüssels beschleunigen kann.

Darmstadt,
im Mai 2008. Ulrich Seyfarth

# Inhaltsverzeichnis

# 1. Einführung

Die Quantenkryptographie beschäftigt sich mit der Frage, in welchem Rahmen es möglich ist, dass zwei Parteien Alice und Bob einen Schlüssel nachweisbar sicher austauschen können. Die Theorie der Quantenkryptographie lässt sich zwischen der Quantenmechanik und der Informationstheorie ansiedeln. Die Sicherheitsbeweise benötigen jedoch den Bezug zwischen der Quantenmechanik und der Relativitätstheorie, beruhend auf der Arbeit von Einstein, Podolky und Rosen [EPR35].

Klassische, kryptographische Verfahren konnten immer wieder kompromittiert werden. Für eine sichere Verschlüsselung, wie das 1918 von Gilbert Vernam und Joseph Oswald Mauborgne entwickelte One-Time-Pad, stellt das sichere Austauschen des Schlüssels ein massives Problem dar. Das sichere Austauschen eines Schlüssels für dieses Verfahren ist v. a. deshalb schwierig, weil dieser genauso lang sein muss, wie die zu verschlüsselnde Nachricht. Quantenkryptographische Verfahren akzeptieren die Anwesenheit eines Lauschers Eve, dem kompletter Zugriff auf den Schlüsselaustausch und damit alle verwendeten Kanäle erlaubt ist. 1983 wurde zum ersten Mal öffentlich von Stephen Wiesner [Wie83] über eine Theorie zur Quantenkryptographie nachgedacht. Mit dem BB84-Protokoll [BB84] wurde 1984 von C. H. Bennett, G. Brassard ein Protokoll entwickelt, welches beweisbar sicher ist [SP00]. Für dessen praktische Umsetzung bedarf es der Realisierung von 1-Photonen-Quellen, weil es auf dem *No-Cloning-Theorem* beruht [WZ82]. Da diese Quellen nicht effektiv genug sind, hat man sich nach alternativen Wegen umgeschaut, quantenkryptographie Protokolle realisieren zu können.

Durch die Betrachtung des Quantenklonens hat man sie in der Möglichkeit gefunden, Informationen in kontinuierlichen Variablen zu speichern, beispielsweise den Quadraturamplituden des quantisierten Lichtes [BK98, FSBF98]. Die ersten Ansätze effektiver Protokolle beruhten meist auf der experimentell schwierig realisierbaren Verwendung gequetschter Zustände [Hil00, GP01, CLA01b]. Um die Sicherheit dieser Protokolle zu betrachten, wurden verschiedene Attacken diskutiert [GC04, NA05]. Im weitesten Sinne unterscheidet man zwischen individuellen Attacken [CLA01b, GG02, SRLL02, WLB$^{+}$04, NH04, GC04], kollektiven Attacken [Gro05, NA05] und kohärenten Attacken [GP01, IVAC04]. Die am häufigsten diskutierte ist der Strahlteiler-Angriff [CBL01]. Aufgrund von Symmetriebetrachtungen bestand jedoch das Problem, dass ein Lauscher Eve bei einem verlustbehafteten Kanal von mehr als $50\%$ Verlustrate mehr Information von Alice erhält, als Bob. Für diese Grenze wurden Protokolle mit einem klassischen Abgleich nötig [SRLL02]. Mit dem 2002 von F. Grosshans und P. Grangier diskutierten Protokoll [GG02] mit kohärenten Zuständen wurde das Problem der experimentellen Realisierbarkeit von Systemen mit gequetschten Zuständen umgangen. Schliesslich haben 2006 M. Heid und N. Lütkenhaus 2006 [HL06] die sicheren Schlüsselraten dieses Protokolles ausführlich diskutiert.

Die vorliegende Arbeit soll an dieser Stelle anknüpfen und Wege aufzeigen, mit welchen

Methoden ein weiteres Problem hin zur praktischen Anwendung von Quantenkryptographischen Schlüsselaustauschprotokollen gelöst werden kann. Die sichere Übertragungskapazität ist für eine praktische Anwendung immer noch zu gering. Die Idee dieser Arbeit besteht nun darin, dass Alice die Wahl zwischen mehr als zwei Zuständen hat, die sie an Bob übertragen kann. Somit könnte mehr als ein klassisches Bit mit einer einzigen Übertragung eines kohärenten Zustandes an Bob übermittelt werden. Quantenmechanische Zustände werden in dieser Arbeit als Qudits bezeichnet, wobei diese Terminologie nicht der gewöhnlichen Bedeutung von Qudits entspricht, sondern ein $d$-Qudit einer Auswahl zwischen $d$ Quantenzuständen entspricht (es sich also nicht um ein $d$-dimensionales Quantensystem handelt). Weiterhin ist zu beachten, dass als einziger Angriff ein Strahlteiler-Angriff betrachtet wird.

Kapitel 2 gibt kurze Einführungen in die Theorien, die benötigt werden, diese Fragestellung zu diskutieren. Weiterhin werden die mathematisch jeweils wichtigen Hilfmittel beschrieben. Begonnen wird mit der Einführung in die klassischen Theorien, beginnend mit der Wahrscheinlichkeitstheorie (Abschnitt 2.1), der Kryptographie (Abschnitt 2.1) und schließlich der Informationstheorie (Abschnitt 2.3). Danach folgt eine kurze Erklärung der Quantentheorie (Abschnitt 2.4) und letztendlich der Quanteninformationstheorie (Abschnitt 2.5).

Mithilfe dieser grundlegenden Basis werden im dritten Kapitel einige Protokolle zur Quantenkryptographie vorgestellt, u. a. auch das von F. Grosshans und P. Grangier.

Im vierten Kapitel folgt der wissenschaftlich neue Inhalt dieser Arbeit. Aufbauend auf dem o. a. Protokoll wird eine Verallgemeinerung der Zustände diskutiert. Dabei werden in der Phasenraumdarstellung der kohärenten Zustände nicht mehr nur zwei Zustände auf der reellen Achse präpariert werden können, sondern $n$ Zustände, die auf einem Kreis um den Ursprung äquidistant verteilt sind.

Aufgrund von Symmetrieüberlegungen folgt nun die Erkenntniss, dass es sinnvoll sein könnte, entweder die Präparation oder die Messung zu verändern. Zuerst wird dieses Protokoll ein zweites Mal verallgemeinert, hin zu gequetschten Zuständen, also einer Änderung der Präparation (Abschnitt 4.6). Weiterhin wird in Abschnitt 4.7 eine Änderung der Messung vorgeschlagen, um bessere Resultate der sicheren Schlüsselrate zu erreichen.

# 2. Grundlagen

## 2.1. Wahrscheinlichkeitstheorie

Die Wahrscheinlichkeitstheorie ist ein mathematischer Teilbereich der Stochastik, welche dem Vorhersagen von Ereignissen aufgrund vergangener Ereignisse oder theoretischer Annahmen dient. Hierbei muss beachtet werden, dass alle Versuche unter den gleichen Bedingungen stattfinden.

### 2.1.1. Wahrscheinlichkeit

Treten zwei Ereignisse bei einem Zufallsexperiment im Mittel gleich häufig auf, so haben sie die gleiche Wahrscheinlichkeit. Die Summe über die Wahrscheinlichkeiten aller möglichen Ereignisse ist stets $1$. Die Wahrscheinlichkeit eines Ereignisses $A$ beschreibt somit das Verhältnis zwischen der Häufigkeit des Ereignisses $A$ und der Gesamtzahl der Versuche $n \in \mathbb{N}$ im experimentellen Sinne. Dieses Verhältnis wird umso besser, je mehr Zufallsexperimente man durchführt, da man den Ereignisraum immer weiter abdeckt.

Lässt sich das Zustandekommen des Ereignisses mathematisch beschreiben, so kann man eine theoretische Wahrscheinlichkeit berechnen, die man experimentell nur durch das Abdecken des gesamten Ereignisraumes bekommen würde.

Es gelten für Ereignisse die Regeln der Booleschen Algebra (aktuell von Giuseppe Peano in „calculo geometrico“).

### 2.1.2. Gemeinsame Wahrscheinlichkeit

Betrachtet man zwei Ereignisse $A$ und $B$, so ist die gemeinsame Wahrscheinlichkeit $P(A, B)$ als diejenige Wahrscheinlichkeit definiert, dass die Ereignisse $A$ und $B$ gemeinsam auftreten. Außerdem gilt $P(A, B) = P(B, A)$.

### 2.1.3. Bedingte Wahrscheinlichkeit

Betrachtet man zwei Ereignisse $A$ und $B$, so definiert man die bedingte Wahrscheinlichkeit $P(B|A)$ als diejenige Wahrscheinlichkeit, mit der das Ereignis $B$ unter der Bedingung auftritt, dass zuvor das Ereignis $A$ aufgetreten ist.

Mathematisch kann man dies auch formulieren als:

$$P(B|A) = \frac{P(B, A)}{P(A)} \quad , \tag{2.1}$$

solange $P(A) \neq 0$ ist.

Sind beide Ereignisse unabhängig, also $P(A, B) = P(A)P(B)$, so gilt:

$$P(B|A) = P(A) \quad \text{und} \quad P(A|B) = P(B) \quad . \tag{2.2}$$

### 2.1.4. Satz von Bayes

Betrachtet man nun alle möglichen Ereignisse $B_l$ aus der Ereignismenge $Z$ mit der Grösse $n = |Z| \in \mathbb{N}$, so gilt der Satz der *vollständigen Wahrscheinlichkeit*:

$$P(A) = \sum_{l=1}^{n} P(A|B_l)P(B_l) \quad . \tag{2.3}$$

Daraus und mit $P(A, B) = P(B, A)$ folgt durch die Definition der *bedingten Wahrscheinlichkeit* der *Satz von Bayes*:

$$P(B_k|A) = \frac{P(A|B_k)P(B_k)}{\sum_{l=1}^{n} P(A|B_l)P(B_l)} \quad . \tag{2.4}$$

### 2.1.5. A-Posteriori-Wahrscheinlichkeit

Möchte man nun im Nachhinein feststellen, wie wahrscheinlich es ist, dass das Ereignis $A_k$ vor dem Ereignis $B$ auftritt, so definiert man die *A-Posteriori-Wahrscheinlichkeit*. Wie der Name schon sagt kann man sie erst nach der Messung durchführen, man benötigt alle *A-Priori-Wahrscheinlichkeiten* $P(A_l)$:

$$P(A_k|B) = \frac{P(B|A_k)P(A_k)}{\sum_{l=1}^{n} P(B|A_l)P(A_l)} = \frac{P(B|A_k)P(A_k)}{P(B)} \quad . \tag{2.5}$$

### 2.1.6. Zufallsvariable

Unter einer Zufallsvariablen $X$ beschreibt man die Grösse, welche aufgrund von Wahrscheinlichkeiten $p_i$ den zugehörigen Ereigniswert $x_i$ annimmt. Es gilt also $P(X = x_i) = p_i$ und $X(P = p_i) = x_i$.

## 2.2. Kryptographie

Die Verwendung kryptographischer Methoden geht mindestens auf die Zeit der alten Ägypter zurück, die ungewöhnliche Hieroglyphen nutzten, um den Inhalt von Nachrichten unwissenden Dritten gegenüber zu verbergen. Kryptographische Methoden werden meist beim Übermitteln von Nachrichten verwendet. Man kann analog zur Informationstheorie (Kapitel 2.3) die Begriffe des *Senders* (meist *Alice*), *Empfängers* (meist *Bob*), der *Nachricht*, des *Kanals* und des *Alphabets* beschreiben. Allerdings gibt es sowohl eine *verschlüsselte*, als auch eine *unverschlüsselte Nachricht*, welche die Gleiche wie die *entschlüsselte Nachricht* sein sollte. Zudem gibt es verschiedene Ver- und Entschlüsselungsverfahren, sodass der Schlüssel zum Entschlüsseln und der zum Verschlüsseln nicht immer der Gleiche sein müssen. Wichtig zur Sicherheit eines Verfahrens ist auch das Diskutieren der Anwesenheit eines *Angreifers*, meist als *Eve* bezeichnet.

### 2.2.1. Ver- und Entschlüsselung

Sehr einfache Verschlüsselungsfahren sind die der *mono-* und *polyalphabetischen Substitution*.

Im ersten Fall bedeutet dies, dass das vorhandene Alphabet durch ein anderes Alphabet ersetzt wird. Da die beiden Alphabete in der Regel gleich sind und damit auch die gleiche Länge besitzen, kann man den Abstand zwischen einem Buchstaben aus dem Ausgangsalphabet und dem dazugehörigen Buchstaben aus dem verschlüsselten Alphabet wiederum durch einen Buchstaben ausdrücken. Ersetzt man beispielsweise den Buchstaben $A$ durch den Buchstaben $D$, so ist der Abstand dazwischen $3$, man würde als Schlüssel also ein $C$ schreiben.

Bei der *polyalphabetischen Verschlüsselung* wird nicht nur ein verschlüsseltes Alphabet verwendet, sondern mehrere. Ein Schlüssel besteht dann aus mehreren Buchstaben.

Es folgen einige Beispiele zum besseren Verständnis dieser Methoden.

### 2.2.2. Monoalphabetische Verschlüsselung

Eine Nachricht wird verschlüsselt, indem das verwendete Alphabet durch ein zweites ersetzt wird. Jeder Buchstabe des alten Alphabetes wird also immer auf denselben Buchstaben des neuen Alphabetes abgebildet. Allerdings kann es sein, dass auch zwei Buchstaben auf den gleichen abgebildet werden (häufig bei $i$ und $j$ wegen deren Ähnlichkeit).

#### Caesar-Verschlüsselung

Eine der ältesten und wahrscheinlich auch einfachsten Verschlüsselungsmethoden ist die *Caesar-Verschlüsselung*. Das Alphabet wird durch ein verschobenes Alphabet (daher manchmal auch als *Verschiebechiffre* bezeichnet) ersetzt. Einzig wählbarer Parameter ist also die Größe der Verschiebung. Wählt man z. B. $5$, so würde aus einem $A$ ein $F$, aus einem $B$ ein $G$ usw. Ein $Z$ wäre dann wieder ein $E$. Eine mit dieser Methode verschlüsselte Nachricht kann ebenso leicht wieder entschlüsselt werden. Man muss nur die Verschiebung kennen und diese im Alphabet rückwärts gehen. Da es nur $25$ mögliche Verschiebungen gibt, ist diese Verschlüsselungsmethode sehr unsicher - man kann alle $25$ Möglichkeiten sehr leicht ausprobieren.

#### Substitution

Eine ebenso alte, aber sehr viel effektivere Verschlüsselungsmethode ist die Substitution. Jeder Buchstabe wird durch einen anderen Buchstaben (meist des gleichen Alphabetes) ersetzt. Aus einem $A$ könnte dann z. B. ein $E$ werden, aus einem $B$ aber ein $L$. Es gibt a priori keinen Zusammenhang zwischen beiden Alphabeten, wobei man sich, um sich die Verschlüsselung besser merken zu können, häufig ein Schlüsselwort merkt und spätere Buchstaben im Alphabet rät. Als Schlüsselwort wird das Wort bezeichnet, welches den Abstand der Buchstaben zwischen den Alphabeten in Form von Buchstaben bezeichnet. Wählt man das Wort $DIPLOMARBEIT$ als Schlüssel, würde man ein $A$ in ein $E$ chiffrieren, da $D$ der $4$. Buchstabe im Ausgangsalphabet ist und $E$ der $5$. nach dem $A$. Dementsprechend werden ($B \rightarrow K$, $C \rightarrow S$, $D \rightarrow P$, $E \rightarrow T$, $F \rightarrow S$, usw.). Wie man hier schon sieht, würden sowohl $C$, als auch $F$ auf $S$ abgebildet. Es wäre sicher sinnvoll ein anderes Schlüsselwort zu wählen, um die Entschlüsselung zu gewährleisten, da man ein $S$ nicht mehr eindeutig entschlüsseln kann.

Durch solche Überlegungen wird allerdings die Anzahl sinnvoller Schlüsselwörter geringer und man macht es einem potentiellen Angreifer leichter, den Schlüssel zu erraten. Am sinnvollsten wäre es also, jedem Buchstaben einen zufälligen Buchstaben aus dem neuen Alphabet zuzuweisen und diese auch nur einmal auszuwürfeln, sodass die Abbildung bijektiv und die Entschlüsselung damit sichergestellt ist.

Selbst, wenn die Substitution der einzelnen Buchstaben absolut zufällig ist, so ist es doch die Verteilung der Buchstaben in Sprachen nicht. Weiß man z. B., dass es sich um einen deutschen Text handelt, so kann man eine Häufigkeitsanalyse durchführen und sehen, dass der Buchstabe $E$ im Mittel eine Häufigkeit von ca. $17\%$ hat. Bei einer Gleichverteilung sollte jeder Buchstabe eine Häufigkeit von $1/26 \approx 4\%$ besitzen. Andere Buchstaben treten sehr viel seltener auf. Mit einer solchen Analyse kann man einen Text, der auf diese Art verschlüsselt wurde, nach und nach entschlüsseln. Irgendwann erkennt man durch bloßes Hinschauen auch die restlichen Ersetzungen. Je länger der Text ist, desto eher entspricht die Häufigkeitsverteilung der gemittelten Häufigkeitsverteilung der deutschen Sprache und desto leichter ist ein Text zu entschlüsseln. Lange Texte kann man somit auch maschinell sehr zuverlässig entschlüsseln lassen. Neben der Analyse der Häufigkeiten einzelner Buchstaben (Monogramme) kann man auch die Häufigkeit von Buchstabenkombinationen betrachten (Bi-, Tri-, Tetra-, $N$-Gramme). Dies ermöglicht es dann, auch kürzere Texte besser entschlüsseln zu können.

### 2.2.3. Polyalphabetische Verschlüsselung

Ist zusätzlich die Position eines Zeichens entscheidend für deren Verschlüsselung, so kann man dies auch als eine positionsabhängige Wahl des Alphabetes ansehen.

#### Vigenère-Verschlüsselung

Die wahrscheinlich bekannteste *polyalphabetische Verschlüsselung* ist die *Vigenère-Verschlüsselung*, welche Blaise de Vigenère in der zweiten Hälfte des 16. Jahrhunderts (um 1580) aus den Ansätzen von Leon Battista Alberti (um 1460) entwickelte. Dieses Verfahren basiert auf der *Caesar-Verschlüssung*. Man wählt bei dieser Verschlüsselung nicht nur ein verschobenes Alphabet, sondern mehrere. Ist der Schlüssel $KEY$ und der Text $TEXT$, so wird das $T$ mit einem $K$ verschlüsselt, $E$ mit $E$, $X$ mit $Y$ und $T$ wieder mit $K$. $A$ bewirkt eine Verschiebung um $0$, $B$ eine von $1$, etc. Der verschlüsselte Text wäre also $DIVD$. Erst ca. 300 Jahre später konnte es von Charles Babbage gebrochen werden. Schaut man sich die Korellation des verschlüsselten Textes mit einem verschobenen Text an, so ist diese am größten bei einem Vielfachen der Schlüssellänge. Nun reicht eine Häufigkeitsanalyse, um die Nachricht zu entschlüsseln. Je länger ein Schlüssel ist, desto sicherer ist dieses Verfahren.

Die *Autokey-Vigenère-Verschlüsselung* setzt den Schlüssel nicht periodisch fort, sondern nutzt am Ende des Schlüssels die Nachricht zur Verschlüsselung. Sie ist deutlich sicherer und meist nur durch kombinierte Verfahren kompromittierbar.

### Vernam-Verschlüsselung

Die *Vernam-Verschlüsselung* kann man als einen Spezialfall der *Vigenère-Verschlüsselung* ansehen. Sie wurde Anfang des 20. Jahrhunderts von *Gilbert Vernam* diskutiert und hat im Gegensatz zur *Vigenère-Verschlüsselung* einen Schlüssel, der genauso lang ist, wie die Nachricht selbst. Diesen Schlüssel kann man meist nicht mehr per Hand kompromittieren. Es müssen Häufigkeitsanalysen des Schlüssels angestellt werden, wenn man vermuten kann, dass er selbst ein Text ist oder andere statistisch bekannte Strukturen ausweist. Bei einer manuellen Wahl eines Schlüssels ist z. B. das mehrfache Auftreten des gleichen Buchstabens hintereinander statistisch gesehen zu selten. Auch andere Eigenschaften liegen einem solchen Schlüssel zugrunde, auf die man schließen kann.

### One-Time-Pad

Gilbert Vernam und Joseph Oswald Mauborgne sind 1918 nun der Idee nachgegangen, diesen Schlüssel zufällig erzeugen zu lassen und so jegliche statistische Analyse des Schlüssels auszuschließen. Dieses informationstheoretisch sichere symmetrische Verschlüsselungsverfahren bezeichnet mal als One-Time-Pad. Eine Nachricht wird durch einen Schlüssel der gleichen Länge codiert, wobei dieser absolut zufällig erzeugt werden muss. Wie der Name schon sagt, darf das One-Time-Pad, also der Schlüssel, auch nur einmal verwendet werden, sonst wäre es wieder nur eine *Vigenère-Verschlüsselung*.

Bisherige Ansätze zur Implementierung dieses Verfahrens hatten immer das Problem, einen rein zufälligen Schlüssel erzeugen zu können, da alle klassischen Generatoren eine gewisse Regelmäßigkeit aufweisen. Sollte allerdings die Theorie der Quantenmechanik korrekt sein, so kann man sich beispielsweise vorstellen, einen reinen Zustand in einer Spin-$1/2$-Basis mit $|e_1\rangle$ und $|e_2\rangle$ zu erzeugen und in einer Basis mit $|e_1+e_2\rangle$ und $|e_1-e_2\rangle$ zu messen, in der das Ergebnis dann maximal unbestimmt ist. Solange der präparierte Zustand also nicht zerfällt, ist es gleich wahrscheinlich, welchen der beiden möglichen Zustände man misst. Dies wäre ein perfekter Zufallsgenerator.

Ein weiteres Problem des One-Time-Pads liegt im sicheren Austausch des Schlüssels. Im Wesentlichen beschäftigen sich alle Ansätze mit eben diesem Problem. Der Sender (Alice) präpariert einen Zustand und der Empfänger (Bob) versucht, diesen zu messen. Ein Lauscher (Eve) hat ebenfalls daran Interesse, wobei Alice und Bob mit Sicherheit wissen wollen, dass nur sie den späteren Schlüssel besitzen. Dem Angreifer ist es erlaubt, vollständigen Zugang zu allen Übertragungskanälen zu erhalten.

## 2.2.4. Weitere Verschlüsselungsverfahren

Alle bisher vorgestellen Verfahren bezeichnet man auch als symmetrische Verschlüsselungsverfahren, da sowohl zur Ver- als auch Entschlüsselung derselbe Schlüssel verwendet wird. Ist dies nicht der Fall, so spricht man von asymmetrischer Verschlüsselung. Bekannt sind z. B. Verfahren wie *RSA*, die als Public-Key-Verfahren gelten und im Zusammenhang z. B. von *PGP* genutzt werden oder wurden. Diese Verfahren beruhen meist auf der Tatsache, dass das Zerlegen grosser Zahlen in ihre Primfaktoren einen sehr viel höheren Aufwand darstellt, als das Ausmultiplizieren zweier grosser Primzahlen. Man veröffentlicht einen Schlüssel zum Verschlüsseln von

Nachrichten, kann allerdings nur mit einem anderen, privaten Schlüssel die Nachricht wieder entschlüsseln. Die Idee solcher Verfahren stammt vor allem aus dem Schutz von Informationen vor größeren Organisationen, da durch sie eine sehr sichere, geschützte Kommunikation möglich wird und keine zusätzliche Infrastruktur benötigt wird.

Öffentlich bekannt sind Verschlüsselungsverfahren auch aus Kriminalromanen v. a. von Edgar Allan Poe.

## 2.3. Informationstheorie

Die Informationstheorie geht im Wesentlichen zurück auf die Arbeit von Claude Elwood Shannon Mitte des 20. Jahrhunderts (A Mathematical Theory of Communications) [Sha48]. In der Nachrichtentechnik ist es meist wichtig, Daten möglichst schnell übertragen zu können. Zudem sind Kanäle nicht absolut rauschfrei und Daten müssen nach der Übertragung korrigiert werden können. Claude Shannon hat deshalb mathematische Begriffe wie *Information*, *Kanal*, *Kompression*, *Kodierung* und *Informationsgehalt* diskutiert, um die Nachrichtenübertragung mathematisch beschreiben zu können.

### 2.3.1. Shannon-Entropie

Um den Informationsgehalt einer Nachricht beschreiben zu können - unabhängig vom gewählten Alphabet - hat Shannon eine Entropie eingeführt, die später nach ihm benannt wurde:

$$H(X) := -\sum_{i=1}^{n} P(X = x_i) \log P(X = x_i) \quad , \tag{2.6}$$

wobei $p_i$ die Wahrscheinlichkeit ist, dass das Zeichen $z_i$ aus dem Alphabet $Z = z_1, z_2, \ldots, z_n$ auftritt und gilt, dass $\sum_{i=1}^{n} p_i = 1$.

Die Entropie kann man nun deuten als den mittleren Informationsgewinn pro Zeichen, wobei als Basis des Logarithmus meist 2 gewählt wird, also als Informationseinheit das Bit, welches eine einzige atomare Entscheidung darstellt und als kleinste Informationseinheit angesehen werden kann. Beschränkt man sich nun auf eine solche Entscheidung, so ist die Entropie maximal 1 an dem Punkt, an welchem beide Ereignisse gleich wahrscheinlich sind.

Die Shannon-Entropie gibt also die kleinste Menge an Informationseinheiten z. B. in Bit an, in die sich die betrachtete Information kodieren lässt.

Analog kann man nun auch die Entropie für eine kontinuierliche Verteilung mit der Dichtefunktion $p(x)$ definieren als

$$H(X) = -\int_{-\infty}^{\infty} p(x) \log p(x) \mathrm{d}x \quad , \tag{2.7}$$

und analog gilt auch:

$$H(X_1, \ldots, X_n) = -\int \cdots \int p(x_1, \ldots, x_n) \log p(x_1, \ldots, x_n) \; \mathrm{d}x_1 \cdots \mathrm{d}x_n \quad . \tag{2.8}$$

### 2.3.2. Bedingte Entropie

Passend zum Begriff der *bedingten Wahrscheinlichkeit* (Abschnitt 2.1.3) wird der Begriff der *bedingten Entropie* formuliert:

$$H(X|A) = -\sum_{i=1}^{n} P(X = x_i|A) \log P(X = x_i|A) \quad . \tag{2.9}$$

Somit gilt für zwei kontinuierliche Zufallsvariablen $X$ und $Y$:

$$H(X|Y) = H(X,Y) - H(Y) = -\int\int p(x,y) \log \frac{p(x,y)}{p(y)} \mathrm{d}x\mathrm{d}y \quad . \tag{2.10}$$

Die bedingte Entropie beschreibt also die Information, die über eine Grösse $X$ noch nicht bestimmt ist, wenn man über die Grösse $Y$ bereits alle Information in Erfahrung gebracht hat.

### 2.3.3. Transinformation

Die Transinformation (engl.: mutual information, häufig auch als gemeinsame Information bezeichnet) beschreibt die gegenseitige Information zwischen mehreren Zufallsvariablen. Also die Information, die man über eine Variable $Y$ erhält, wenn man etwas über die Variable $X$ erfahren hat. Sie muss also genau die Information sein, die übrig bleibt, wenn man von der gemeinsamen Entropie beide bedingten Entropien abzieht:

$$I(X;Y) = H(X,Y) - H(X|Y) - H(Y|X) \quad . \tag{2.11}$$

Somit gilt natürlich auch:

$$I(X;Y) = H(X) - H(X|Y) \quad \text{und} \tag{2.12}$$
$$I(X;Y) = H(Y) - H(Y|X) \quad . \tag{2.13}$$

## 2.4. Quantenphysik

Die Quantenphysik ist begründet auf Arbeiten von Planck [Pla01], Einstein [Ein05], Heisenberg [Hei26, Hei27], Schrödinger [Sch26a, Sch26b], Pauli [Pau24, Pau25, Pau27, Pau26], Bohr [Boh35], Dirac [Dir27a, Dir26, Dir27b, Dir28, Dir58], von Neumann [vN32] , Born [BJ25, BHJ26], Sommerfeld, Feynman [Fey48] und vielen anderen.

Dachte man noch Ende des 19. Jahrhunderts, die Physik würde bald alles beschreiben können und so als Wissenschaft in absehbarer Zeit aussterben, ist es fast schon dem Mut von Wissenschaftlern wie Albert Einstein oder Max Planck zu verdanken, dass die Physik durch Theorien wie der Relativitätstheorie [Ein05] oder der Quantenmechanik eine Renaissance erleben durfte und heutzutage einen unschätzbaren Wert für die Weiterentwicklung moderner Technologien und das Fortbestehen der gesamten Menschheit besitzt.

Die Quantenphysik beschreibt die Wechselwirkung zwischen Teilchen auf der Ebene von Bosonen und Fermionen. Im Gegensatz zur klassischen Physik sind die physikalischen Regeln weniger intuitiv und konnten erst mithilfe mathematischer Werkzeuge wie dem Hilbertraum ausreichend verstanden werden. Der Zustand eines quantenmechanischen Systems ist nicht mehr im klassischen Sinne komplett beschreibbar, die Heisenbergsche Unschärferelation verbietet dies. Nach Dirac werden Zustände eines Systems durch einen Ket-Vektor ($|\cdot\rangle$) im Hilbertraum beschrieben.

### 2.4.1. Elektromagnetisches Feld

Das *elektromagnetische Feld* lässt sich mithilfe der 2. Quantisierung [Dir27b] durch harmonische Oszillatoren in verschiedenen Feldmoden beschreiben, also einer Fourierzerlegung in verschiedene Frequenzen. Betrachtet man nur eine Mode, so lässt sich deren Hamilton-Operator als harmonischer Oszillator in der Form

$$H = \frac{1}{2}\hbar\omega a a^\dagger \tag{2.14}$$

mit den *Erzeugungs*- und *Vernichtungsoperatoren* $a^\dagger$ bzw. $a$ schreiben. Definiert man nun zwei Operatoren $X_1$, $X_2$ mit

$$X_1 := a + a^\dagger \quad \text{und} \tag{2.15}$$

$$X_2 := i(a - a^\dagger) \quad , \tag{2.16}$$

so ergibt sich für den Vernichtungsoperator:

$$a = \frac{X_1 + iX_2}{2} \quad . \tag{2.17}$$

Führt man also eine Basistransformation aus, so beschreiben die dimensionslosen Operatoren $X_1$ und $X_2$ die Amplituden der beiden Quadraturphasen dieser Mode. Die Varianzen von Phase und Amplitude des *elektrischen Feldes* lassen sich beispielsweise durch die Unschärfe dieser Feldquadraturen angeben.

Anschaulich werden diese Quadraturen in der Gauss'schen Zahlenebene dargestellt, um eben die Unschärfe von Zuständen zu visualisieren.

### 2.4.2. Kohärente Zustände

Durch die Arbeiten von Erwin Schrödinger 1926 [Sch26a] und Roy J. Glauber 1963 [Gla63] hat man einen Quantenzustand gefunden, dessen Erwartungswerte $\langle X\rangle$, $\langle P\rangle$ zu jedem Zeitpunkt den Variablen $x$ und $p$ der klassischen Bewegung des harmonischen Oszillator gleichen, sich also klassisch verhalten. Man spricht deshalb auch von *quasiklassischen Zuständen*. Da allerdings nach Heisenberg zwei kanonisch konjugierte Größen, also nichtkommutierende Observable, nicht gleichzeitig beliebig genau gemessen werden können [Hei27] und auch das perfekte Klonen eines unbekannten Zustandes nach dem *Quantum no-cloning theorem* nicht möglich ist [WZ82], lassen sich keine quantenmechanischen Zustände mit exaktem Impuls und exaktem

Ort erzeugen (jedoch kann man bis auf die Unschärfe einen Zustand präparieren bzw. messen). Schrödinger hat nun nach einem Zustand minimaler Unschärfe gesucht und diesen gefunden, wenn man die Unschärfe in Ort und Impuls gleich wählt (dies sind jedoch nicht alle möglichen Zustände mit minimaler Unschärfe und nur ein Spezialfall der gequetschten Zustände (siehe Kapitel 2.4.3)). Man spricht von einem kohärenten Zustand $|\alpha\rangle$.

Für einen solchen Zustand gilt:

$$(P - \langle P \rangle)|\alpha\rangle = i(X - \langle X \rangle)|\alpha\rangle \quad . \tag{2.18}$$

Dies kann man sehr anschaulich verdeutlichen, indem man die Feldquadraturen (Kapitel 2.4.1) als Real- und Imaginärteil mit deren jeweiliger Unschärfe in die Gauss'sche Zahlenebene projeziert. Die mögliche Position des Aufenthaltes in diesem Phasenraum lässt sich durch einen Kreis abgrenzen.

Ein kohärenter Zustand kann erreicht werden über eine Verschiebung des *Vakuum-Zustandes* $|0\rangle$. Man führt dazu den Verschiebeoperator $D(\alpha)$ ein mit:

$$D(\alpha) = \mathrm{e}^{\alpha a^\dagger - \alpha^* a} \quad , \tag{2.19}$$

welcher nun jeden kohörenten Zustand aus dem Vakuum-Zustand erzeugen kann.

$$|\alpha\rangle = D(\alpha)|0\rangle \quad . \tag{2.20}$$

Kohärente Zustände sind auch die Eigenzustände des Vernichtungsoperators.

$$a|\alpha\rangle = \alpha|\alpha\rangle \quad . \tag{2.21}$$

Für den Vernichtungsoperator gilt:

$$a|n\rangle = \sqrt{n}|n-1\rangle \quad \text{für } n \neq 0 \quad \text{und} \tag{2.22}$$

$$a|0\rangle = 0 \quad . \tag{2.23}$$

Also kann man die Eigenwertgleichung

$$a|\alpha\rangle = \alpha|\alpha\rangle \quad \text{mit} \tag{2.24}$$

$$|\alpha\rangle = \sum_{n=0}^{\infty} c_n |n\rangle \tag{2.25}$$

lösen:

$$\alpha|\alpha\rangle = \sum_{n=0}^{\infty} c_n a|n\rangle \tag{2.26}$$

$$= \sum_{n=1}^{\infty} c_n \sqrt{n}|n-1\rangle \tag{2.27}$$

$$= \sum_{n=0}^{\infty} \underbrace{c_{n+1}\sqrt{n+1}}_{\alpha c_n}|n\rangle \quad , \tag{2.28}$$

woraus folgt:

$$c_1 = \alpha c_0 \quad , \tag{2.29}$$

und durch Rekursion:

$$c_n = c_0 \frac{\alpha^n}{\sqrt{n!}} \quad . \tag{2.30}$$

Durch Normierung erhält man:

$$\sum_{n=0}^{\infty} |c_n|^2 = 1 \tag{2.31}$$

$$\Rightarrow |c_0|^2 = \mathrm{e}^{-|a|^2} \quad . \tag{2.32}$$

Man kann nun den Koeffizienten $c_0$ bis auf eine Phase bestimmen, diese aber frei festlegen und wählt o. B .d .A. 0. Somit ergibt sich für die Darstellung eines kohärenten Zustandes in der Fock-Basis:

$$|\alpha\rangle = \exp^{-\frac{\alpha^2}{2}} \sum_{n=0}^{\infty} \frac{\alpha^n}{\sqrt{n!}} |n\rangle \quad . \tag{2.33}$$

### Wahrscheinlichkeitsverteilung

Die Wahrscheinlichkeitsverteilung der Photonenanzahl eines kohärenten Zustandes ist gegeben durch:

$$P(n) = |\langle n|\alpha\rangle|^2 = \frac{|\alpha|^{2n} \mathrm{e}^{-|\alpha|^2}}{n!} \quad , \tag{2.34}$$

und entspricht damit einer Poissonverteilung. Für den Erwartungswert der Photonenzahl gilt $\bar{n} = \langle \alpha | a^\dagger a | \alpha \rangle = |\alpha|^2$, also:

$$P(n) = \frac{\bar{n}^n \mathrm{e}^{-\bar{n}}}{n!} \quad . \tag{2.35}$$

### Überlapp

Der Überlapp zweier kohärenter Zustände $\langle \beta | \alpha \rangle$ ist gegeben durch:

$$\langle \beta | \alpha \rangle = \mathrm{e}^{-\frac{1}{2}(|\alpha|^2 + |\beta|^2)} \sum_{n,m=0}^{\infty} \frac{(\beta^*)^m}{\sqrt{m!}} \frac{\alpha^n}{\sqrt{n!}} \langle m|n \rangle \tag{2.36}$$

$$= \mathrm{e}^{-\frac{1}{2}(|\alpha|^2 + |\beta|^2)} \sum_{n=0}^{\infty} \frac{(\alpha\beta^*)^n}{\sqrt{n!}} \quad \text{mit} \quad \langle m|n \rangle = \delta_{mn} \tag{2.37}$$

$$= \mathrm{e}^{-\frac{1}{2}(|\alpha|^2 + |\beta|^2 - 2\alpha\beta^*)} \neq \delta(\alpha - \beta) \quad . \tag{2.38}$$

Kohärente Zustände sind also nicht orthogonal. Für das Betragsquadrat gilt:

$$|\langle \beta | \alpha \rangle|^2 = \mathrm{e}^{-|\alpha - \beta|^2} \quad . \tag{2.39}$$

### Übervollständigkeit

Um nun zu zeigen, dass die kohärenten Zustände eine Basis bilden, zeigt man deren Vollständigkeit:

$$\int |\alpha\rangle\langle\alpha| \mathrm{d}^2\alpha = \int \mathrm{d}^2\alpha \mathrm{e}^{-|\alpha|^2} \sum_{n,m=0}^{\infty} \frac{\alpha^n(\alpha^*)^m}{\sqrt{n!m!}} |n\rangle\langle m| \tag{2.40}$$

$$= \sum_{n,m=0}^{\infty} \frac{|n\rangle\langle m|}{\sqrt{n!m!}} \int_0^{\infty} r\mathrm{d}r\mathrm{e}^{-r^2} r^{n+m} \underbrace{\int_0^{2\pi} \mathrm{d}\Theta \mathrm{e}^{i(n-m)}}_{2\pi\delta_{nm}} \tag{2.41}$$

$$= 2\pi \sum_{n=0}^{\infty} \frac{|n\rangle\langle n|}{n!} \int_0^{\infty} \mathrm{d}r r^{2n+1} \mathrm{e}^{-r^2} \tag{2.42}$$

$$= \pi \sum_{n=0}^{\infty} \frac{|n\rangle\langle n|}{n!} \underbrace{\int_0^{\infty} \mathrm{d}x \mathrm{e}^{-x} x^n}_{n!} \quad \text{mit } x = r^2 \tag{2.43}$$

$$= \pi \sum_{n=0}^{\infty} |n\rangle\langle n| \tag{2.44}$$

$$\int |\alpha\rangle\langle\alpha| \mathrm{d}^2\alpha = \pi \mathbb{1} \quad . \tag{2.45}$$

Die kohärenten Zustände bilden also eine Basis, die vollständig ist, d. h. jeder Vektor des Vektorraumes kann aus dieser Basis erzeugt werden. Allerdings sind die Basisvektoren nicht orthogonal und damit nicht linear unabhängig voneinander, eine Zerlegung in diese Basis ist also nicht eindeutig. Man spricht deshalb von einer Übervollständigkeit.

### Übergangswahrscheinlichkeit

Aus der Übervollständigkeit folgt:

$$\frac{1}{\pi} \int \mathrm{d}^2\alpha |\alpha\rangle\langle\alpha| = \mathbb{1} \quad . \tag{2.46}$$

Multipliziert man nun von rechts einen beliebigen, normierten Zustand $|\Psi\rangle$ und von links $\langle\Psi|$, so erhält man:

$$\frac{1}{\pi} \int \mathrm{d}^2\alpha |\langle\alpha|\Psi\rangle|^2 = |\Psi|^2 = 1 \tag{2.47}$$

und damit

$$\int \mathrm{d}^2\alpha \left| \frac{\langle\alpha|\Psi\rangle}{\sqrt{\pi}} \right|^2 = 1 \quad . \tag{2.48}$$

Dieser Ausdruck lässt sich nun als eine Normierung von Übergangswahrscheinlichkeiten deuten. Möchte man also den Ursprung des Zustandes $|\Psi\rangle$ betrachten und integriert über alle möglichen bedingten Wahrscheinlichkeiten von kohärenten Ausgangszuständen $|\alpha\rangle$, so muss dieses Integral 1 ergeben.

Damit ist die bedingte Wahrscheinlichkeit $P(\Psi|\alpha)$ allergings gegeben durch:

$$P(\Psi|\alpha) = \left|\frac{\langle\alpha|\Psi\rangle}{\sqrt{\pi}}\right|^2 \quad , \tag{2.49}$$

wobei der Faktor $\sqrt{\pi}$ aus der Übervollständigkeit kommt. Die Wahrscheinlichkeitsfunktion $P$ bezeichnet man auch als die Q-Funktion von Husimi-Kano (Kapitel 2.4.4).

### Gruppeneigenschaften

Man zählt den Verschiebeoperator zur *Heisenberg-Gruppe*, einer *Lie-Gruppe*.

Es gilt für den Verschiebeoperator in Normal- und Antinormalform:

$$D(\alpha) = \mathrm{e}^{-\frac{|\alpha|^2}{2}}\mathrm{e}^{\alpha a^\dagger}\mathrm{e}^{-\alpha^* a} \qquad \text{Normalform} \tag{2.50}$$

$$D(\alpha) = \mathrm{e}^{\frac{|\alpha|^2}{2}}\mathrm{e}^{-\alpha^* a}\mathrm{e}^{\alpha a^\dagger} \qquad \text{Antinormalform} \tag{2.51}$$

Es ist klar, dass für die inverse Operation $D^\dagger(\alpha) = D(-\alpha)$ gilt. Es folgt für die multiplikative Gruppenoperation:

$$D(\alpha)D(\beta) = \mathrm{e}^{\frac{1}{2}\left(|\alpha|^2-|\beta|^2\right)}\mathrm{e}^{-\alpha^* a}\mathrm{e}^{(\alpha+\beta)a^\dagger}\mathrm{e}^{-\beta^* a} \quad . \tag{2.52}$$

Mit der *Baker-Campbell-Hausdorff-Formel* $\mathrm{e}^A\mathrm{e}^B = \mathrm{e}^B\mathrm{e}^A\mathrm{e}^{[A,B]}$ und der Vertauschungsrelation $[a, a^\dagger] = 1$ lassen sich die letzten beiden Operatoren vertauschen mit:

$$D(\alpha)D(\beta) = \mathrm{e}^{\frac{1}{2}\left(|\alpha|^2+|\beta|^2+2\beta^*\alpha\right)}\mathrm{e}^{-(\alpha^*+\beta^*)a}\mathrm{e}^{(\alpha+\beta)a^\dagger} \quad . \tag{2.53}$$

Um nun wieder ein Gruppenelement zu erhalten, muss nur noch der Vorfaktor an den der Antinormalform angepasst werden, es gilt:

$$D(\alpha)D(\beta) = \mathrm{e}^{\frac{1}{2}(\beta^*\alpha-\beta\alpha^*)}\mathrm{e}^{\frac{1}{2}\left(|\alpha+\beta|^2\right)}\mathrm{e}^{-(\alpha^*+\beta^*)a}\mathrm{e}^{(\alpha+\beta)a^\dagger} \tag{2.54}$$

$$= \mathrm{e}^{\frac{1}{2}(\beta^*\alpha-\beta\alpha^*)}D(\alpha+\beta) \quad . \tag{2.55}$$

## 2.4.3. Gequetschte Zustände

Neben den kohärenten Zuständen haben auch die gequetschten (kohärenten) Zustände (engl.: *squeezed states*) minimale Unschärfe, allerdings ist diese nicht in allen Richtungen dieselbe. *Gequetsche Zustände* erzeugt man experimentell durch nichtlineare Effekte z. B. in einem optisch parametrischen Oszillator [JZY$^+$03, KTY$^+$06, TYAF05]. Erzeugt werden sie theoretisch durch den Quetschoperator:

$$S(\xi) = \mathrm{e}^{\frac{1}{2}\xi^* a^2-\frac{1}{2}\xi a^{\dagger 2}} \quad , \tag{2.56}$$

mit $\xi = r\mathrm{e}^{2i\phi}$, wobei die Amplitude die Quetschstärke und die Phase die Quetschrichtung angeben (siehe auch Abb. 2.1(b)).

### Gequetschtes Vakuum

In der Literatur wird unter einem gequetschten Zustand häufig nur gequetsches Vakuum verstanden, also ein Zustand in der Art $S(\xi)|0\rangle = |\xi\rangle$. Diese Zustände lassen sich nach [DF04], in der *Fock-Basis* schreiben als:

$$|\xi\rangle = \frac{1}{\sqrt{\cosh r}} \sum_{n=0}^{\infty} \frac{\sqrt{(2n)!}}{2^n n!} \left(-\mathrm{e}^{i2\phi} \tanh r\right)^n |2n\rangle \quad . \tag{2.57}$$

Wie man sieht, werden nur geradzahlige Photonenzahlen benötigt. Der im Vergleich zur Literatur zusätzliche Ausdruck $(-1)^n$ folgt daher, dass in [DF04] der Quetschoperator als $S'(\xi) = \exp\left((\xi a^{\dagger 2} - \xi^* a^2)/2\right)$ definiert ist. Der gequetschte Vakuumzustand ist ein Eigenzustand des Vernichtungsoperators $\beta$, den man durch eine *Bogoliubov-Transformation* aus den Erzeugungs- und Vernichtungsoperatoren $\alpha$ und $\alpha^*$ erhält [LK87]:

$$b = S^\dagger(\xi) a S(\xi) = a \cosh r - a^\dagger \mathrm{e}^{i2\phi} \sinh r \quad , \tag{2.58}$$

$$b^\dagger = S^\dagger(\xi) a^\dagger S(\xi) = a^\dagger \cosh r - a\mathrm{e}^{-i2\phi} \sinh r \quad . \tag{2.59}$$

Für den Überlapp zwischen zwei gequetschten Vakuumzuständen gilt damit [DF04]:

$$\langle \xi' | \xi \rangle = \left(\cosh r \cosh r' - \mathrm{e}^{i2(\phi - \phi')} \sinh r \sinh r'\right)^{-1/2} \quad . \tag{2.60}$$

### Gequetschter Zustand

Als *gequetschter Zustand* soll im Folgenden ein Zustand bezeichnet werden, der nach einer Quetschung des Vakuums verschoben wird. Man könnte auch von einem verschobenen, *gequetschten Vakuum* oder einem *kohärenten, gequetschten Zustand* sprechen. Allerdings bedarf es hier keiner Unterscheidung, ob der Zustand zuerst gequetscht oder zuerst verschoben wurde, nur ersteres wird betrachtet. In der Literatur findet sich leider keine eindeutige Definition. Häufig werden alle diese Zustände als gequetschte kohärente Zustände zusammengefasst aufgrund der folgenden Vertauschungsrelation. Allerdings soll hier sehr präzise unterschieden werden.

Ein gequetschter Zustand kann also erzeugt werden durch:

$$D(\alpha) S(\xi) |0\rangle \quad . \tag{2.61}$$

Die Quetschung ist eine unitäre Operation, es gilt $S(\xi) S^\dagger(\xi) = 1$.

Häufig wird die Vertauschung des Quetsch- mit dem Verschiebeoperator benötigt, es gilt für $D(\alpha)S(\xi)$ mit $r \geq 0$:

$$D(\alpha) S(\xi) = S(\xi) S^\dagger(\xi) D(\alpha) S(\xi) \tag{2.62}$$

$$= S(\xi)\, \mathrm{e}^{S^\dagger(\xi)(\alpha a^\dagger - \alpha^* a) S(\xi)} \tag{2.63}$$

$$= S(\xi)\, \mathrm{e}^{(\alpha \cosh r + \alpha^* \mathrm{e}^{i2\phi} \sinh r) a^\dagger - (\alpha^* \cosh r + \alpha \mathrm{e}^{-i2\phi} \sinh r) a} \tag{2.64}$$

$$= S(\xi)\, \mathrm{e}^{\beta a^\dagger - \beta^* a} \tag{2.65}$$

$$= S(\xi) D(\beta) \quad \text{mit } \beta = \alpha \cosh r + \alpha^* \mathrm{e}^{i2\phi} \sinh r \quad . \tag{2.66}$$

### 2.4.4. Quasi-Wahrscheinlichkeitsverteilungen

Eugene Wigner hat 1932 erfolgreich versucht [Wig32] die Wellenfunktion aufgrund der physikalischen Interpretation, durch eine Phasenraumdichte zu ersetzen. Später haben auch Glauber [Gla63] bzw. Sudarshan [Sud63] und Husimi [Hus40] bzw. Kano noch weitere dieser Wahrscheinlichkeitverteilungen entdeckt. Man spricht von *Quasi-Wahrscheinlichkeitsverteilungen*, weil sie Eigenschaften haben, die normale Wahrscheinlichkeitsverteilungen nicht besitzen dürfen. Zudem handelt es sich um Distributionen. Diese Verteilungen sind allerdings gewöhnlich auf 1 normiert.

Man kann sie analog zu normalen Wahrscheinlichkeitsverteilungen als die *Fouriertransformation* der *charakteristischen Funktion* auffassen. Die zugehörige *charakteristische Funktion* ist eine Zerlegung des Dichteoperators. Genauer: betrachtet man ein quantenmechanisches System aus vielen Teilchen, z. B. ein *elektromagnetisches Feld*, so ist die Dichtematrix in der *Fock-Basis* nur sehr schwer zu bestimmen. Allerdings können mit einer passenden Wahrscheinlichkeitsverteilung statistische Aussagen getroffen werden, solange man sich nur für die Verteilungen der Teilchenzahlen interessiert. Für die Wahrscheinlichkeitsverteilung $P_n$ lassen sich die Diagonalelemente des Dichteoperators mit $\rho = \sum_{n=0}^{\infty} P_n |n\rangle\langle n|$ angeben. Geht man nun von einem Feld mit einer Poisson-Verteilung der Teilchenzahlen pro Mode aus, so kann man sich der Basis der kohärenten Zustände bedienen, die genau diese Bedingung erfüllt (2.35). Zerlegt man nun den Dichteoperator in diese Basis, so erhält man die *charakteristische Funktion*. Die einzelnen Verteilungen unterscheiden sich mathematisch nur darin, ob die zugehörige *charakteristische Funktion* normal, symmetrisch oder antinormal geordnet ist. Für die *charakteristischen Funktionen* gilt:

$$\chi_N(\eta) = \mathrm{Spur}\left(\rho \mathrm{e}^{\eta a^\dagger} \mathrm{e}^{-\eta^* a}\right) \qquad \text{normal geordnet} \tag{2.67}$$

$$\chi_S(\eta) = \mathrm{Spur}\left(\rho \mathrm{e}^{\eta a^\dagger - \eta^* a}\right) \qquad \text{symmetrisch geordnet} \tag{2.68}$$

$$\chi_A(\eta) = \mathrm{Spur}\left(\rho \mathrm{e}^{-\eta^* a} \mathrm{e}^{\eta a^\dagger}\right) \qquad \text{antinormal geordnet} \quad , \tag{2.69}$$

wobei man die (anti-)normal geordneten Funktionen durch die *Baker-Campbell-Hausdorff-Formel* $\mathrm{e}^{A+B} = \mathrm{e}^A \mathrm{e}^B \mathrm{e}^{-[A,B]/2}$ erhält (für $[A,[A,B]] = [B,[A,B]] = 0$).

#### Wigner-Funktion

Die von Wigner entdeckte Distribution erhält man durch die *Fouriertransformation* der symmetrisch geordneten charakteristischen Funktion:

$$W(\alpha) = \frac{1}{\pi^2} \int \mathrm{e}^{\eta^* \alpha - \eta \alpha^*} \chi_S(\eta) \mathrm{d}^2\eta \quad . \tag{2.70}$$

Eine besondere Eigenschaft der Wigner-Funktion, die sie eben auch zu einer Quasi-Wahrscheinlichkeitsverteilung macht, ist die Tastsache, dass sie auch negativ sein kann.

#### Kohärenter Zustand

Die Wignerfunktion wird häufig zur anschaulichen Darstellung von Zuständen im Phasenraum verwendet, sie lässt sich nach [WM95] für einen kohärenten Zustand mit $|\alpha\rangle = |\frac{1}{2}(X_1 + iX_2)\rangle$

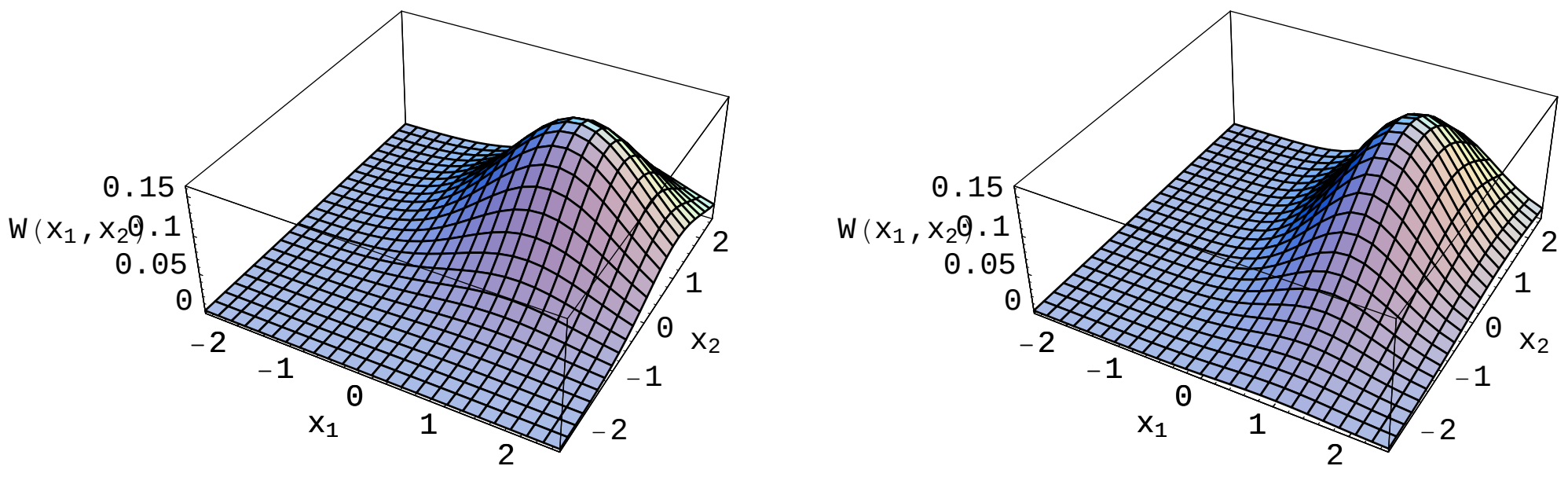

(a) (b)

Abbildung 2.1.: Abb. (a) zeigt einen Wigner-Plot für einen kohärenten Zustand bei $(1,1)$, in Abb. (b) wurde dieser Zustand zusätzlich gequetscht mit $r = 0,3$ und $\phi = 0$.

schreiben, als:

$$W(x_1, x_2) = \frac{1}{2\pi} \mathrm{e}^{-\frac{1}{2}\left({x'_1}^2 + {x'_2}^2\right)} \quad , \tag{2.71}$$

mit $x'_i = x_i - X_i$, wobei $|x_i\rangle$ die Eigenzustände der $X_i$ bezeichnen. Abb. 2.1(a) zeigt einen Plot der Wigner-Funktion für einen kohärenten Zustand bei $(1,1)$.

### Gequetschter Zustand

Für einen *gequetschten Zustand* gilt, ebenfalls nach [WM95]:

$$W(x_1, x_2) = \frac{1}{2\pi} \mathrm{e}^{-\frac{1}{2}\left({x'_1}^2 \mathrm{e}^{2r} + {x'_2}^2 \mathrm{e}^{-2r}\right)} \quad . \tag{2.72}$$

Abb. 2.1(b) zeigt einen Plot der Wigner-Funktion für einen *gequetschten Zustand* bei $(1,1)$ mit einer Quetschung von $r = 0,3$. Man sieht, dass die Wahrscheinlichkeitsverteilung in $X_1$-Richtung schmaler wird, in $X_2$-Richtung dafür aber breiter.

### P-Funktion

Die von Glauber und Sudarshan [Gla63, Sud63] eingeführte *P-Funktion* erhält man durch die Darstellung eines Zustandes in der übervollständigen, nicht-orthogonalen Basis der kohärenten Zustände:

$$\rho = \int P(\alpha) |\alpha\rangle\langle\alpha| \mathrm{d}^2\alpha \quad . \tag{2.73}$$

Da die Projektoren $|\alpha\rangle\langle\alpha|$ nicht orthogonal sind, kann diese Funktion nicht als klassische Wahrscheinlichkeitsfunktion interpretiert werden. Zudem kann man Quantenzustände erzeugen, für deren Darstellung die P-Funktion negative Werte annimmt.

Die P-Funktion lässt sich als Fourier-Transformation der Normal-geordneten *charakteristischen Funktion* des Dichteoperators darstellen mit:

$$P(\alpha) = \frac{1}{\pi^2}\int e^{\eta^*\alpha-\eta\alpha^*}\chi_N(\eta)d^2\eta \quad . \tag{2.74}$$

### Q-Funktion

Betrachtet man nun die Diagonalelemente des Dichteoperators in der Basis der kohärenten Zustände, so erhält man die von Husimi [Hus40] und Kano beschriebene Q-Funktion:

$$Q(\alpha) = \frac{\langle\alpha|\rho|\alpha\rangle}{\pi} \quad . \tag{2.75}$$

Da der Dichteoperator positiv ist, ist die Q-Funktion die einzige dieser drei Funktionen, die nicht negativ sein kann. Analog lässt diese sich wiederum erzeugen aus der Fourier-Transformation der Antinormal-geordneten *charakteristischen Funktion* des Dichteoperators:

$$Q(\alpha) = \frac{1}{\pi^2}\int e^{\eta^*\alpha-\eta\alpha^*}\chi_A(\eta)d^2\eta \quad . \tag{2.76}$$

## 2.5. Quanteninformation

Um Kommunikation auf Quantenebene sinnvoll ermöglichen zu können, muss eine Theorie ähnlich der *Informationstheorie* beschrieben werden. Dies ist die Theorie der *Quanteninformation*. Man kann also Begriffe wie *Quanteninformation*, *Quantenkanal* oder *„Quanteninformationsgehalt“* definieren. Es ergeben sich interessante neue Möglichkeiten, die nur aufgrund quantenmechnischer Effekte, wie der Verschränkung, dem Superpositionsprinzip oder der Messunschärfe zweier konjugierter Observablen funktionieren. Die Informationseinheit in der Quanteninformation ist das *Qubit*. Weiterhin ist die Arbeit von Bell, v. a. die *Bellsche Ungleichung* zu nennen [Bel64], welche die *Quantenkryptographie* erst im heutigen Verständnis ermöglicht.

### 2.5.1. Von-Neumann-Entropie

Das Analogon zur *Shannon-Entropie* in der Informationstheorie ist die *von-Neumann-Entropie* [vN32] in der Quanteninformation. Die Aufgabe der Zufallsvariablen übernimmt der Dichteoperator, und es gilt:

$$\rho = \sum_{i=1}^{n} p_i\rho_i \quad . \tag{2.77}$$

Man wählt also mit einer gewissen Wahrscheinlichkeit einen Systemzustand aus. Durch die abstrakte, basisunabhängige Natur des Dicheoperators lässt sich nun die *von-Neumann-Entropie* für jeden Dichteoperator schreiben, als:

$$S(\rho) = -\operatorname{Spur}(\rho\log\rho) \quad . \tag{2.78}$$

Für eine orthonormale Basis $\{|x\rangle\}$ mit der darin diagonalen Dichtematrix $\rho = \sum_x c_x |x\rangle\langle x|$ gilt:

$$S(\rho) = H(X) \quad . \tag{2.79}$$

Für den Fall, dass das Alphabet eine orthogonale Basis darstellt, reduziert sich die *von-Neumann-Entropie* also auf den klassischen Fall. Interessant wäre es nun, die Möglichkeiten zu betrachten, wenn eben dies nicht zutrifft und man ein Alphabet als Basis wählt, durch welches nicht mehr alle Zustände eindeutig voneinander unterschieden werden können. Eine solche nicht-orthogonale Basis ist im klassischen Fall nicht realisierbar.

### 2.5.2. Holevoinformation

Um die maximal zugängliche klassische Information auf einem Quantenkanal anzugeben, hat Alexander S. Holevo 1973 eine obere Schranke angegeben [Hol73]. Um seine Überlegung nachvollziehen zu können, wird hier ein Beweis in etwas abgeschwächter Form dargestellt. Da Alice und Bob nicht wissen, welche technischen Möglichkeiten Eve besitzt, wird diese Schranke häufig verwendet, um die Eve zugängliche Information des Quantenkanals zu bestimmen. Für die Entropie des von Alice präparierten Zustandes in einer orthogonalen Basis $\{|x\rangle\}$ gilt allgemein:

$$H(A) = S(\rho_A) = S\left(\sum_{x=0}^{\infty} p_x |x\rangle\langle x|\right) \quad . \tag{2.80}$$

Nimmt man nun an, der Quantenkanal $Q$ befinde sich in einem Zustand $\rho_x$ und Eve hat sein System bestenfalls mit $|0\rangle\langle 0|$ präpariert, so gilt für den Dichteoperator des Gesamtsystems:

$$\rho_{AQE} = \sum_{x=0}^{\infty} p_x |x\rangle\langle x|) \otimes \rho_x \otimes |0\rangle\langle 0| \quad . \tag{2.81}$$

Auch die Entropie des Quantenkanals lässt sich bestimmen zu

$$S(\bar{\rho}) := S(\rho_Q) = S(\rho_{QE}) = S\left(\sum_{x=0}^{\infty} p_x \rho_x\right) \quad , \tag{2.82}$$

wobei $\bar{\rho}$ als der gemittelte Dichteoperator des Quantenkanals interpretiert werden kann. Für die Entropie des Gesamtoperators gilt nun:

$$S(\rho_{AQE}) = S(\rho_{AQ}) \tag{2.83}$$

$$= -\sum_{x=0}^{\infty} \operatorname{Spur}(p_x \rho_x \log(p_x \rho_x)) \tag{2.84}$$

$$= -\sum_{x=0}^{\infty} (p_x \log(p_x) \operatorname{Spur}(\rho_x) + p_x \operatorname{Spur}(\rho_x \log(\rho_x))) \tag{2.85}$$

$$= H(A) + \sum_{x=0}^{\infty} p_x S(\rho_x) \quad . \tag{2.86}$$

Wird nun ein unitärer Messoperator in der Form $U(|\phi\rangle_Q \otimes |0\rangle_E) = \sum_y E_y|\phi_Q\rangle \otimes |y\rangle_E$ und der orthogonalen Messbasis $\{|y\rangle\}$ auf dieses System angewendet (dieser wirkt natürlich nicht auf die Präparation), so gilt für den Systemzustand:

$$U(\rho_{AQE}) = \rho'_{AQE} = \sum_{x,y,y'=0}^{\infty} p_x|x\rangle\langle x| \otimes E_y\rho_x E_{y'} \otimes |y\rangle\langle y'| \quad . \tag{2.87}$$

Da die von-Neumann Entropie invariant unter Basistransformationen ist, gilt

$$S(\rho'_{AQE}) = S(\rho_{AQE}) \quad , \tag{2.88}$$

und ebenso

$$S(\rho'_{QE}) = S(\rho_{QE}) = S(\bar{\rho}) \quad . \tag{2.89}$$

Bildet man jetzt die Teilspur über den Quantenkanal, so erhält man

$$S(\rho'_{AE}) = S\left(\sum_{x,y} p_x \operatorname{Spur}(E_y\rho_x)|x\rangle\langle x| \otimes |y\rangle\langle y|\right) \tag{2.90}$$

$$= S\left(\sum_{x,y} p(x,y)|x\rangle\langle x| \otimes |y\rangle\langle y|\right) \tag{2.91}$$

$$= H(A,E) \quad . \tag{2.92}$$

Analog zu $\rho_A$ gilt nun $\rho'_E = H(E)$. An dieser Stelle wird der Satz im Vergleich zu Holevos Beweisführung stark verkürzt, indem die starke Subadditivität der Entropie verwendet wird, denn dies bedeutet:

$$S(\rho'_{AQE}) + S(\rho'_E) \leq S(\rho'_{AE}) + S(\rho'_{QE}) \quad , \tag{2.93}$$

was durch Einsetzen der berechneten Entropien auf

$$H(A) + \sum_{x=0}^{\infty} p_x S(\rho_x) + H(E) \leq H(A,E) + S(\bar{\rho}) \tag{2.94}$$

führt, und mit $I(A:E) = H(A) + H(E) - H(A,E)$ schließlich die Holevo-Schranke zeigt:

$$I(A:E) \leq S(\bar{\rho}) - \sum_{x=0}^{\infty} p_x S(\rho_x) =: \chi \quad . \tag{2.95}$$

### 2.5.3. Strahlteiler

Ein Strahlteiler (engl.: *beam-splitter*) ist in aller Regel ein halbdurchlässiger Spiegel, so wie er auch in Interferometern Verwendung findet. Ein Teil des Strahles wird reflektiert, ein anderer transmittiert (Abbildung 2.2).

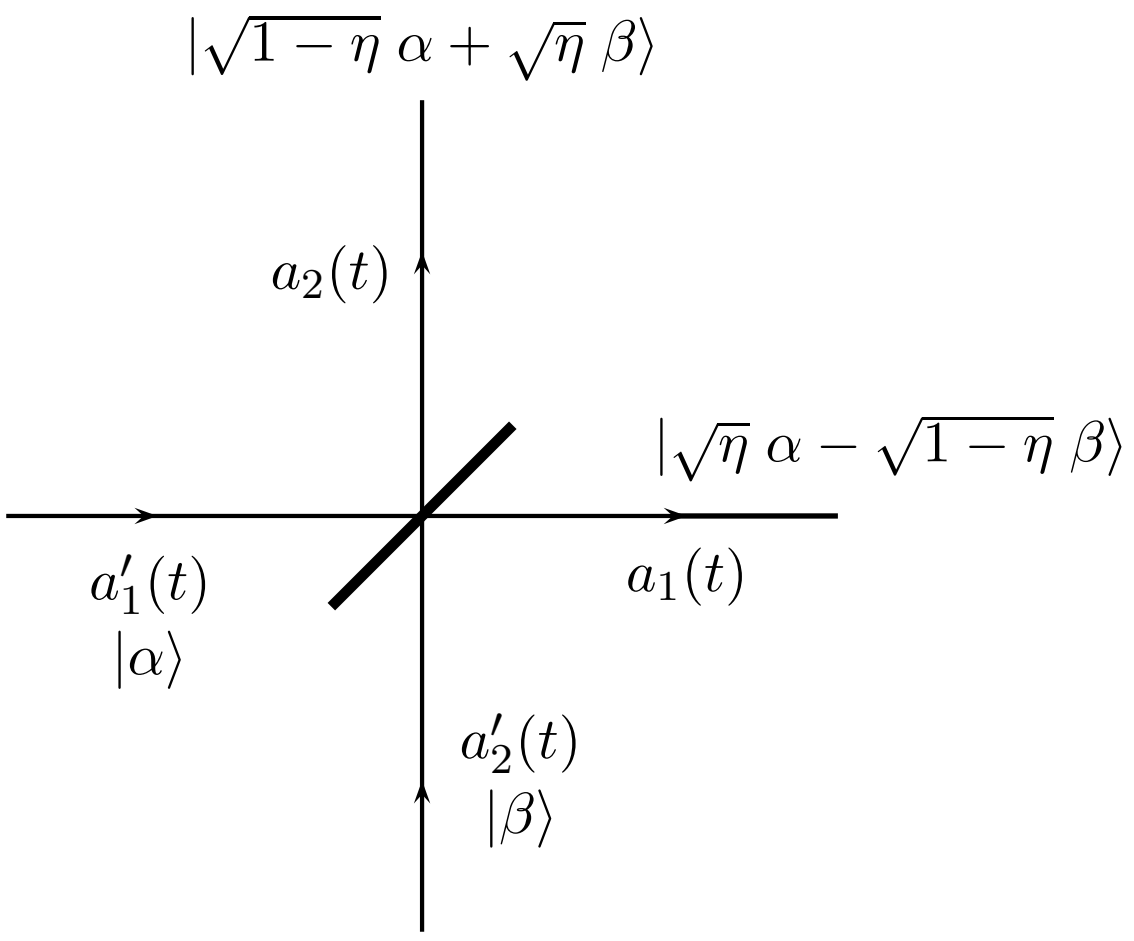

Abbildung 2.2.: Darstellung eines Strahlteilers.

Allgemein lässt sich ein Strahlteiler durch eine lineare Operation darstellen:

$$\begin{pmatrix} a_1 \\ a_2 \end{pmatrix} = \begin{pmatrix} t_1 & r_2 \\ r_1 & t_2 \end{pmatrix} \begin{pmatrix} a_1{}' \\ a_2{}' \end{pmatrix} \quad , \tag{2.96}$$

mit den Eingängen $a_1{}'$ und $a_2{}'$ und den Ausgängen $a_1$ und $a_2$. $t_i$ und $r_i$ sind jeweils die Transmissions- und Reflexionskoeffizienten. Da Energieerhaltung für beliebige Amplituden gelten muss,

$$|a_1|^2 + |a_2|^2 = |a_1{}'|^2 + |a_2{}'|^2 \quad , \tag{2.97}$$

erhält man für $t_i$ und $r_i$:

$$|t_1|^2 + |r_1|^2 = |t_2|^2 + |r_2|^2 = 1 \tag{2.98}$$

$$t_1 r_2^* + t_2^* r_1 = t_1^* r_2 + t_2 r_1^* = 0 \quad . \tag{2.99}$$

Interessant sind v. a. symmetrische Strahlteiler, für die $|t_1|^2 = |t_2|^2$ und $|r_1|^2 = |r_2|^2$ gilt. Betrachtet man nun den Transmissionskoeffizienten $\sqrt{\eta}$, so erhält man:

$$\begin{pmatrix} a_1 \\ a_2 \end{pmatrix} = \begin{pmatrix} \sqrt{\eta} & -\sqrt{1-\eta} \\ \sqrt{1-\eta} & \sqrt{\eta} \end{pmatrix} \begin{pmatrix} a_1{}' \\ a_2{}' \end{pmatrix} \quad . \tag{2.100}$$

Die Wahl des Transmissionkoeffizienten ergibt sich aus der Tatsache, dass man die Intensität betrachtet, für die gilt:

$$I_1 \propto |a_1|^2 = \eta a_1{}' + (1-\eta) a_2{}' \quad , \tag{2.101}$$

und analog für $I_2$. Das Minus bei $r_2$ folgt aus (2.99). Betrachtet man nun den Quantenmechanischen Zustand $|\Psi'\rangle = |\alpha\rangle_{1'}|\beta\rangle_{2'}$, aus zwei kohärenten Zuständen, so geht dieser durch den Strahlteiler in den Zustand $|\Psi\rangle$ über:

$$|\alpha\rangle_{1'}|\beta\rangle_{2'} \rightarrow |\sqrt{\eta}\,\alpha - \sqrt{1-\eta}\,\beta\rangle_1|\sqrt{1-\eta}\,\alpha + \sqrt{\eta}\,\beta\rangle_2 = |\Psi\rangle \quad . \tag{2.102}$$

Für den Sonderfall des 50:50-Strahlteilers, d. h. $50\%$ werden transmittiert und $50\%$ werden reflektiert, gilt:

$$|\Psi\rangle = \left|\frac{\alpha-\beta}{\sqrt{2}}\right\rangle_1 \left|\frac{\alpha+\beta}{\sqrt{2}}\right\rangle_2 \quad . \tag{2.103}$$

Diese Art des Strahlteilers findet z. B. bei der Dual-Homodynmessung Einsatz (Kapitel 2.5.5).

#### Strahlteiler-Angriff

Alle Kanäle zwischen Alice und Bob sind öffentlich zugänglich. Ein Lauscher Eve hat also verschiedene Möglichkeiten, diese Kanäle zu benutzen, um Informationen über den sicheren Schlüssel in Erfahrung zu bringen. Auch ein aktives Beeinflussen der Messung oder der Präparation über diese Kanäle wäre denkbar. Im weitesten Sinne unterscheidet man zwischen individuellen Attacken [CLA01b, GG02, SRLL02, WLB$^+$04, NH04, GC04], kollektiven Attacken [Gro05, NA05] und kohärenten Attacken [GP01, IVAC04]. Der Strahlteiler-Angriff wird als kollektive Attacke bezeichnet.

Überträgt man Information über einen verlustbehafteten Quantenkanal, so kann man sich folgendes Angriffsszenario für einen Lauscher Eve vorstellen: man erlaubt Eve, eine bessere Technologie zu besitzen, mit der er in der Lage ist, verlustfreie Quantenkanäle herzustellen. Nun können Alice und Bob keinen Unterschied darin feststellen, ob sie einen verlustbehafteten Kanal nutzen oder ob Eve diesen Kanal durch einen verlustfreien ersetzt und so mit einem Strahlteiler versehen hat, dass die Transmissionsrate die Gleiche ist.

Die Information aus dem zweiten Ausgang des Strahlteilers geht nun an Eve, und er kann Messungen am System anstellen, ohne dass Alice oder Bob etwas davon wissen.

Dieser Strahlteiler wäre also aus dem kohärenten Zustand $|\alpha\rangle$, der von Alice präpariert wird und einem kohärenten Vakuumzustand $|0\rangle$ aufgebaut. Es ergibt sich somit:

$$|\alpha\rangle_{1'}|0\rangle_{2'} \rightarrow |\sqrt{\eta}\,\alpha - \sqrt{1-\eta}\,0\rangle_1|\sqrt{1-\eta}\,\alpha + \sqrt{\eta}\,0\rangle_2 = |\sqrt{\eta}\,\alpha\rangle_B|\sqrt{1-\eta}\,\alpha\rangle_E \quad , \tag{2.104}$$

d. h. Bob erhält den Zustand $|\sqrt{\eta}\,\alpha\rangle_B$ und Eve den Zustand $|\sqrt{1-\eta}\,0\rangle_E$.

### 2.5.4. Homodynmessung

Mit der Homodynmessung wird eine Feldquadratur, projeziert auf eine Achse im Phasenraum, gemessen. Sie besteht in der Regel aus einem 50:50-Strahlteiler und zwei Photodetektoren. Auf einen Eingang wird das zu messende Signal gelegt, auf den anderen ein lokaler Oszillator. Die Amplitude der Feldquadratur ist proportional zur Differenz der beiden Photodetektoren. Mit einer Phase des Strahlteilers wird das Variieren der Feldachse erlaubt.

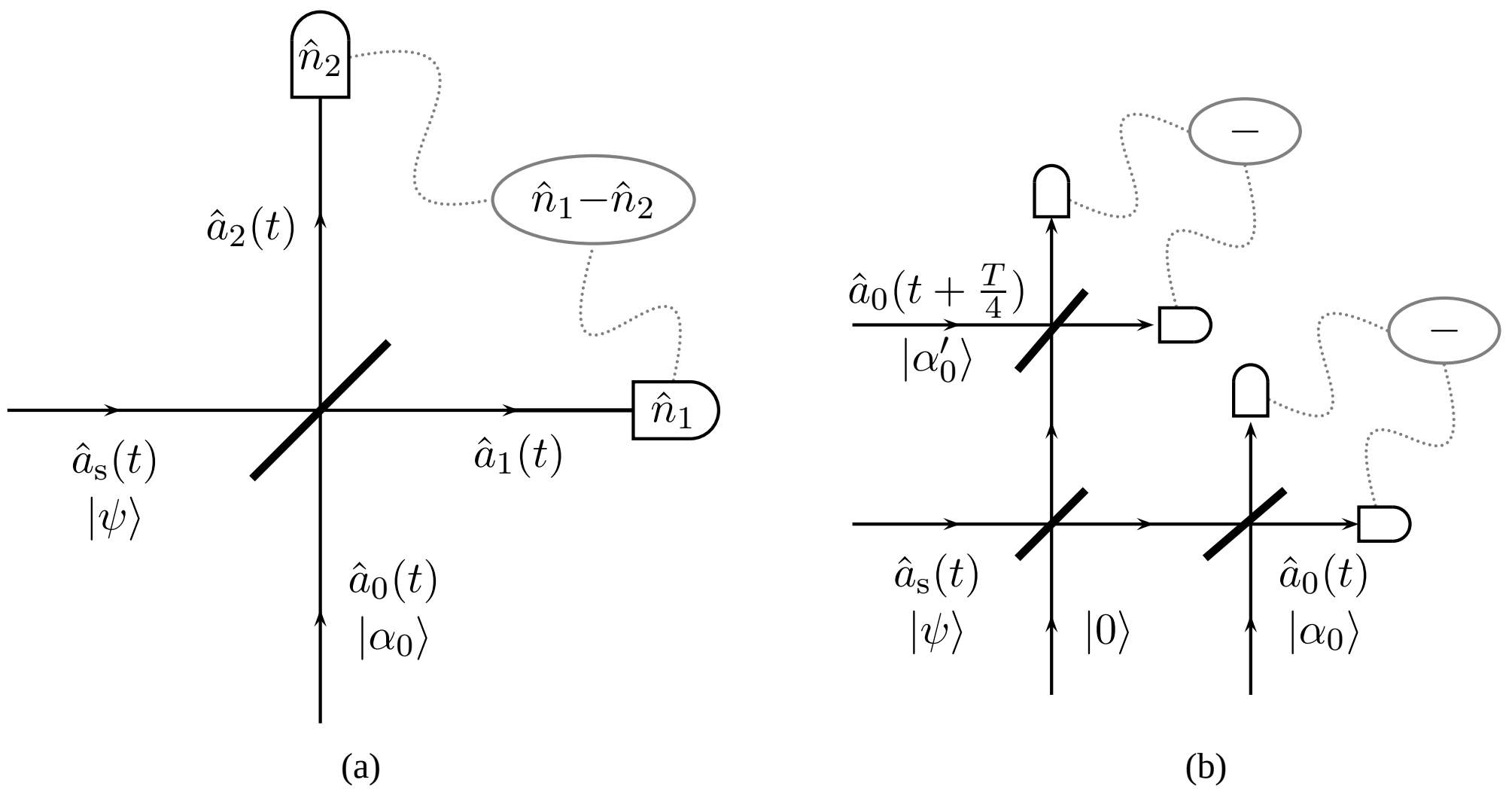

Abbildung 2.3.: Abb. (a) zeigt ein Homodynexperiment, Abb. (b) ein Dual-Homodynexperiment.

Fällt nun ein poissonverteiles Feld, wie z. B. ein kohärentes Feld (2.35) auf einen Photodetektor, der jedes einfallende Photon mit einer Wahrscheinlichkeit $\zeta$ misst, so gilt nach [Man] bzw. [KK64]

$$P(n) = \left\langle : \frac{(\zeta\hat{n})^n}{n!} e^{-\zeta\hat{n}} : \right\rangle \tag{2.105}$$

für den Zustand $\rho$, wobei :: Normalordnung bedeutet.

Betrachtet man nun die beiden Zustände $\rho_1$ und $\rho_2$ als Eingänge eines Strahlteilers, so kann man auch die Wahrscheinlichkeit für zwei Detektoren aufstellen:

$$P(n_1, n_2) = \left\langle : \frac{(\zeta\hat{n}_1)^{n_1}}{n_1!} e^{-\zeta\hat{n}_1} \frac{(\zeta\hat{n}_2)^{n_2}}{n_2!} e^{-\zeta\hat{n}_2} : \right\rangle \quad . \tag{2.106}$$

Ein schematischer Aufbau des Homodynexperimentes findet sich in Abb. 2.3(a). Da man nun an der Differenz der beiden Photodetektoren interessiert ist, also $n = n_1 - n_2$, muss man über alle möglichen Photonenzahlen $n_2$ addieren, um die Wahrscheinlichkeit einer bestimmen Differenz zu berechnen:

$$P(n) = \sum_{k=0}^{\infty} P(n+k, k) = \left\langle : \mathrm{e}^{-\zeta(\hat{n}_1+\hat{n}_2)} \sum_{k=0}^{\infty} \frac{(\zeta\hat{n}_1)^{n+k}(\zeta\hat{n}_2)^k}{(n+k)!k!} : \right\rangle \quad . \tag{2.107}$$

Man erhält nach [YS80] für $n \ll n_1, n_2$ mit $\rho_1 = |\alpha\rangle\langle\alpha|$ und $\alpha = |\alpha|\mathrm{e}^{i\vartheta}$:

$$P_\vartheta(n) \to \mathrm{Spur}_2 \left( \rho_2 : \frac{1}{\sqrt{2\pi\zeta|\alpha|^2}} \exp\left( -\frac{(n + \zeta a_2(\vartheta)|\alpha|)^2}{2\zeta|\alpha|^2} \right) : \right) \quad , \tag{2.108}$$

mit $a_2(\vartheta) := a_2\mathrm{e}^{-i\vartheta} + a_2^\dagger\mathrm{e}^{i\vartheta}$.

### 2.5.5. Dual-Homodynmessung

Die Dual-Homodynmessung erlaubt das gleichzeitige Messen zweier konjugierter Feldquadraturen. Ein schematischer Aufbau des Dual-Homodynexperimentes findet sich in Abb. 2.3(b). Das zu messende Signal wird dafür mit einem $50{:}50$-Strahlteiler in zwei Signale zerlegt, wobei als zweiter Eingang nur der quantenmechanische Vakuumzustand gewählt wird. Ein Ausgangssignal wird dann mithilfe eines Homodynexperimentes auf einer Achse des Phasenraumes gemessen. Das zweite Ausgangssignal geht zu einem zweiten Homodynexperiment, an welchem die dazu senkrechte Achse als Messachse gewählt wird, indem der lokale Oszillator um $\pi/2$ phasenverschoben wird. Aufgrund des Vakuumrauschens am Strahlteiler wird die Heisenberg'sche Unschärferelation nicht verletzt, es können beide Feldquadraturen gleichzeitig gemessen werden. Natürlich ist es auch möglich, einen asymetrischen Strahlteiler zu verwenden.

# 3. Protokolle - status quo

## 3.1. Bekannte Protokolle

### 3.1.1. BB84-Protokoll

Das erste und bekannteste quantenkryptographische Protokoll ist das 1984 von Charles Bennett und Gilles Brassard entwickelte *BB84*-Protokoll [BB84]. Alice benötigt eine 1-Photonen-Quelle, die zufällig polarisierte Photonen erzeugt. Nun misst sie dieses Photon zufällig in einer von zwei Basen (einer Basis $\{|0\rangle, |1\rangle\}$ und einer $45°$ dazu verdrehten Basis $\{|+\rangle, |-\rangle\}$). Alice erzeugt auf diese Weise einen der vier Zustände

$$|0\rangle \tag{3.1}$$

$$|1\rangle \tag{3.2}$$

$$|+\rangle = \frac{1}{\sqrt{2}}\left(|0\rangle + |1\rangle\right) \tag{3.3}$$

$$|-\rangle = \frac{1}{\sqrt{2}}\left(|0\rangle - |1\rangle\right) \quad . \tag{3.4}$$

Nun schickt Alice das präparierte Photon über den Quantenkanal an Bob. Bob wählt ebenfalls bei jedem Photon zufällig zwischen einer der beiden Basen aus und misst den Zustand des Photons. P. Shor und J. Preskill konnten im Jahre 2000 zeigen, dass das BB84-Protokoll sicher ist [SP00].

#### Postprocessing

Nachdem genügend Photonen ausgetauscht wurden, beginnt eine zweite Phase. Alice und Bob vergleichen bei jeder Messung ihre Basen und behalten nur die Messungen, bei denen sie sich zufällig für die Gleiche entschieden haben, also im Mittel $50\%$. Danach vergleichen sie ebenfalls öffentlich stichprobenartig einige Messungen, um zu sehen, ob der Quantenkanal funktioniert bzw. um außschliessen zu können, dass es einen Lauscher geben kann. Danach führen Alice und Bob häufig noch verschiedene Fehlerkorrekturverfahren durch.

Wenn es einen Lauscher Eve gibt, so weiß er nicht, in welcher der beiden Basen das Photon präpariert wurde. Da Eve durch das Messen des Photons, dessen eigentlichen Zustand zerstört, kann er nicht sicher sein, ob sie wirklich den richtigen Zustand gemessen hat. Falls Alice und Bob hinterher feststellen, dass sie häufig in der gleichen Basis gemessen haben, aber nur sehr selten den gleichen Zustand hatten, können sie darauf schließen, dass es einen Lauscher gab, der in der falschen Basis gemessen hat - logischerweise würde auch dies im Mittel in $50\%$ der Fälle auftreten.

### 3.1.2. 6-State-Protokoll

Das 6-State-Protokoll ist eine Erweiterung des BB84-Protokolls. Es wurde 1998 von D. Bruß entwickelt [Bru98] und nutzt die Eigenschaft, dass die dreidimensionale Bloch-Sphäre den Zustand eines Qubits vollständig beschreibt und deren Grundlage die Pauli-Matritzen bilden. Da es sich aber um drei konjugierte Basen handelt, lassen sich 6 verschiedene Zustände beschreiben und so auch höhere Datenraten erreichen.

Es werden die 4 Zustände des BB84-Protokolls verwendet und noch zwei weitere:

$$|\underline{+}\rangle = \frac{1}{\sqrt{2}}\left(|0\rangle + i|1\rangle\right) \tag{3.5}$$

$$|\underline{-}\rangle = \frac{1}{\sqrt{2}}\left(|0\rangle - i|1\rangle\right) \quad . \tag{3.6}$$

Ein Sicherheitsbeweis des Protokolls stammt von H. Lo [Lo01], eine Erweiterung mit besseren Raten bisher nur als eprint von J. Renes und O. Kern [KR07].

### 3.1.3. Ausblick

Das Problem beider Protokolle besteht vor allem in der experimentellen Umsetzbarkeit. Alice benötigt eine 1-Photonen-Quelle, welche beim derzeitigen Stand der Wissenschaft nur mit äußerst präzisen und damit auch teuren Bauteilen zu realisieren ist. Um sicherzustellen, dass wirklich immer nur ein Photon erzeugt wird (sonst könnte Eve einfach ein zweites entnehmen, und der Schlüssel wäre unsicher), werden nur die Fälle betrachtet, bei denen man sich sicher ist, dass es kein zweites Photon gab. Somit wird die Ausbeute sehr gering und das Verfahren zu langsam und damit unpraktikabel. Allerdings wird auch heute noch auf dem Gebiet der 1-Photonen-Quelle geforscht ([KMZW00, KHR02, CPKK04]), denn für die vorhandenen Protokolle existieren Sicherheitsbeweise.

## 3.2. Kontinuierliche Variablen

Im Gegensatz zu den klassischen Verfahren der Quanten-Schlüsselverteilung (engl.: *quantum key distribution* (QKD)), welche die Informationen stets in diskreten Variablen wie der Polarisation (horizontal / vertikal) oder einem Energieniveau (angeregt / nicht angeregt) speichern, gibt es neuere Verfahren, die sich kontinuierlicher Variablen bedienen. Konkret bedeutet dies, dass man z. B. eine Phase oder eine Amplitude zum Speichern von Informationen nutzt, welche kontinuierlich geändert werden kann. Ein klarer Vorteil besteht in der experimentellen Durchführbarkeit und Fehlertoleranz.

Die experimentelle Realisierung eines 1-Photonen-Lasers schreitet immer weiter voran. Wenn man aber z. B. sicherstellen will, dass nie mehr als ein Photon den Laser verlässt, so hat man sehr große Intervalle zwischen zwei Impulsen. Ähnlich sieht es auf der Seite der Messung aus. Während das Nachweisen von einzelnen Photonen, z. B. auch über sehr lange Quantenkanäle sehr schwierig ist, gelingt das Messen von kontinuierlichen Parametern mittels Homodyn- oder Heterodyn-Experimenten sehr viel zuverlässiger.

Natürlich ergeben sich durch diese Methoden völlig neue Angriffsszenarien, die diskutiert werden müssen, wie die Strahlteiler-Attacke.

Die ersten QKD Protokolle mit kontinuierlichen Variablen wurden Ende der 1990er Jahre [BK98, FSBF98] mit nichtklassischem Licht, häufig mit *gequetschen Zuständen* oder mit EPR-Paaren, entwickelt [GP01, Hil00, CLA01a]. Da das experimentelle Arbeiten mit *gequetschtem Licht* schwierig ist (höhere Beugungsordnungen gehen z. B. durch Linsenoptik verloren und das Licht wird kohärent), haben Frédéric Grosshans und Philippe Grangier [GG02] untersucht, ob man auch mithilfe von *kohärentem Licht* sichere QKD-Protokolle entwickeln kann.

Die im späteren Verlauf diskutierte Betrachtungsweise des Protokolles bezieht sich v. a. auf die Arbeit von Matthias Heid und Norbert Lütkenhaus [HL06].

### 3.2.1. Beschreibung des Protokolls von Grosshans und Grangier

Das Protokoll von Grosshans und Grangier behandelt nun folgende Situation: Alice präpariert einen der beiden *kohärenten Zustände*

$$|\underline{0}\rangle = |+\alpha\rangle \quad , \tag{3.7}$$

$$|\underline{1}\rangle = |-\alpha\rangle \quad , \tag{3.8}$$

mit reellem $\alpha$ und gleicher Wahrscheinlichkeit $p = 1/2$. Dieses Signal schickt Alice an Bob. Bob misst nun dieses Signal mittels Homodynmessung (Kapitel 2.5.4) auf der reellen Achse und bildet es somit auf einen *kohärenten Zustand* $|\beta\rangle = |\beta_x + i\beta_y\rangle$ ab. Weiterhin ist der Quantenkanal verlustbehaftet, allerdings gibt es keinerlei Rauschen.

### 3.2.2. Strahlteiler-Angriff

Erlaubt man Eve einen Strahlteiler-Angriff (Kapitel 2.5.3), so führt dies zu folgendem Szenario:

$$|\pm\alpha\rangle_A \rightarrow |\pm\sqrt{\eta}\alpha\rangle_B \otimes |\pm\sqrt{1-\eta}\alpha\rangle_E \quad . \tag{3.9}$$

Nun gilt allerdings für einen von Alice präparierten Zustand $|\phi_i\rangle$, welcher von Bob als $|\Psi_i\rangle$ und von Eve als $|\epsilon_i\rangle$ gemessen wird:

$$U|\psi_i\rangle|\epsilon^0\rangle = |\Psi_i\rangle|\epsilon_i\rangle \quad , \tag{3.10}$$

mit einem Grundzustand von Eve $|\epsilon^0\rangle$. Damit gilt:

$$\langle\phi_i|\phi_j\rangle = \langle\Psi_i|\Psi_j\rangle\langle\epsilon_i|\epsilon_j\rangle \quad . \tag{3.11}$$

Da sowohl die Präparation als auch die Messung bekannt sind, ist es auch der Überlapp der für Eve zugänglichen Informationen. Deshalb kann hier die Holevoinformation (Kapitel 2.5.2) als obere Schranke betrachtet werden. Es gilt also für die sichere Schlüsselrate zwischen Alice und Bob:

$$G \geq I(A:B) - \chi \tag{3.12}$$

### 3.2.3. Transinformation zwischen Alice und Bob

Die Transinformation zwischen Alice und Bob lässt sich nach der Definition (2.12) berechnen mittels:

$$I(A:B) = H(A) - H(A|B) \quad . \tag{3.13}$$

Da Alice beide Zustände mit der gleichen Wahrscheinlichkeit $p = 1/2$ präpariert, gilt für deren Entropie:

$$H(A) = -p(\underline{0})\log_2 p(\underline{0}) - p(\underline{1})\log_2 p(\underline{1}) = -2p\log_2 p = 1 \quad . \tag{3.14}$$

Die bedingte Entropie $H(A|B)$ berechnet sich aus den *a-posteriori*-Wahrscheinlichkeiten. Zudem werden gemäß der Homodynmessung alle Zustände auf die reelle Achse abgebildet:

$$H(A|B) = -\int \mathrm{d}\beta_x p(\beta_x) \sum_{k=0}^{1} p(\alpha_k|\beta_x)\log_2 p(\alpha_k|\beta_x) \quad . \tag{3.15}$$

Wegen der Übergangswahrscheinlichkeit (2.49) zweier *kohärenter Zustände* und dem Überlapp (2.39) gilt für $p(\beta|\underline{0})$ und $p(\beta|\underline{1})$, da $\alpha$ reell ist:

$$p(\beta|\underline{0}) = \frac{1}{\pi}\mathrm{e}^{-\left((\sqrt{\eta}\alpha-\beta_x)^2+\beta_y^2\right)} \tag{3.16}$$

$$p(\beta|\underline{1}) = \frac{1}{\pi}\mathrm{e}^{-\left((\sqrt{\eta}\alpha+\beta_x)^2+\beta_y^2\right)} \quad . \tag{3.17}$$

Weiterhin kann man $p(\beta_x)$, nach dem Satz der *vollständigen Wahrscheinlichkeit* (2.3) und dem Integral $\int \mathrm{e}^{-y^2}\mathrm{d}y = \sqrt{\pi}$, berechnen zu:

$$p(\beta_x) = \int \mathrm{d}\beta_y p(\beta) = \frac{1}{2\sqrt{\pi}}\left(\mathrm{e}^{-(\sqrt{\eta}\alpha-\beta_x)^2} + \mathrm{e}^{-(\sqrt{\eta}\alpha+\beta_x)^2}\right) \quad . \tag{3.18}$$

Wie man sieht, sind beide Messungen für $\beta_x$ gleich wahrscheinlich. Man kann nun für jeden Absolutwert von $\beta_x$ einen effektiven Informationskanal beschreiben. Damit besteht für jeden Kanal die Wahl zwischen zwei verschiedenen Zuständen. Für spätere Methoden wie der *Postselektion* ist diese Betrachtungsweise wichtig. Aufgrund der mathematischen Symmetrie bedeutet sie keinerlei Einschränkung. Wichtig ist sie auch in der folgenen Überlegung: da Alice und Bob nach der Übertragung der Quantenbits von Alice zu Bob ihre Informationen mithilfe klassischer Bits abgleichen, macht es durchaus Sinn, einen Kanal auf diese Weise zu definieren. Abhängig davon, ob nun Alice diese klassischen Bits an Bob überträgt oder umgekehrt, spricht man von *reverse* oder *direct reconciliation*. Übertragen werden jeweils der Betrag von $\beta_x$ und $\beta_y$. Im Falle der *direct reconciliation* kann Bob nun mithilfe seiner Messung feststellen, welche Zustand Alice mit größerer Wahrscheinlichkeit präpariert hat. Im Falle der *reverse reconciliation* entscheidet Alice aufgrund dieser Parameter von Bob, welchen Zustand Bob vermutlich gemessen hat und passt dementsprechend ihren gespeicherten Wert für diese Übertragung an.

Nun lassen sich die *a-posteriori*-Wahrscheinlichkeiten schreiben als:

$$e^{+} = \frac{p(\beta_x|\underline{1})}{p(\beta_x|\underline{0}) + p(\beta_x|\underline{1})} \quad \text{und} \tag{3.19}$$

$$e^{-} = \frac{p(\beta_x|\underline{0})}{p(\beta_x|\underline{0}) + p(\beta_x|\underline{1})} \quad , \tag{3.20}$$

wobei leicht ersichtlich ist, dass gilt:

$$e := e^{+} = e^{-} = \frac{1}{1 + \mathrm{e}^{4\sqrt{\eta}\alpha|\beta_x|}} \quad , \tag{3.21}$$

und so für die bedingte Entropie eines Informationskanals folgt:

$$H(A|B) = -e \log_2(e) - (1-e)\log_2(1-e) \quad . \tag{3.22}$$

Die Wahrscheinlichkeit, dass ein solcher Kanal genutzt wird ergibt sich aufgrund der Symmetrie zu $p_c(\beta_x) = 2p(\beta_x)$, und damit gilt für die Transinformation:

$$I(A,B) = \int_0^\infty \mathrm{d}\beta_x p_c(\beta_x)(1 - H(A|B)) \quad . \tag{3.23}$$

### 3.2.4. Direct reconciliation

Im Falle des direkten Abgleichs (engl.: *direct reconciliation*) schickt Alice klassische Korrekturinformationen zu Bob, damit dieser seine Messung verbessern kann. Die Zustände von Eve nach dem Strahlteiler sind gegeben durch:

$$|\epsilon_i\rangle = |\pm\sqrt{1-\eta}\,\alpha\rangle \quad , \tag{3.24}$$

wobei dies reine Zustände sind und in der Betrachtung der Holeveinformation der Mischterm entfällt. Für den gemittelten Dichteoperator gilt nun:

$$\bar{\rho} = \frac{1}{2}\left(|\epsilon_0\rangle\langle\epsilon_0| + |\epsilon_1\rangle\langle\epsilon_1|\right) \quad . \tag{3.25}$$

Aufgrund der Symmetrie lassen sich diese Zustände zerlegen in eine orthonormale Basis der Form:

$$|\epsilon_0\rangle = c_0|\Phi_0\rangle + c_1|\Phi_1\rangle \quad \text{und} \tag{3.26}$$

$$|\epsilon_1\rangle = c_0|\Phi_0\rangle - c_1|\Phi_1\rangle \quad . \tag{3.27}$$

Damit ergibt sich für die Holevoinformation im Falle der *direct reconciliation*:

$$\chi^{\mathrm{DR}} = S(\bar{\rho}) = -\sum_{i=0}^{1} |c_i|^2 \log_2(|c_i|^2) \quad . \tag{3.28}$$

Mit der Normalisierung von $\rho$ ($|c_0|^2 + |c_1|^2 = 1$) und dem Überlapp beider Zustände von Eve, $\langle\epsilon_0|\epsilon_1\rangle = |c_0|^2 - |c_1|^2$ folgt für die Koeffizienten:

$$|c_0|^2 = \frac{1}{2}(1 + \langle\epsilon_0|\epsilon_1\rangle) \quad \text{und} \tag{3.29}$$

$$|c_1|^2 = \frac{1}{2}(1 - \langle\epsilon_0|\epsilon_1\rangle) \quad . \tag{3.30}$$

Damit lässt sich schliesslich wegen $\langle\epsilon_0|\epsilon_1\rangle = \mathrm{e}^{-2(1-\eta)\alpha^2}$ die Holevoinformation berechnen. Anzumerken sei noch, dass sie nicht von $\beta_x$ abhängig ist.

### 3.2.5. Reverse reconciliation

Der Unterschied in der *reverse reconciliation* besteht nun darin, dass die Zustände nicht mehr von Alice bestimmt werden, sondern von Bobs Messung abhängig sind. Weil Bob die Wahl des Informationskanals über den klassischen Kanal an Alice bekannt gibt, erhält auch Eve diese Information und kann nun annehmen, dass sie einen der beiden Zustände

$$\rho_+ = (1-e)|\epsilon_0\rangle\langle\epsilon_0| + e|\epsilon_1\rangle\langle\epsilon_1| \quad \text{oder} \tag{3.31}$$

$$\rho_- = (1-e)|\epsilon_1\rangle\langle\epsilon_1| + e|\epsilon_0\rangle\langle\epsilon_0| \tag{3.32}$$

gemessen hat. Da nun beide Zustände mit der gleichen Wahrscheinlichkeit auftreten und diese durch $p(\beta_x)$ bestimmt wird, lässt sich die Holveinformation als

$$\chi^{\mathrm{RR}}(\beta_x) = S(\bar{\rho}) - \frac{1}{2}\left(S(\rho_+) + S(\rho_-)\right) \quad , \tag{3.33}$$

schreiben. Durch eine unitäre Transformation (in diesem Fall eine Phasendrehung um $\pi$) können die beiden möglichen Zustände ineinander überführt werden $\rho_+ = U\rho_- U^\dagger$, und so gilt:

$$\chi^{\mathrm{RR}}(\beta_x) = S(\bar{\rho}) - S(\rho_+) \quad . \tag{3.34}$$

Die Basis $\{|\Phi_i\rangle\}$ lässt sich nun erneut nutzen und man erhält:

$$\rho_+ = \begin{pmatrix} |c_0|^2 & (1-2e)c_1^* c_0 \\ (1-2e)c_0^* c_1 & |c_1|^2 \end{pmatrix} \quad . \tag{3.35}$$

Zur Berechnung der Entropie $S(\rho_+) = \sum_{i=0}^{1} -\lambda_i \log_2 \lambda_i$ benötigt man nun noch die Eigenwerte dieser Matrix, die sich im zweidimensionalen Fall einfach ablesen lassen zu:

$$\lambda_{0,1} = \lambda_\pm = \frac{1}{2}\left(1 \pm \sqrt{1 - 4e(1-e)(1 - \mathrm{e}^{-4(1-\eta)\alpha^2})}\right) \quad . \tag{3.36}$$

Somit ist die Holevoinformation der *reverse reconciliation* gegeben durch:

$$\chi^{\mathrm{RR}}(\beta_x) = S(\bar{\rho}) - S(\rho_+) \quad . \tag{3.37}$$

### 3.2.6. Postselektion

Die sichere Schlüsselrate zwischen Alice und Bob ist nun in beiden Fällen gegeben durch

$$G = \int_0^\infty \mathrm{d}\beta_x p_c(\beta_x) \left(1 - H(A|B) - \chi(\beta_x)\right) \quad , \tag{3.38}$$

wobei sich der Ausdruck $1 - H(A|B) - \chi(\beta_x)$ als der Informationsgewinn von Bob über Eve ansehen lässt. Ist dieser Term allerdings negativ, so ist die Transinformation zwischen Eve und Alice größer, als die zwischen Bob und Alice. Aufgrund der Definition des Informationskanals, die wegen der Symmetrie der Präparation möglich ist, können Alice und Bob diese Kanäle ignorieren. Wird also ein solcher Kanal benutzt, werden Alice und Bob ihre Werte für diese Messung verwerfen. Über diese Kanäle wird also nicht integriert, um die sichere Schlüsselrate berechnen zu können. Dies bezeichnet man als *Postselektion*.

### 3.2.7. Fehlerkorrektur

Alice und Bob korrigieren ihre Fehler durch das Übertragen klassischer Information. Allerdings sind die dazu verwendeten Protokolle nicht ideal. Um die Effizienz eines solchen Protokolls realistisch abzuschätzen, wurde in der Arbeit von Heid und Lütkenhaus [HL06] die Effizienz des Protokolls CASCADE [BS93] betrachtet und linear gefittet. Beschreibt man deren Effizienz mit $f(e) \geq 0$, so gilt

$$G = \int_0^\infty \mathrm{d}\beta_x p_c(\beta_x) \left(1 - f(e)H(A|B) - \chi(\beta_x)\right) \quad . \tag{3.39}$$

### 3.2.8. Detektorrauschen

Betrachtet man das Verhältnis zwischen der experimentell gemessenen Varianz, $\Delta^2_{\mathrm{obs}}\beta_x$, und der durch Vakuumrauschen (shot-noise) limitierten Varianz $\Delta^2_{\mathrm{SNL}}\beta_x$, so lässt sich ein Detektorrauschen in Einheiten des Vakuumrauschens angeben mit

$$\delta = \frac{\Delta^2_{\mathrm{obs}}\beta_x}{\Delta^2_{\mathrm{SNL}}\beta_x} - 1 \quad . \tag{3.40}$$

Das Rauschen beeinflusst den von Alice präparierten Zustand

$$|\alpha\rangle \rightarrow \left|\frac{\alpha}{1+\delta}\right\rangle \tag{3.41}$$

genauso, wie den von Bob gemessenen

$$|\beta_x + i\beta_y\rangle \rightarrow \left|\frac{\beta_x + i\beta_y}{1+\delta}\right\rangle \quad . \tag{3.42}$$

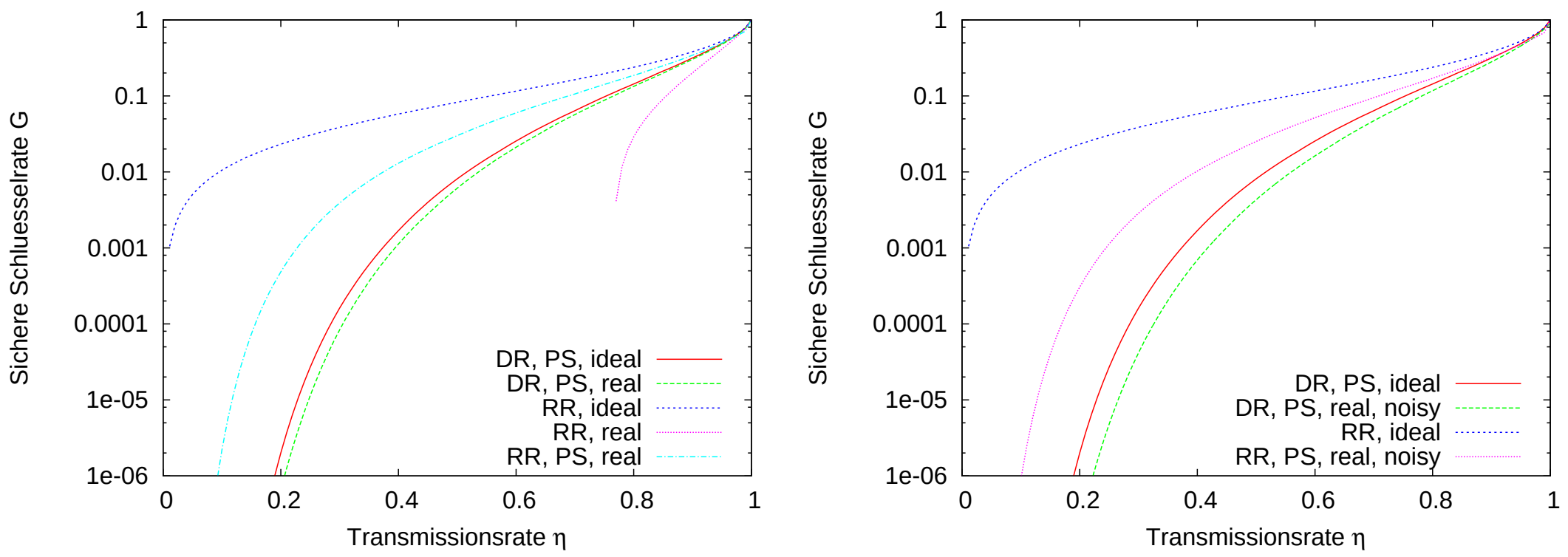

(a) (b)

Abbildung 3.1.: Abb. (a) zeigt die sichere Schlüsselrate $G$ über der Transmissionsrate $\eta$. Es wurde über die Amplitude $|\alpha|$ optimiert, und es werden sowohl Protokolle mit *direct* als auch mit *reverse reconciliation* gezeigt. Die Protokolle mit realistischer Fehlerkorrektur beziehen sich auf die Effektivität von CASCADE. Abb. (b) zeigt nun den Einfluss von Detektorrauschen mit $\delta = 0.1$ für Protokolle mit realistischer Fehlerkorrektur.

Für die Wahrscheinlichkeitsverteilung $p(\beta_x)$ ergibt sich nun:

$$p^{\mathrm{NOISE}}(\beta_x) = \int \mathrm{d}\beta_y p(\beta) = \frac{1}{2\sqrt{\pi(1+\delta)}} \left( \mathrm{e}^{-(\sqrt{\eta}\alpha-\beta_x)^2/(1+\delta)} + \mathrm{e}^{-(\sqrt{\eta}\alpha+\beta_x)^2/(1+\delta)} \right) \quad , \tag{3.43}$$

und analog

$$e^{\mathrm{NOISE}} = \frac{1}{1+\mathrm{e}^{4\sqrt{\eta}\alpha|\beta_x|/(1+\delta)}} \tag{3.44}$$

### 3.2.9. Numerische Resultate

Betrachtet man nun die sichere Schlüsselrate $G$ über der Transmissionsrate $\eta$, so ist die entsprechende Berechnung nur numerisch durchführbar, damit man ein *Postselektion* durchführen kann. Weiterhin kann man über die Amplitude $\alpha$ optimieren, sodass die optimale Schlüsselrate betrachtet werden kann.

Abb. 3.1(a) zeigt die sichere Schlüsselrate $G$ über der Transmissionsrate $\eta$. Man sieht, dass die *reverse-reconciliation*-Protokolle deutlich bessere Schlüsselraten erlauben, als Protokolle mit einer *direct reconciliation*. Jedoch ist der Einfluss einer realistischen Fehlerkorrektur (also das Betrachten der Effizient von CASCADE) deutlich stärker für Letztere. Der Plot der sicheren Schlüsselrate mit *reverse reconciliation* ohne *Postselektion* zeigt deutlich den Vorteil dieser Methode, welche Verlustraten über $50\%$ erlaubt [SRLL02].

Abb. 3.1(b) zeigt für die Protokolle mit realisterischer Fehlerkorrektur nun den Einfluss von Detektorrauschen mit $\delta = 0.1$.

# 4. Erweiterung des Protokolls von Grosshans und Grangier zu Qudits

Um Schlüsselaustauschverfahren (engl.: *Quantum key distribution*) sinnvoll zur Anwendung bringen zu können, bedarf es in der heutigen Zeit bei immer größeren Datenmengen auch immer höherer sicherer Schlüsselraten. Eben hier muss man die Vorteile einer optimalen Sicherheit durch das One-Time-Pad mit einer geringen Übertragungsgeschwindigkeit bezahlen. Die einfachste Möglichkeit bestünde wohl darin, mehrere unabhängige Quantenkanäle mit unabhängigen Präparations- und Messeinheiten zu benutzen. Dies würde allerdings die Kosten linear zum Geschwindigkeitsgewinn steigern.

Eine alternative Möglichkeit besteht in einer Verbesserung der Technologie auf experimenteller oder theoretischer Basis. Der Kern dieser Arbeit besteht im Wesentlichen aus einer Verallgemeinerung des Protokolls von Grosshans und Grangier, welche es ermöglicht, eine Verbesserung der sicheren Übertragungsleistung zu erreichen.

Der erste Teil dieses Kapitels beschäftigt sich mit der einfachsten Verallgemeinerung dieses Protokolls zu mehr als zwei Zuständen. Es wird die sichere Schlüsselrate berechnet (Abschnitt 4.2), für den *direct*- (Abschnitt 4.4), und den *reverse-reconciliated*-Fall (Abschnitt 4.5). Danach folgen jeweils numerische Ergebnisse.

Im zweiten Teil des Kapitels steht Änderung der Präparation hin zu *gequetschten Zuständen*, also einer weiteren Verallgemeinerung des Protokolls (Abschnitt 4.6). Im letzten Teil wird eine Änderung der Messmethode, eine Dual-Homodynmessung diskutiert (Abschnitt 4.7). Es folgen jeweils die Berechnungen des *direct*- und des *reverse-reconciliated*-Fall und anschließend Ergebnisse.

## 4.1. Generelle Idee und Diskussion

Hat man bisher einen von zwei Zuständen $|\pm\alpha\rangle$ betrachtet, so erlaubt man nun äquidistant verteilte Zustände auf einem Kreis um die Gauss'sche Zahlenebene. Im einfachsten Fall also komplexe Wurzeln. Man könnte also die Zustände $|\pm\alpha\rangle$ und $|\pm i\alpha\rangle$ betrachten.

Betrachtet man $d$ Zustände $\{|\alpha_0\rangle \dots |\alpha_{d-1}\rangle\}$, wobei man o. B. d. A. einen Zustand als $|\alpha_0\rangle$ bezeichnet, so kann man den Zustand $|\alpha_k\rangle$ schreiben als:

$$|\alpha_k\rangle = ||\alpha| \cdot \mathrm{e}^{i\frac{2\pi}{d}k}\rangle \quad . \tag{4.1}$$

Misst man nun weiterhin mittels einer Homodynmessung, so kann die Messachse auch variabel gestaltet werden. Theoretisch gesehen ist es natürlich egal, ob man eine andere Messachse wählt oder jeden möglichen Zustand mit einer konstanten Phase versieht. Bezeichnet man den Winkel von der reellen Achse zur Messachse mit $\chi$, so muss man entsprechend den präparierten Zustand um $-\chi$ drehen, um das gleiche Resultat zu erhalten, solange man weiterhin auf der reellen Achse misst. Der Zustand $|\alpha_k(\chi)\rangle$ wäre also:

$$|\alpha_k(\chi)\rangle = ||\alpha| \cdot \mathrm{e}^{i\left(\frac{2\pi}{d}k-\chi\right)}\rangle \quad . \tag{4.2}$$

Eine Veranschaulichung dieser Zustände im Phasenraum mit der jeweils zugehörigen Wahrscheinlichkeitsverteilung $p(\beta_x)$ findet sich in Abb. 4.1. Die Berechnung von $p(\beta_x)$ findet sich in Kapitel 4.3.

## 4.2. Sichere Schlüsselrate

Wird nun wieder allgemein den Zustand des gesamten Systems betrachtet, so handelt es sich immer noch um eine unitäre Transformation des präparierten Zustandes $|\phi_i\rangle$ und eines Grundzustandes $|\epsilon^0\rangle$ von Eve in einen Produktzustand von Bob $|\Psi_i\rangle$ und Eve $|\epsilon_i\rangle$:

$$U|\phi_i\rangle|\epsilon^0\rangle = |\Psi_i\rangle|\epsilon_i\rangle \quad . \tag{4.3}$$

Daraus ergibt sich durch Präparation und Messung:

$$\langle\phi_i|\phi_j\rangle = \langle\Psi_i|\Psi_j\rangle\langle\epsilon_i|\epsilon_j\rangle \quad . \tag{4.4}$$

Den Strahlteiler-Angriff kann man also weiterhin schreiben als

$$|\alpha_k(\chi)\rangle \to |\sqrt{\eta}\cdot\alpha_k(\chi)\rangle_B \otimes |\sqrt{1-\eta}\cdot\alpha_k(\chi)\rangle_E \tag{4.5}$$

mit der Transmittivität des Quantenkanals $\eta$.

Die weiteren Protokollparameter gleichen denen des Protokolls von Grosshans und Grangier. Alice wählt einen von $k$ Zuständen zufällig und mit gleicher Wahrscheinlichkeit aus ($p_k = 1/k$) und Bob projeziert diesen auf einen *kohärenten Zustand* $|\beta\rangle = |\beta_x + i\beta_y\rangle$. Alice veröffentlicht auf einem klassischen Kanal die gewählte Amplitude.

Die sichere Schlüsselrate $G$ ergibt sich nun erneut aus der von Devetak und Winter [DW05] berechneten Differenz der Transinformationen

$$G \geq I(A:B) - I(A:C) \quad , \tag{4.6}$$

was durch die obere Schranke der Holevoinformation führt zu:

$$G \geq I(A:B) - \chi \quad . \tag{4.7}$$

## 4.3. Transinformation zwischen Alice und Bob

Um nun die sichere Schlüsselrate mittels (4.7) berechenen zu können, benötigt man für den ersten Summanden die Transinformation zwischen Alice und Bob. Diese lässt sich nach der Definition (2.12) berechnen durch:

$$I(A:B) = H(A) - H(A|B) \quad . \tag{4.8}$$

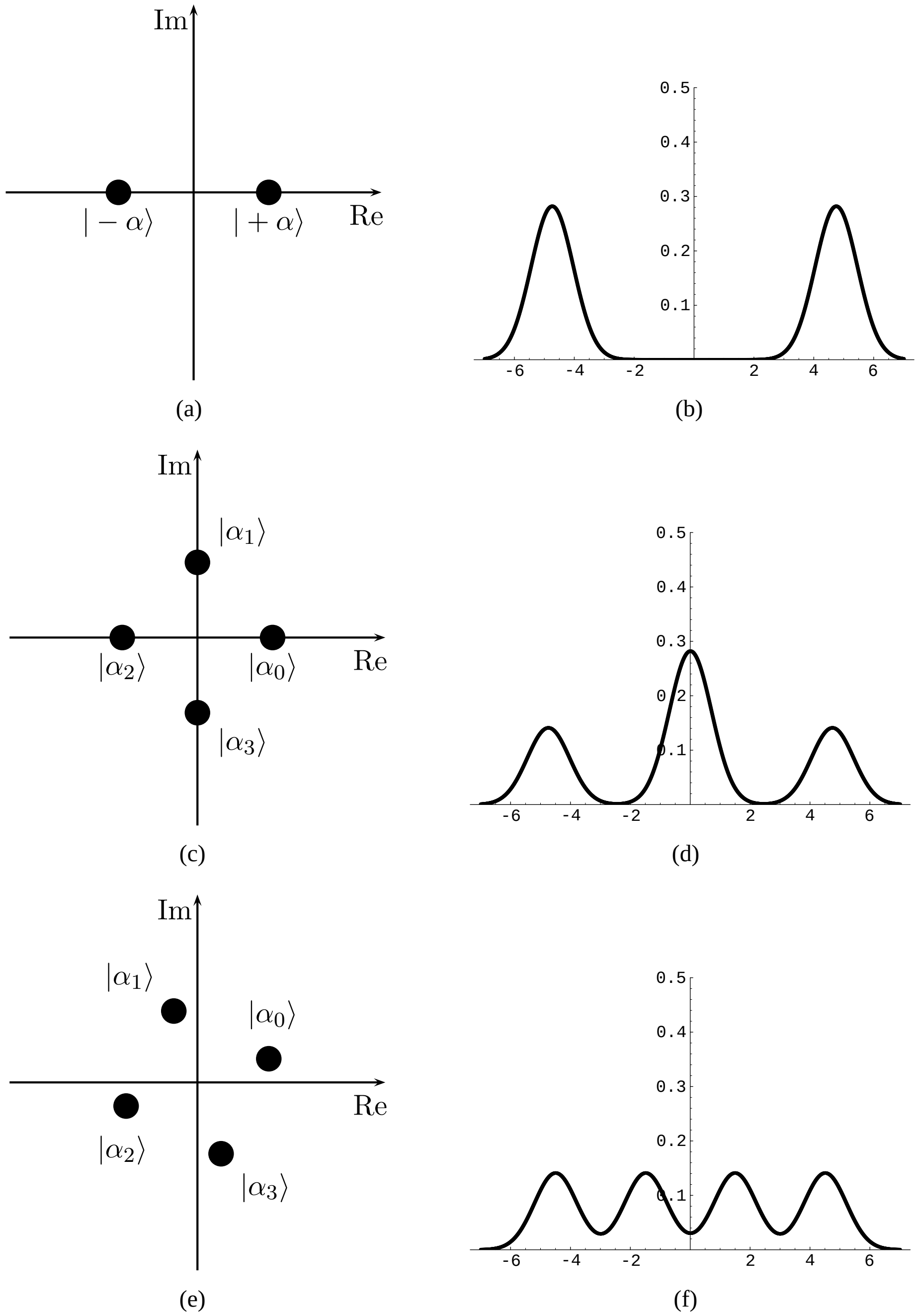

Abbildung 4.1.: Abb. (a) zeigt eine Skizze der zwei *kohärenten Zustände* im Phasenraum in einem 2-Qudit-System, und Abb. (b) die zugehörige Wahrscheinlichkeitsverteilung $p(\beta_x)$ für eine Messung auf der reellen Achse mit $|\alpha| = 5$, $\eta = 0.9$ und $\chi = 0$.
Die Abbildungen (c) und (d) zeigen die Zustände und die Wahrscheinlichkeitsverteilung für ein System mit 4 Qudits, so wie es im Weiteren hergeleitet werden wird.
Abb. (e) und Abb. (f) beziehen sich auf eine verdrehte Messachse mit $\chi = 0.32$, die für eine Gleichverteilung der Abbildungen der Zustände auf der reellen Achse sorgt.

Es wird zuerst die Entropie der von Alice präparierten Zustände betrachtet. Sie lässt sich berechnen als

$$H(A) = -\sum_{k=0}^{d-1} p(\alpha_k) \log_d p(\alpha_k) \quad , \tag{4.9}$$

wobei der Logarithmus hier zur Basis $d$ berechnet wird, da die Information in Informationseinheiten zur Basis $d$ übermittelt wird [Sha48]. Nach den Logarithmenregeln überträgt man also $n$ Bits, wenn man als Einheit $2^n$ wählt, im Falle von $n = 1$ also genau 1 Bit.

Da Alice aus einem der $d$ Zustände zufällig auswählt, gilt mit $p(\alpha_k) = 1/d$ und $-\log(1/d) = \log(d)$:

$$H(A) = \log_d d = 1 \quad . \tag{4.10}$$

Weiterhin muss die bedingte Entropie zwischen Alice und Bob $H(A|B)$ betrachtet werden. Da theoretisch die mögliche Drehung der Messachse im Präparationsvorgang von Alice betrachtet wird, misst Bob formal auf der reellen Achse. Durch die Kommunikation über den Quantenkanal findet also ein Übergang von einem von Alice präparierten Zustand $|\alpha_k\rangle$ zu einem von Bob gemessenen Zustand $|\beta_x\rangle$ statt. $H(A|B)$ lässt sich nun schreiben als:

$$H(A|B) = -\int \mathrm{d}\beta_x p(\beta_x) \sum_{k=0}^{d-1} p(\alpha_k|\beta_x) \log_d p(\alpha_k|\beta_x) \quad . \tag{4.11}$$

Wegen $\int \mathrm{d}\beta_x p(\beta_x) = 1$ gilt somit für die Transinformation:

$$I(A:B) = \int \mathrm{d}\beta_x p(\beta_x) \left( 1 + \sum_{k=0}^{d-1} p(\alpha_k|\beta_x) \log_d p(\alpha_k|\beta_x) \right) \quad . \tag{4.12}$$

Um nun die *a-posteriori*-Wahrscheinlichkeiten $p(\alpha_k|\beta_x)$ zu berechnen, lässt sich der *Satz von Bayes* (2.4) verwenden, um

$$p(\alpha_k|\beta_x) = \frac{p(\beta_x|\alpha_k)p(\alpha_k)}{\sum_{l=0}^{d-1} p(\beta_x|\alpha_l)p(\alpha_l)} = \frac{p(\beta_x|\alpha_k)p(\alpha_k)}{p(\beta_x)} \tag{4.13}$$

zu erhalten. Die bedingten Wahrscheinlichkeiten $p(\alpha_k|\beta_x)$ lassen sich über die in Gleichung (2.49) eingeführte *Q-Funktion* berechnen zu:

$$p(\beta_x|\alpha_k) = \frac{1}{\pi} |\langle \beta_x | \sqrt{\eta} \alpha_k \rangle|^2 \quad . \tag{4.14}$$

Der Faktor $\sqrt{\eta}$ stammt aus der Tatsache, dass zum Zeitpunkt der Messung bei Bob der Zustand $|\alpha\rangle$ von Alice bereits den verlustbehafteten Kanal bzw. den Strahlteiler passiert hat und so nur noch der Zustand $|\sqrt{\eta}\,\alpha\rangle$ vorhanden ist. Mit $p(\alpha_k) = 1/d$ ergibt sich nun:

$$p(\alpha_k|\beta_x) = \frac{1}{d\pi} \cdot \frac{|\langle \beta_x | \sqrt{\eta} \alpha_k \rangle|^2}{p(\beta_x)} \quad . \tag{4.15}$$

Nun folgt aus $p(\beta_x) = \sum_{l=0}^{d-1} p(\beta_x|\alpha_l)p(\alpha_l) = \pi^{-1} \sum_{l=0}^{d-1} |\langle\beta_x|\alpha_l\rangle|^2 p(\alpha_l)$ mit $p(\alpha_l) = 1/d \forall l$ und $p(\beta_x) = \int p(\beta)\mathrm{d}\beta_y$:

$$p(\beta_x) = \frac{1}{d\pi} \int \mathrm{d}\beta_y \sum_{l=0}^{d-1} \mathrm{e}^{-(\sqrt{\eta}\alpha_l - \beta_x - i\beta_y)(\sqrt{\eta}\alpha_l^* - \beta_x + i\beta_y)} \tag{4.16}$$

$$= \frac{\mathrm{e}^{-(\eta|\alpha|^2+\beta_x^2)}}{d\pi} \sum_{l=0}^{d-1} \mathrm{e}^{2\sqrt{\eta}|\alpha|\beta_x \cos(\frac{2\pi}{d}l-\chi)} \int \mathrm{d}\beta_y \mathrm{e}^{-(\beta_y^2 - 2\sqrt{\eta}|\alpha|\beta_y \sin(\frac{2\pi}{d}l-\chi))} \quad . \tag{4.17}$$

Der Integrand lässt sich durch quadratische Ergänzung umformen in

$$\int \mathrm{d}\beta_y \mathrm{e}^{-(\beta_y^2 - 2\sqrt{\eta}|\alpha|\beta_y \sin(\frac{2\pi}{d}l-\chi))} = \mathrm{e}^{-(\beta_y - \sqrt{\eta}|\alpha|\sin(\frac{2\pi}{d}l-\chi))^2} \cdot \mathrm{e}^{\eta|\alpha|^2 \sin^2(\frac{2\pi}{d}l-\chi)} \quad , \tag{4.18}$$

und durch Substitution und Anwenden von $\int \mathrm{e}^{-y^2}\mathrm{d}y = \sqrt{\pi}$ folgt daraus:

$$p(\beta_x) = \frac{\mathrm{e}^{-(\eta|\alpha|^2+\beta_x^2)}}{d\sqrt{\pi}} \sum_{l=0}^{d-1} \mathrm{e}^{2\sqrt{\eta}|\alpha|\beta_x \cos(\frac{2\pi}{d}l-\chi) + \eta|\alpha|^2 \sin^2(\frac{2\pi}{d}l-\chi)} \quad . \tag{4.19}$$

Sortiert man nun die Terme um und wendet die Identität $\sin^2\phi + \cos^2\phi = 1$ an, so folgt:

$$p(\beta_x) = \frac{1}{d\sqrt{\pi}} \sum_{l=0}^{d-1} \mathrm{e}^{-(\sqrt{\eta}|\alpha|\cos(\frac{2\pi}{d}l-\chi) - \beta_x)^2} \quad . \tag{4.20}$$

Analog lässt sich nun auch $p(\beta_x|\alpha_k)$ berechnen zu:

$$p(\beta_x|\alpha_k) = \frac{1}{\sqrt{\pi}} \mathrm{e}^{-(\sqrt{\eta}|\alpha|\cos(\frac{2\pi}{d}k-\chi) - \beta_x)^2} \quad . \tag{4.21}$$

Die Ähnlichkeit dieser Wahrscheinlichkeiten zum zweidimensionalen Protokoll ist offensichtlich und auch die Rückführung auf diesen Spezialfall (3.18).

Schließlich lässt sich die Transinformation zwischen Alice und Bob schreiben als:

$$I(A:B) = \int \mathrm{d}\beta_x \frac{1}{d\sqrt{\pi}} \sum_{l=0}^{d-1} \mathrm{e}^{-(\sqrt{\eta}|\alpha|\cos(\frac{2\pi}{d}l-\chi)-\beta_x)^2} \times \left(1 + \sum_{k=0}^{d-1} \frac{d\mathrm{e}^{-(\sqrt{\eta}|\alpha|\cos(\frac{2\pi}{d}k-\chi)-\beta_x)^2}}{\sum_{l=0}^{d-1} \mathrm{e}^{-(\sqrt{\eta}|\alpha|\cos(\frac{2\pi}{d}l-\chi)-\beta_x)^2}} \log_d \frac{d\mathrm{e}^{-(\sqrt{\eta}|\alpha|\cos(\frac{2\pi}{d}k-\chi)-\beta_x)^2}}{\sum_{l=0}^{d-1} \mathrm{e}^{-(\sqrt{\eta}|\alpha|\cos(\frac{2\pi}{d}l-\chi)-\beta_x)^2}}\right) \tag{4.22}$$

## 4.4. Direct reconciliation

Für den Fall, dass die Fehlerkorrekturinformationen auf einem klassischen Kanal von Alice zu Bob geschickt werden, spricht man von *direct reconciliation*. Alice schickt den Betrag des gesendeten Zustandes, also $|\alpha|$ an Bob, der mithilfe dieser Information und der Kanal-Transmittivität $\eta$ aus dem gemessenen Wert für $\beta_x$ und auf den präparierten Zustand $\alpha_k$ schliessen kann.

Da diese Information für Eve frei zugänglich ist, hat sie Auswirkung auf die von Eve erhaltene Erkenntnis des von Alice präparierten Zustandes. Der für Eve zugängliche Quantenzustand am zweiten Ausgang des Strahlteilers ist

$$|\epsilon_k\rangle = |\sqrt{1-\eta}\ \alpha_k\rangle \quad . \tag{4.23}$$

### 4.4.1. Holevoinformation

Die daraus maximal gewinnbare Information lässt sich nun wieder durch die Holevoinformation nach oben beschränken. Da es sich um einen reinen Zustand handelt, ist der gemischte Term der Holevoinformation gleich null und es bleibt:

$$\chi^{\mathrm{DR}} = S(\bar{\rho}) \quad . \tag{4.24}$$

Weil Alice alle Zustände mit gleicher Wahrscheinlichkeit präpariert, ergibt sich für den gemittelten Dichteoperator:

$$\bar{\rho} = \frac{1}{d}\sum_{k=0}^{d-1} |\epsilon_k\rangle\langle\epsilon_k| \quad . \tag{4.25}$$

Um nun die Entropie des gemittelten Dichteoperators zu bestimmen, benötigt man seine Eigenwerte. Allgemein könnte man dies numerisch lösen, allerdings würden entsprechende Algorithmen von höherdimensionalen Systemen eine sehr lange Rechenzeit erfordern. Um diesen Umstand zu vermeiden, müssen Symmetriebetrachtungen angestellt werden, um eine orthonormale Basis zu finden, in welcher der Dichteoperator diagonal ist.

#### Orthonormalisierung des Dichteoperators

Ein *kohärenter Zustand* $|\alpha\rangle$ ist allgemein gegeben durch:

$$|\alpha\rangle = \mathrm{e}^{-\frac{1}{2}|\alpha|^2} \sum_{n=0}^{\infty} \frac{\alpha^n}{\sqrt{n!}}|n\rangle \quad . \tag{4.26}$$

Zur Erklärung des Orthogonalisierungsschemas wird nun o. B. d. A. ein System mit vier möglichen Zuständen und ohne eine Phase $\chi$ betrachtet:

Definiert man nun die $4$ möglichen Zustände als $|\alpha_0\rangle := |+\alpha\rangle$, $|\alpha_1\rangle := |i\alpha\rangle$, $|\alpha_2\rangle := |-\alpha\rangle$ und $|\alpha_3\rangle := |-i\alpha\rangle$, so kann man diese in Fock-Zuständen entwickeln in:

$$|\alpha_0\rangle = \mathrm{e}^{-\frac{1}{2}|\alpha|^2}\Big(|0\rangle + \alpha|1\rangle + \alpha^2|2\rangle + \alpha^3|3\rangle + \alpha^4|4\rangle + \dots\Big) \tag{4.27}$$

$$|\alpha_1\rangle = \mathrm{e}^{-\frac{1}{2}|\alpha|^2}\Big(|0\rangle + i\alpha|1\rangle - \alpha^2|2\rangle - i\alpha^3|3\rangle + \alpha^4|4\rangle + i\dots\Big) \tag{4.28}$$

$$|\alpha_2\rangle = \mathrm{e}^{-\frac{1}{2}|\alpha|^2}\Big(|0\rangle - \alpha|1\rangle + \alpha^2|2\rangle - \alpha^3|3\rangle + \alpha^4|4\rangle - \dots\Big) \tag{4.29}$$

$$|\alpha_3\rangle = \mathrm{e}^{-\frac{1}{2}|\alpha|^2}\Big(|0\rangle - i\alpha|1\rangle - \alpha^2|2\rangle + i\alpha^3|3\rangle + \alpha^4|4\rangle - i\dots\Big) \tag{4.30}$$

Da die Fock-Zustände orthonormal sind, sind die Summen mit unterschiedlichen Fock-Zuständen als Summanden orthogonal. Durch Betrachtung der Gleichungen (4.27) können wir nun eine orthogonale Basis $\{|b\rangle\}$ definieren:

$$|b_0\rangle = \sum_{n=0}^{\infty} \frac{\alpha^{4n}}{\sqrt{(4n)!}} |4n\rangle \tag{4.31}$$

$$|b_1\rangle = \sum_{n=0}^{\infty} \frac{\alpha^{4n+1}}{\sqrt{(4n+1)!}} |4n+1\rangle \tag{4.32}$$

$$|b_2\rangle = \sum_{n=0}^{\infty} \frac{\alpha^{4n+2}}{\sqrt{(4n+2)!}} |4n+2\rangle \tag{4.33}$$

$$|b_3\rangle = \sum_{n=0}^{\infty} \frac{\alpha^{4n+3}}{\sqrt{(4n+3)!}} |4n+3\rangle \quad . \tag{4.34}$$

Somit gilt allgemein für eine $d$-dimensionale Basis für $d$ äquidistant verteilte Zustände mit gleicher Amplitude für den Basisvektor $|b_l^d\rangle$, $l \in \{0, \dots, d-1\}$:

$$|b_l^d\rangle := \sum_{n=0}^{\infty} \frac{\alpha^{d \cdot n + l}}{\sqrt{(d \cdot n + l)!}} |d \cdot n + l\rangle \quad . \tag{4.35}$$

Um diese Basis zu normieren, wird eine verallgemeinerte Exponentialfunktion über die Potenzreihe

$$\mathfrak{e}_{d,k}^{x} := \sum_{n=0}^{\infty} \frac{x^{dn+k}}{(dn+k)!} \tag{4.36}$$

definiert, die für $d = 1$ und $k = 0$ in die normale Exponentialfunktion $e^x$ übergeht. Somit gilt für die orthonormierte Basis $\{|\Phi\rangle\}$,

$$|\Phi_l^d(|\alpha|)\rangle := \frac{1}{\mathfrak{e}_{d,l}^{|\alpha|^2}} \cdot |b_l^d\rangle = \frac{1}{\mathfrak{e}_{d,l}^{|\alpha|^2}} \cdot \sum_{n=0}^{\infty} \frac{\alpha^{d \cdot n + l}}{\sqrt{(d \cdot n + l)!}} |d \cdot n + l\rangle \quad . \tag{4.37}$$

Ein Zustand $|\alpha_k^d\rangle$ würde sich in dieser Basis also schreiben lassen, als:

$$|\alpha_k^d\rangle = \mathrm{e}^{-\frac{1}{2}|\alpha|^2} \sum_{l=0}^{d-1} \mathrm{e}^{i\frac{2\pi}{n}kl} \mathfrak{e}_{d,l}^{|\alpha|^2} |\Phi_l^d(|\alpha|)\rangle \quad . \tag{4.38}$$

Für die Zustände $|\epsilon_k^d\rangle = |\sqrt{1-\eta}\alpha_k^d\rangle$ von Eve gilt damit:

$$|\epsilon_k^d\rangle = \mathrm{e}^{-\frac{1}{2}|\epsilon|^2} \sum_{l=0}^{d-1} \mathrm{e}^{i\frac{2\pi}{d}kl} \mathfrak{e}_{d,l}^{|\epsilon|^2} |\Phi_l^d(|\epsilon|)\rangle = \sum_{l=0}^{d-1} c_l \mathrm{e}^{i\frac{2\pi}{d}kl} |\Phi_l^d(|\epsilon|)\rangle \quad , \tag{4.39}$$

womit sich nun die Holevoinformation ausdrücken lässt mit:

$$\chi^{\mathrm{DR}} = S(\bar{\rho}) = -\sum_{l=0}^{d-1} |c_l|^2 \log_d(|c_l|^2) \quad , \tag{4.40}$$

denn die Koeffizienten $|c_l|^2$ lassen sich mithilfe der Überlappe $\langle\epsilon_l^d|\epsilon_k^d\rangle$ bestimmen. Zur Bestimmung von $d$ Koeffizienten benötigt man $d$ unabhängige Bestimmungsgleichungen, o. B. d. A. wird $\langle\epsilon_0^d|\epsilon_k^d\rangle$ gewählt. Es gilt:

$$\langle\epsilon_0^d|\epsilon_k^d\rangle = \exp\left(-|\alpha|^2(1-\eta)\cdot(1-\mathrm{e}^{i\frac{2\pi}{d}(k-0)})\right) = \sum_{l=0}^{d-1}|c_l|^2\mathrm{e}^{i\frac{2\pi}{d}kl} \quad . \tag{4.41}$$

Man erkennt sehr leicht, dass $\langle\epsilon_0^d|\epsilon_k^d\rangle = \langle\epsilon_m^d|\epsilon_{(k+m)\mathrm{mod}d}^d\rangle$ gilt und die Vorraussetzung damit wirklich nicht die Allgemeinheit beschränkt.

Betrachtet man diese Gleichung als eine Vektortransformation in der Form $\epsilon_k = E_{kl}\cdot c_l$, so lässt sich die Matrix $E$ mittels Normierung durch die *Fourier-Matrix* $\mathcal{F}$ beschreiben, welche unitär ist, $E=\sqrt{d}\mathcal{F}$. Also gilt $c_l = (1/\sqrt{d})\ \mathcal{F}_{kl}^\dagger\cdot\epsilon_k = (1/d)\ E_{kl}^\dagger\cdot\epsilon_k$.

Gleichung (4.41) lässt sich damit umschreiben in:

$$|c_l|^2 = \frac{1}{d}\sum_{k=0}^{d-1}\mathrm{e}^{-i\frac{2\pi}{d}kl}\langle\epsilon_0^d|\epsilon_k^d\rangle = \frac{1}{d}\sum_{k=0}^{d-1}\mathrm{e}^{-i\frac{2\pi}{d}kl}\mathrm{e}^{-(|\alpha|^2(1-\eta)\cdot(1-\exp i\frac{2k\pi}{d}))} \quad . \tag{4.42}$$

Es folgt zusammenfassend für die Holevoinformation:

$$\chi^{\mathrm{DR}} = -\frac{1}{d}\sum_{k,l=0}^{d-1}\mathrm{e}^{-i\frac{2\pi}{d}kl-(|\alpha|^2(1-\eta)\cdot(1-\exp i\frac{2k\pi}{d}))}\log_d\left(\frac{1}{d}\sum_{k=0}^{d-1}\mathrm{e}^{-i\frac{2\pi}{d}kl-(|\alpha|^2(1-\eta)\cdot(1-\exp i\frac{2k\pi}{d}))}\right) . \tag{4.43}$$

### 4.4.2. Sichere Schlüsselrate

Die sichere Schlüsselrate ergibt sich aus (4.7), (4.12), (4.20), (4.21) (4.40) und (4.42) zu einem Ausdruck in der Form:

$$G = \int \mathrm{d}\beta_x p(\beta_x)\left(1+\sum_{k=0}^{d-1}p(\alpha_k|\beta_x)\log_d p(\alpha_k|\beta_x) - \chi^{\mathrm{DR}}\right) \tag{4.44}$$

$$= \int \mathrm{d}\beta_x p(\beta_x)\left(1+\sum_{k=0}^{d-1}p(\alpha_k|\beta_x)\log_d p(\alpha_k|\beta_x) + \sum_{l=0}^{d-1}|c_l|^2\log_d(|c_l|^2)\right) \quad , \tag{4.45}$$

wobei $\chi^{\mathrm{DR}}$ unabhängig von $\beta_x$ ist.

### 4.4.3. Postselektion

Man kann nun einen Quantenkanal definieren für jeden möglichen Zustand für $\beta_x$. Darin ist der Ausdruck $\left(1+\sum_{k=0}^{d-1}p(\alpha_k|\beta_x)\log_d p(\alpha_k|\beta_x) - \chi^{\mathrm{DR}}\right)$ als der minimal mögliche Informationsgewinn von Bob über Eve anzusehen (die Holevoinformation ist eine obere Schranke). Misst Bob nun einen Wert für $\beta_x$, für den dieser Ausdruck kleiner als null ist, so erhält Eve mehr Information, als Bob. Man spricht von *Postselektion*, wenn man eben diese Messungen verwirft, da sie zu unsicher sind, weil Eve im Durchschnitt mehr Informationsgewinn als Bob hat. Das Integral wird also nur über die Kanäle ausgeführt, die keinen negativen Beitrag leisten.

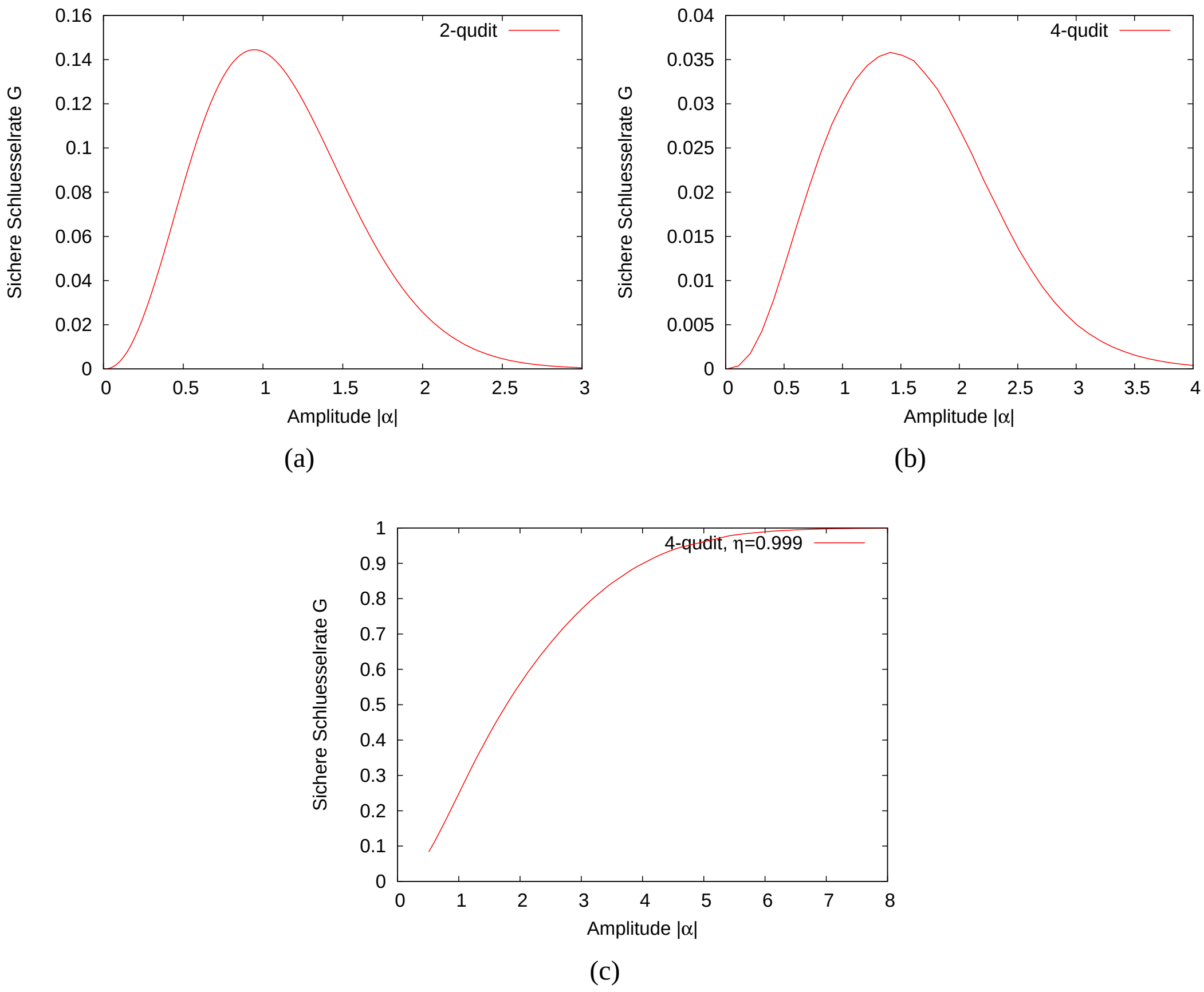

Abbildung 4.2.: Abb. (a) zeigt den Verlauf der sicheren Schlüsselrate im Falle der *direct reconciliation* über der Amplitude $|\alpha|$ für ein 2-Qudit-System bei $\eta = 0.8$, Abb. (b) für ein 4-Qudit System und Abb. (c) mit $\eta = 0.999$ und einer optimierten Phase von $\phi = \arctan 1/3$

### 4.4.4. Optimierung der Amplitude

Die sichere Schlüsselrate $G$ lässt sich noch abhängig von der Transmission $\eta$ des Quantenkanals über den Amplitudenparameter $|\alpha|$ optimieren. Mit einer steigenden Amplitude sind zwei *kohärente Zustände* immer besser unterscheidbar für Bob, allerdings erhält damit auch Eve immer mehr Informationen. Weil man davon ausgeht, dass Eve die maximal zugängliche Information nutzen kann, steigt sie schneller, als die Information, die Bob erhält. Jedoch hat Bob aufgrund der *reconciliation*-Methoden einen Vorsprung, sodass man eine Kurve mit einem Peak erwartet. Genau dies zeigen auch die Betrachtungen für ein 2-Qubit- und ein 4-Qubit-Sytem, jeweils mit $\eta = 0.8$ (Abb. 4.2(a) und Abb. 4.2(b)). Für eine perfekte Übertragung mit $\eta = 1$ allerdings wird Eve keine Information mithilfe eine Strahlteiler-Angriffs erhalten, die Amplitude muss also nur eine gewisse Grenze überschreiten (Abb. 4.2(c)).

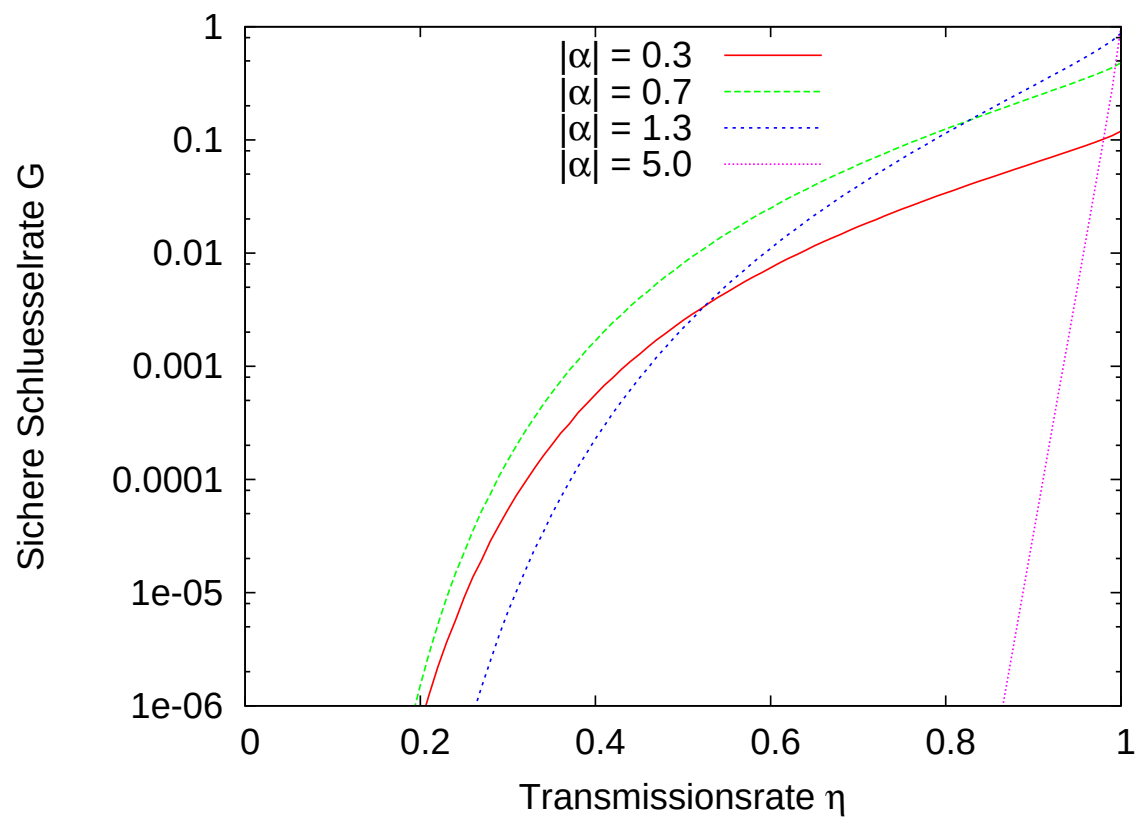

Abbildung 4.3.: Sichere Schlüsselrate (*direct reconciliation*) für 2-Qudit-Zustände, normiert auf 1 für verschiedene Amplituden $|\alpha|$ ($\chi = 0$).

### 4.4.5. Optimierung der Phase

Die Schlüsselrate kann auch über die Phase $\chi$, also der Messachse von Eve bzw. der Präparationsachse von Alice optimiert werden. Allerdings wird auf diese Optimierung verzichtet, aber eine analytische Betrachtung im Bereich nahezu perfekter Transmission durchgeführt (Kapitel 4.4.7).

### 4.4.6. Detektorrauschen

Analog zum Ergebnis des zweidimensionalen Falles (Kapitel 3.2.8) lässt sich das Detektorrauschen betrachten und es gilt:

$$p(\beta_x|\alpha_k) = \frac{1}{\sqrt{\pi}} e^{-(\sqrt{\eta}|\alpha| \cos(\frac{2\pi}{d}k - \chi) - \beta_x)^2/(1+\delta)} \quad . \tag{4.46}$$

### 4.4.7. Resultate

Zur Berechnung der sicheren Schlüsselraten wurde im Rahmen dieser Arbeit ein Programm in C++ implementiert (Anhang B).

#### Betrachtung zur Amplitudenoptimierung

Lässt man die sichere Schlüsselrate für verschiedene Amplituden $|\alpha|$ für ein 2-Qudit-System berechnen und trägt diese Werte über der Transmissionsrate $\eta$ auf (Abbildung 4.3), so lässt sich sehen, dass im Bereich nahezu perfekter Transmission die Ergebnisse für eine hohe Amplitude besser ausfallen.

Dies lässt sich dadurch begründen, dass Eve nahezu keinerlei Information über das System erhalten kann aufgrund der hohen Transmissionsrate, Bob allerdings die Zustände sehr gut voneinander unterscheiden kann, da sie sehr weit voneinander entfernt liegen (ähnlich wie in Abb. 4.1(b) zu sehen).

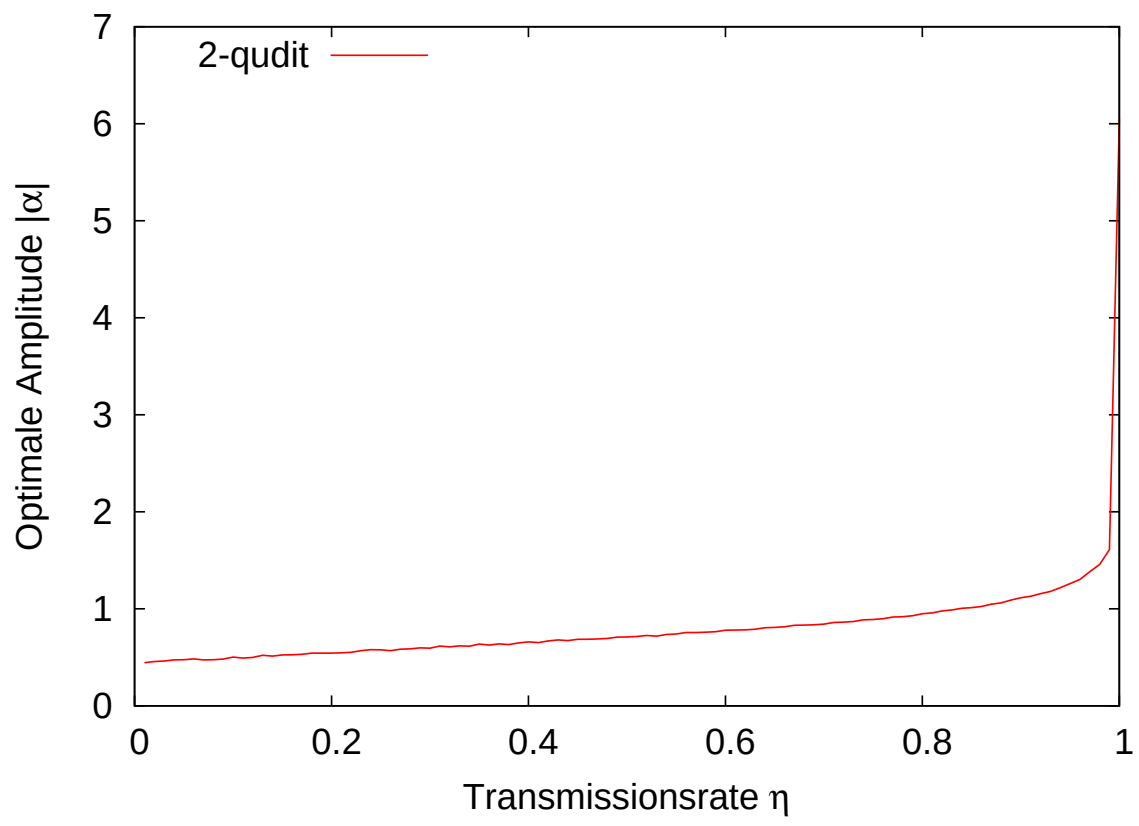

Abbildung 4.4.: Optimale Amplitude $|\alpha|$ für 2-Qudit-Zustände ($\chi = 0$).

Im Bereich niedrigerer Transmissionsrate hingegen erreicht man eine höhere Schlüsselrate durch die Wahl einer immer kleineren Amplitude, die ausreichend klein ist, damit Eve diese nur schwer unterscheiden kann, allerdings so gross, dass Bob die Zustände mithilfe der *reconciliation*-Methoden unterscheiden kann. Vor allem aber auch der Schritt der *Postselektion* sei hier zu erwähnen, da keine für Eve interessanten Kanäle benutzt werden.

Möchte man nun die optimale Amplitude über der Transmissionsrate betrachten, so erwartet man eine monoton fallende, stetige Funktion für eine schlechter werdende Transmissionsrate. Dies ist aus Abbildung 4.4 sehr gut zu erkennen.

### Vergleiche verschiedener Ordnungen

Für Qudit-Systeme höherer Ordnung sind ähnliche Ergebnisse zu erwarten. Dies sieht man eindrucksvoll in einem Plot der optimalen Amplituden über $\eta$ (Abbildung 4.5(a)).

Wiederum nur im Bereich einer sehr guten Kanalqualität sind deutliche Unterschiede zu erkennen. Mit steigender Ordnung des Systems wird hier die optimale Amplitude immer größer, die Begründung ist dieselbe, wie zuvor. Da man immer mehr Zustände voneinander unterscheiden muss, müssen diese im Optimalfall immer weiter voneinander entfernt sein, um sie auch sicher voneinander unterscheiden zu können. Der Überlapp *kohärenter Zustände* ist von deren Abstand im Phasenraum exponentiell abhängig. Man kann also sagen, zwei *kohärente Zustände* werden bei einem grossen Abstand orthogonal.

Betrachtet man nun die sicheren Schlüsselraten aufgetragen über $\eta$ für verschiendene Ordnungen wie in Abbildung 4.5(a) gezeigt, so sieht man, dass die Schlüsselraten für höhere Ordnungen immer schlechter werden. Allerdings scheint es, als könnte sich die Kurve einer Grenzkurve annähern.

### Renormierung

Betrachtet man allerdings nur Zustände mit einer Ordnung zur Basis 2, also $d = 2^n$-Qudits, so kann man diese Informationseinheiten auch neu interpretieren. Nach Shannon wird der Loga-rithmus der Entropien immer in der Dimension des Systems angegeben, also der betrachteten Informationseinheit. Im Falle eines $2^2$-dimensionalen Systems ist die Einheit also $4$. Da klassische Bits übertragen werden, kann man ein *4-digit* auch als 2 *2-digits*, also 2 Bit interpretieren. Genauso wird aus einem Logarithmus zur Basis 4 zweimal der Logarithmus zur Basis 2. Betrachtet man also die sichere Schlüsselrate immer in Einheiten von Bit, so lässt diese sich aus der sicheren Schlüsselrate in der Basis der Informationseinheit für alle $d = 2^n$-dimensionalen Systeme durch die Multiplikation mit $n$ erhalten. Diese Betrachtung wird im Weiteren als Normierung auf ein Bit bezeichnet.

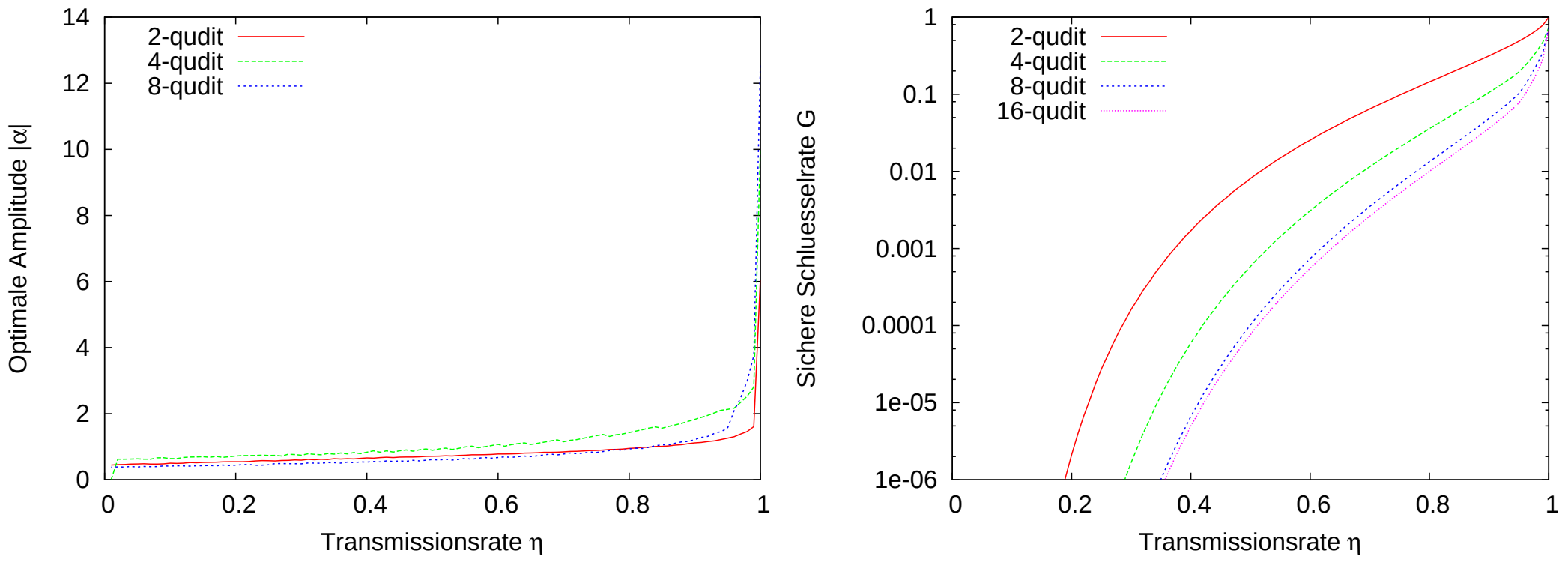

(a) (b)

Abbildung 4.5.: In Abb. (a) wird die optimale Amplitude $|\alpha|$ über der Transmissionsrate $\eta$ gezeigt im Falle der *direct reconciliation* für 2-, 4- und 8-Qudit-Zustände ($\chi = 0$). Abb. (b) zeigt die zugehörigen sicheren Schlüsselraten, zusätzlich auch für ein 16-Qudit-System

Zustände unterschiedlicher Ordnung lassen sich so sehr einfach vergleichen. Es ist natürlich möglich, Zustände jeder Basis auf diese Art zu betrachten. Da aber eine Betrachtung von Zuständen zur Basis 2 keine Einschränkung darstellt, dafür jedoch einen anschaulicheren Vergleich zulässt, reduziert sich die Diskussion hier v. a. auf eben solche Zustände. Der einfache Vergleich von Systemen mit $d = \{2, 4, 8\}$ in Abbildung 4.6(a) zeigt auf den ersten Blick keine wesentliche Änderung zu Abbildung 4.5(b), allerdings fallen die Kurven des $8$-Qudit und $16$-Qudit-Systems nun weitgehend zusammen. Schaut man sich noch höhere Ordnungen an, so zeigen auch diese gleiches Verhalten.

Allerdings ist es interessant, den Verlauf der Kurven im Bereich nahezu perfekter Transmissionsrate zu betrachten. Im Idealfall erwartet man nämlich, dass bei $\eta = 1$ die Übertragung der gesamten Information sicher ist, die Informationsmenge also der Informationseinheit entspricht. Dies entspräche bei einem 2-Qudit einem Bit, bei einem 4-Qudit zwei Bit, usw. Betrachtet man den Plot also noch einmal in einer Vergrößerung (Abbildung 4.6(b)), so wird deutlich, dass zwar die Systeme höherer Ordnung nun mehr Informationen übertragen, jedoch im Falle eines idealen Kanals nicht - wie erwartet - die maximal mögliche. Dieser Frage widmet sich auch der nächste Abschnitt.

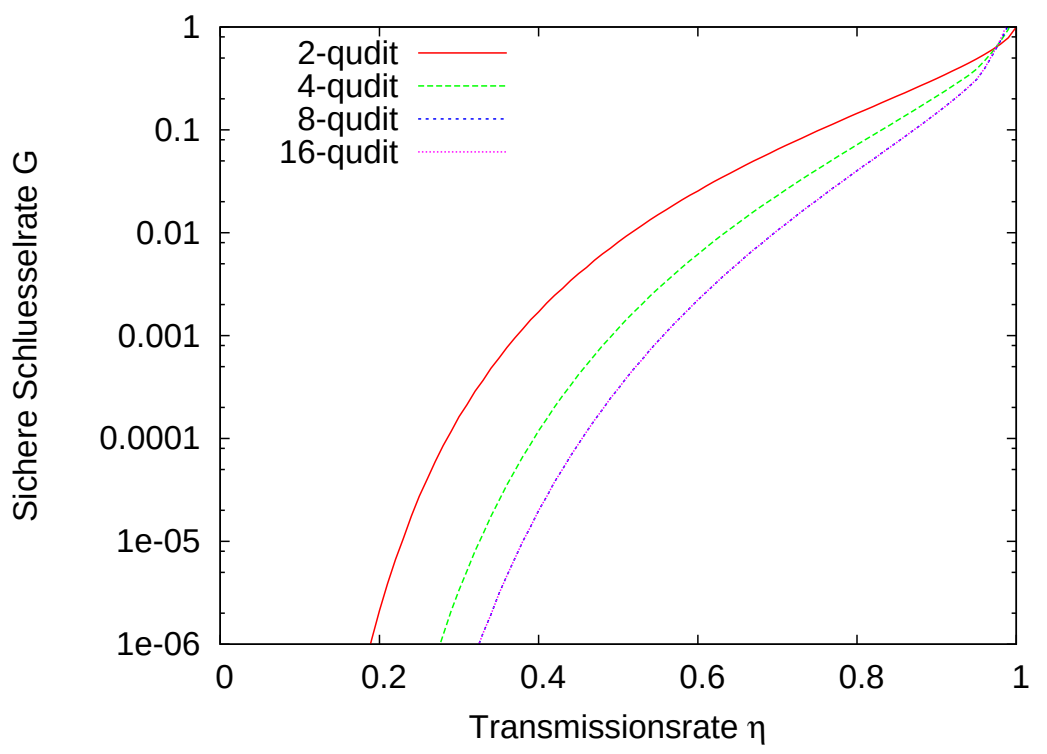

(a)

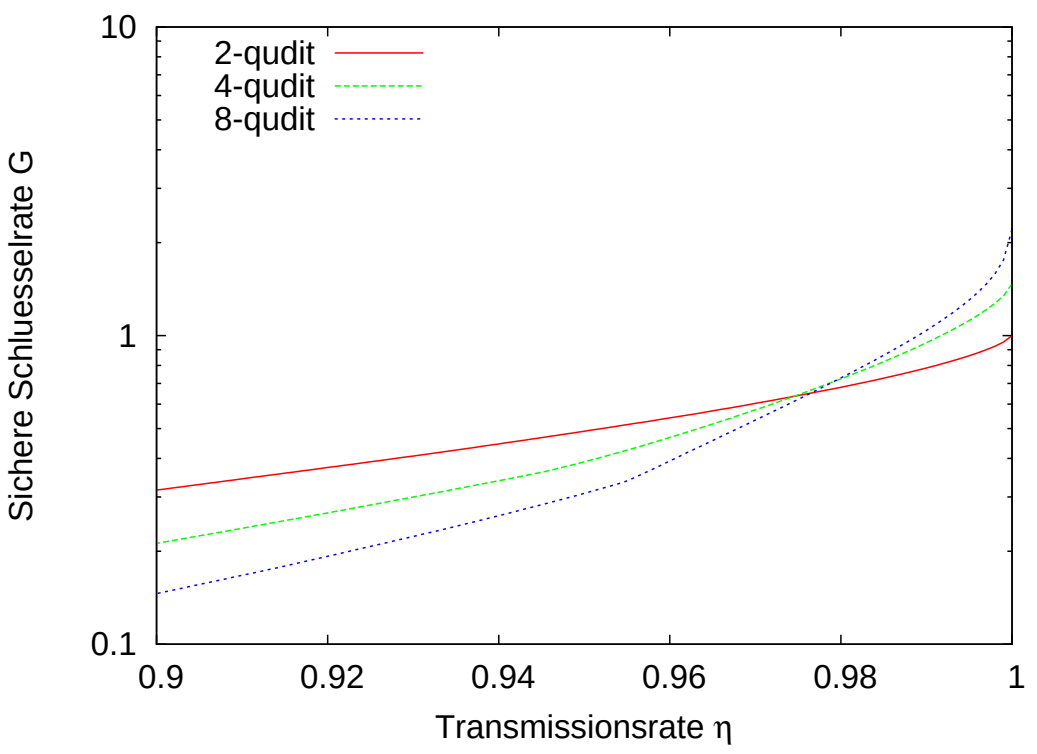

(b)

Abbildung 4.6.: Abb. (a) zeigt die sichere Schlüsselrate (*direct reconciliation*) für 2-, 4-, 8- und 16-Qudit-Zustände, normiert auf ein Bit ($\chi = 0$), Abb. (b) eine Vergrößerung im Bereich nahezu perfekter Transmissionsrate.

### Wahl einer optimalen Phase

Perfekte Transmissionsrate bedeutet, dass es von Vorteil ist, alle Zustände auch perfekt voneinander unterscheiden zu können. Da allerdings bei der bisherigen Betrachtung als Messachse immer die reelle Achse gewählt wurde, sind nur die beiden Zustände eindeutig voneinander unterscheidbar, die auf der reellen Achse liegen. Alle anderen Zustände haben einen Partner mit gleichem Realteil und negativem Imaginärteil. Durch die Art der Messung fallen beide zusammen. Schaut man sich nun ein System mit einer ungeraden Anzahl an Zuständen an, also z. B. ein 3-Qubit-System, so ist es naheliegend, die Messachse um $\pi/2$ zu drehen, sodass die verschiedenen Zustände nach einer Projektion auf die reelle Achse grösstmöglichen Abstand zueinander haben und die Abstände symmetrisch zu den jeweiligen Nachbarn verteilt sind. Dies ist natürlich auch für ein 4-Qubit-System möglich (Abb. 4.1(f)). Die Realteile der 4 Zustände sind genau dann maximal voneinander entfernt, wenn sie alle äquidistant sind. Man muss also nur den Abstand von $|\alpha_0\rangle$ und $|\alpha_3\rangle$ mit dem von $|\alpha_3\rangle$ und $|\alpha_1\rangle$ gleichsetzen. Dies bedeutet aufgrund der Symmetrie dieses Systems keine Beschränkung der Allgemeinheit. Für diese Abstände ergibt sich:

$$\cos(\chi) - \cos\left(\frac{3}{2}\pi + \chi\right) \stackrel{!}{=} \underbrace{\cos\left(\frac{3}{2}\pi + \chi\right)}_{\sin(\chi)} - \underbrace{\cos\left(\frac{1}{2}\pi + \chi\right)}_{-\sin(\chi)} \tag{4.47}$$

$$\Rightarrow \cos(\chi) = 3\sin(\chi) \tag{4.48}$$

$$\Rightarrow \chi = \arctan\left(\frac{1}{3}\right) \quad . \tag{4.49}$$

Es stellt sich die Frage, ob dieser Frage für alle Systeme, also auch solche mit einer geraden Anzahl an Zuständen, verallgemeinert werden kann. Es ergeben sich allerdings zwei Forderungen, die sich nicht gleichzeitig erfüllen lassen:

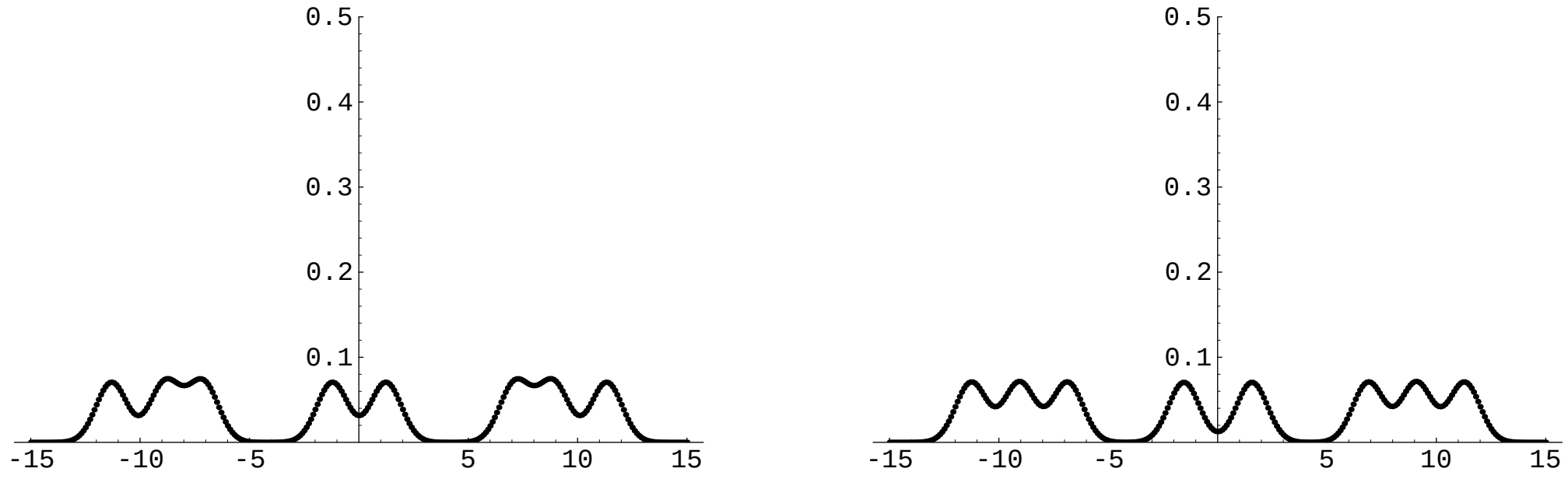

(a) (b)

Abbildung 4.7.: Abb. (a) zeigt die Wahrscheinlichkeitsverteilung von $p(\beta_x)$ für ein 8-Qudit-System mit $\chi = (1 - \cos(\pi/4))/(2 + \sin(\pi/4))$ („globale Optimierung“), Abb. (b) mit $\chi = (1 - \cos(\pi/4))/(3 \sin(\pi/4))$ („lokale Optimierung“).

- Die beste zu erreichende Gleichverteilung ergibt sich, wenn die Differenz auf der reellen Achse zwischen den beiden Zuständen, welche bei $\chi = 0$ rein imaginär sind und der Differenz des positiv reellen Zustandes mit dem „letzten“ Zustand, also $|\alpha_{d-1}\rangle$ gleich sind. In anderen Worten: die Projektion eines Kreises auf die reelle Achse bedeutet eine Abbildung. Diese Abbildung hat bei $x = 0$ keine Steigung und bei $x = r$ maximale Steigung. Betrachtet man zwei benachbarte Zustände im Bereich der minimalen Steigung und zwei im Bereich der maximalen Steigung und setzt deren Abstand gleich, so kann man dies als eine sehr intuitive Methode bezeichnen, die Abbildungen der Zustände möglichst gleich zu verteilen. Um eine wirkliche Gleichverteilung auf der reellen Achse zu erhalten, müsste man die Zustände mit unterschiedlichen Distanzen auf dem Kreis verteilen, wozu allerdings eine weitere Basistransformation nötig wäre. Am Beispiel eines 8-Qubit-Systems ist das Ergebnis in Abb. 4.7(a) zu sehen.

  Aus dieser Argumentation ergibt sich:

$$\cos(\chi) - \cos\left(-\frac{2}{d} + \chi\right) = \underbrace{\cos\left(\frac{3}{2}\pi + \chi\right)}_{\sin(\chi)} - \underbrace{\cos\left(\frac{1}{2}\pi + \chi\right)}_{-\sin(\chi)} \tag{4.50}$$

$$\Rightarrow \tan(\chi) = \frac{1 - \cos\left(\frac{2}{d}\pi\right)}{2 + \sin\left(\frac{2}{d}\pi\right)} \quad . \tag{4.51}$$

- Die zweite Formulierung ergibt sich aus der Betrachtung zweier benachbarter Zustände. Betrachtet man die Differenzen auf der reellen Achse zwischen dem Zustand $|\alpha_0\rangle$ und seinen beiden Nachbarzuständen $|\alpha_1\rangle$ und $|\alpha_{d-1}\rangle$, so ergibt sich als Forderung:

$$\cos(\chi) - \cos\left(-\frac{2\pi}{d} + \chi\right) = \cos\left(-\frac{2\pi}{d} + \chi\right) - \cos\left(\frac{2\pi}{d} + \chi\right) \tag{4.52}$$

$$\Rightarrow \tan\chi = \frac{1 - \cos\left(\frac{2\pi}{d}\right)}{3 \sin\left(\frac{2\pi}{d}\right)} \quad . \tag{4.53}$$

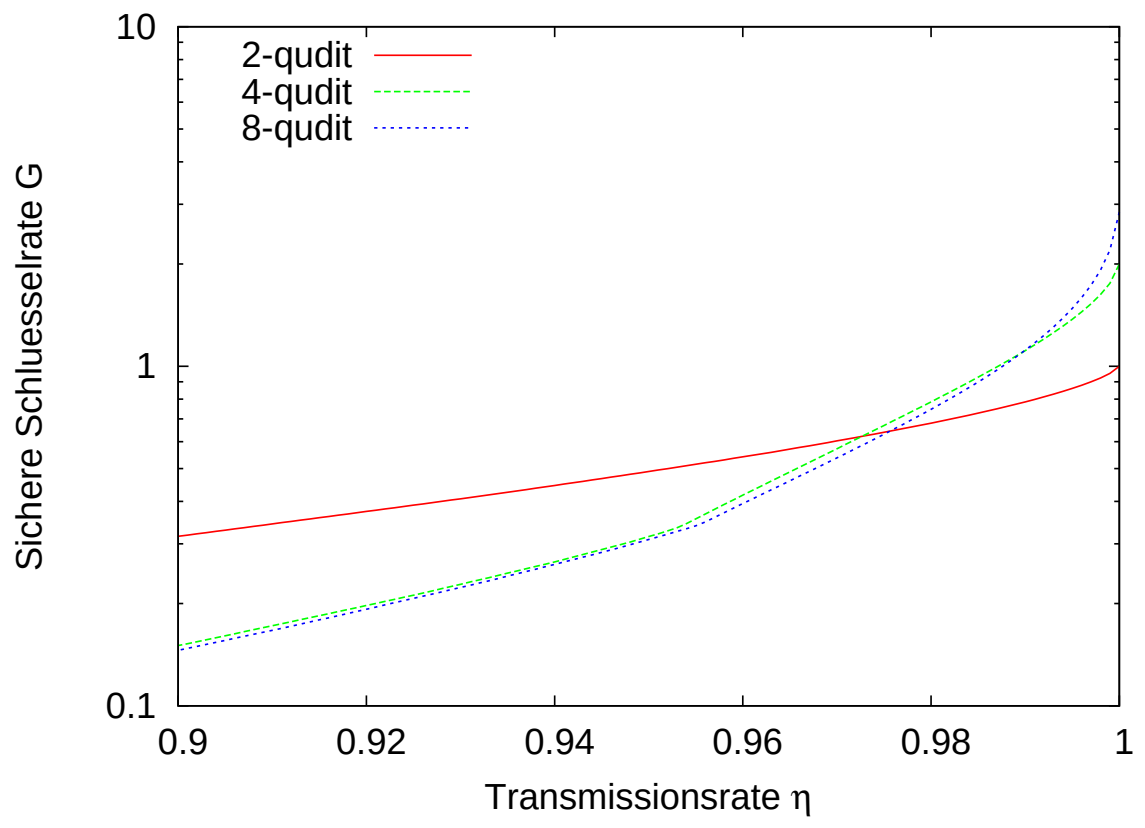

Abbildung 4.8.: Sichere Schlüsselrate (*direct reconciliation*) für 2-, 4- und 8-Qudit-Zustände, wobei $\chi = \arctan(1/3)$ für das 4-Qudit-System und $\chi = \arctan((1 - \cos(\pi/4))/(3\sin(\pi/4)))$ für das 8-Qudit-System gewählt wurde. Wie man sehen kann, ist im optimalen Fall die Übertragung von 2 Bit für das 4-Qudit-System möglich.

Diese Formulierung erzielt ein deutlich besseres Ergebnis, wie man anhand des Beispiels für 8 Dimensionen sehen kann (Abb. 4.7(a)).

Für das 4-dimensionale System fallen diese beiden Aussagen aufgrund der Symmetrie zusammen, in höheren Dimensionen widersprechen sie sich allerdings. Dort versagt die erste Formulierung vollständig, sodass die zweite als die sinnvollere angesehen werden kann und im Folgenden auch Anwendung findet.

In Abbildung 4.8 wird diese Betrachtungsweise für ein 4- und ein 8-Qudit-System angewandt. Das 2-Qubit-System gilt als Vergleich. Es ist zu sehen, dass im Falle des 4-Qudit-Systems die Übertragung von 2 Bit noch möglich ist durch die symmetrische Verteilung der Zustände, sich allerdings im Falle der 8-Qubit-Systems nur $2,85$ Bit übertragen lassen und nicht 3 (mit der ersten Symmetrieüberlegung lassen sich nur $2,82$ Bit übertragen). Dies lässt sich also damit erklären, dass die Distanz zwischen zwei Zuständen nicht immer gleich ist und sich damit benachbarte Zustände nicht immer „symmetrisch“ voneinander unterscheiden lassen.

## 4.5. Reverse reconciliation

Im Falle der *reverse reconciliation* wird keine klassische Information von Alice zu Bob übertragen, um die Ergebnisse aus Präparation und Messung aufeinander abzustimmen, sondern es wird klassische Information von Bob zu Alice übertragen. Solange man davon ausgeht, dass keine quantenmechanischen Informationen über die Messung von Bob auf den Quantenkanal gelangen, hat Eve auch keinerlei Information über dessen Meßergebnisse. Es ist zu vermuten, dass Bob mit klassischen Information in diesem Fall weniger anfangen kann, als mit der klassischen Information im Falle der *direct reconciliation*.

Das Protokoll verändert sich also dahingehend, dass Alice präparierte Quantenzustände an Bob versendet, Bob diese misst und Alice die erwartete Amplitude über einen klassischen Kanal mitteilt. D. h. Bob nimmt an, seine Messung sei perfekt. Da seine Messung aber eine Projektion auf eine Achse ist, von Zuständen, die kreisförmig angeordnet sind, kann er durch den gemessenen Wert einen fiktiven Radius bestimmen. Ausgehend von einem erwarteten Radius sind so manchmal zwei verschiedene Radien möglich, je nach dem, wieviele unterschiedliche Zustände möglich sind. Weil Alice und Bob die Transmittivität des Kanals kennen (z. B. durch (un)regelmässige Messungen), kann daraus natürlich Alice erfahren, welchen Zustand Bob gemessen haben muss. Eve kann mit dieser Information weniger anfangen, da er weder weiss, welchen Zustand Alice wirklich präpariert hat und deshalb auch nicht, welchen Bob gemessen hat. Es kann in diesem Fall nämlich durchaus sein, dass Bob einen anderen Zustand misst als den, den Alice präpariert hat. Alice weiß aber durch die klassische Korrekturinformation von Bob, welchen Zustand Bob gemessen hat und so den Schlüssel an dieser Stelle eben so anpasst, als wäre die Messung von Bob richtig gewesen und die Präparation „falsch“.

### 4.5.1. Sichere Schlüsselrate

Um wiederum die sichere Schlüsselrate zwischen Alice und Bob zu berechnen, geht man analog dem Schema der *direct reconciliation* vor. Für die Präparation ändert sich nichts und obwohl aufgrund der Meßergebnisse von Bob der Schlüssel generiert wird, kann man von einer Gleichverteilung der Zustände ausgehen. Somit kann man die Transinformation zwischen Alice und Bob ohne Änderung betrachten.

Allerdings muss man die *Holevoinformation* neu berechnen, da Eve zwar die gleiche Information über den Quantenkanal erhält, allerdings erst nach der Übertragung der klassischen Information Überlegungen darüber anstellen kann, welchen Zustand Bob gemessen hat. Somit lässt sich der Zustand, den Eve erhält durch eine Dichtematrix in der Form

$$\rho_m^d = \sum_{k=0}^{d-1} \frac{p(\beta_x|\alpha_k^d)}{\sum\limits_{l=0}^{d-1} p(\beta_x|\alpha_l^d)} |\epsilon_k^d\rangle\langle\epsilon_k^d| \tag{4.54}$$

beschreiben, wobei Eve nur mit einer Wahrscheinlichkeit $1/d$ eine dieser $d$ Dichtematritzen enthält. Da sich die verschiedenen Dichtematrizen nur durch eine Phase unterscheiden, kann man sie durch unitäre Transformationen ineinander überführen,

$$\rho_m^d = U\rho_l^d U^\dagger \quad \forall n \quad . \tag{4.55}$$

Dabei ändert sich die Entropie nicht, es muss also nur eine Dichtematrix betrachtet werden.

Im Falle der *direct reconciliation* ließen sich die Zustände $|\epsilon_k^d\rangle$ in der orthonormalen Basis der $|\Phi_k^d\rangle$ diagonalisieren (4.39). Nun kann man wiederum diese Basis nutzen, allerdings gibt es auch Nebendiagonalelemente, es gilt für $\rho_m^d$ nach einer Basistransformation:

$$\rho_m^d = \sum_{k=0}^{d-1}\sum_{l=0}^{d-1}\sum_{l'=0}^{d-1} \frac{p(\beta_x|\alpha_k^d)}{\sum\limits_{m=0}^{d-1} p(\beta_x|\alpha_m^d)} c_l c_{l'}^* \mathrm{e}^{i\frac{2\pi}{d}k(l-l')} |\Phi_l^d(|\epsilon|)\rangle\langle\Phi_{l'}^d(|\epsilon|)| \quad . \tag{4.56}$$

Um die Holevoinformation berechnen zu können, muss der Logarithmus dieser Dichte-Matrix berechnet werden. Dies ist sehr einfach möglich, wenn man die Matrix zuvor diagonalisiert bzw. ihre Eigenwerte berechnet. Da diese Umformung nicht trivial erscheint und auch allgemein für $d$ Dimensionen gelöst werden müsste, liegt eine numerische Betrachtung nahe. Allerdings müssen dazu erst einige Überlegungen angestellt werden.

Aus (4.42) ist ersichtlich, dass man nur eine Aussage über die Betragsquadrate der Koeffizienten $c_l$ treffen kann. Somit kennt man deren Amplitude, aber nicht die Phase. Um (4.56) diagonalisieren können, ohne die Phase der Koeffizienten $c_l$ zu kennen, scheint es sinnvoll, zuerst die Struktur dieser Matrix genauer zu betrachten.

Ähnlich dem Beispiel der *direct reconciliation* wird nun der 4-dimensionale Fall betrachtet. Es gilt

$$\rho_0^4 = \begin{pmatrix} |c_0|^2 & c_0c_1^*\sigma_3 & c_0c_2^*\sigma_2 & c_0c_3^*\sigma_1 \\ c_1c_0^*\sigma_1 & |c_1|^2 & c_1c_2^*\sigma_3 & c_1c_3^*\sigma_2 \\ c_2c_0^*\sigma_2 & c_2c_1^*\sigma_1 & |c_2|^2 & c_2c_3^*\sigma_3 \\ c_3c_0^*\sigma_3 & c_3c_1^*\sigma_2 & c_3c_2^*\sigma_1 & |c_3|^2 \end{pmatrix} \tag{4.57}$$

in der Basis der $|\Phi_l^4\rangle$. Diese Matrix lässt sich interpretieren als eine Basistransformation einer zirkulanten Matrix:

$$\rho_0^4 = \begin{pmatrix} c_0 & 0 & 0 & 0 \\ 0 & c_1 & 0 & 0 \\ 0 & 0 & c_2 & 0 \\ 0 & 0 & 0 & c_3 \end{pmatrix} \begin{pmatrix} \sigma_0 & \sigma_3 & \sigma_2 & \sigma_1 \\ \sigma_1 & \sigma_0 & \sigma_3 & \sigma_2 \\ \sigma_2 & \sigma_1 & \sigma_0 & \sigma_3 \\ \sigma_3 & \sigma_2 & \sigma_1 & \sigma_0 \end{pmatrix} \begin{pmatrix} c_0^* & 0 & 0 & 0 \\ 0 & c_1^* & 0 & 0 \\ 0 & 0 & c_2^* & 0 \\ 0 & 0 & 0 & c_3^* \end{pmatrix} \quad . \tag{4.58}$$

Die $\sigma_k$ berechnen sich wie folgt:

$$\sigma_k = \frac{\sum\limits_{l=0}^{3} \mathrm{e}^{-i\frac{2\pi}{4}k(l+1)} p(\beta_x|\alpha_l^4)}{\sum\limits_{l=0}^{3} p(\beta_x|\alpha_l^4)} \quad . \tag{4.59}$$

Nun kann $\sigma_k$ ebenfalls zerlegt werden durch eine *Fourier-Matrix*. Diese Form gilt für andere Dimensionen analog, man kann also für $(\rho_0^d)_{n,m}$ schreiben, wobei das Matrixelement $n, m$ der Dichtematrix in der Basis der $|\Phi_k^d\rangle$ gemeint ist:

$$(\rho_0^d)_{n,m} = c_n c_m^* \sigma_{(n-m)\mathrm{mod}d} \tag{4.60}$$

$$= c_n c_m^* \frac{\sum\limits_{l=0}^{d-1} \mathrm{e}^{-i\frac{2\pi}{d}(n-m)(l+1)} p(\beta_x|\alpha_l^d)}{\sum\limits_{l=0}^{d-1} p(\beta_x|\alpha_l^d)} \quad . \tag{4.61}$$

Hier gilt unabhängig von der Dimension $d$, $\sigma_0 = 1$. Um die Eigenwerte dieser Matrix numerisch berechen zu können, benötigt man folgenden Satz:

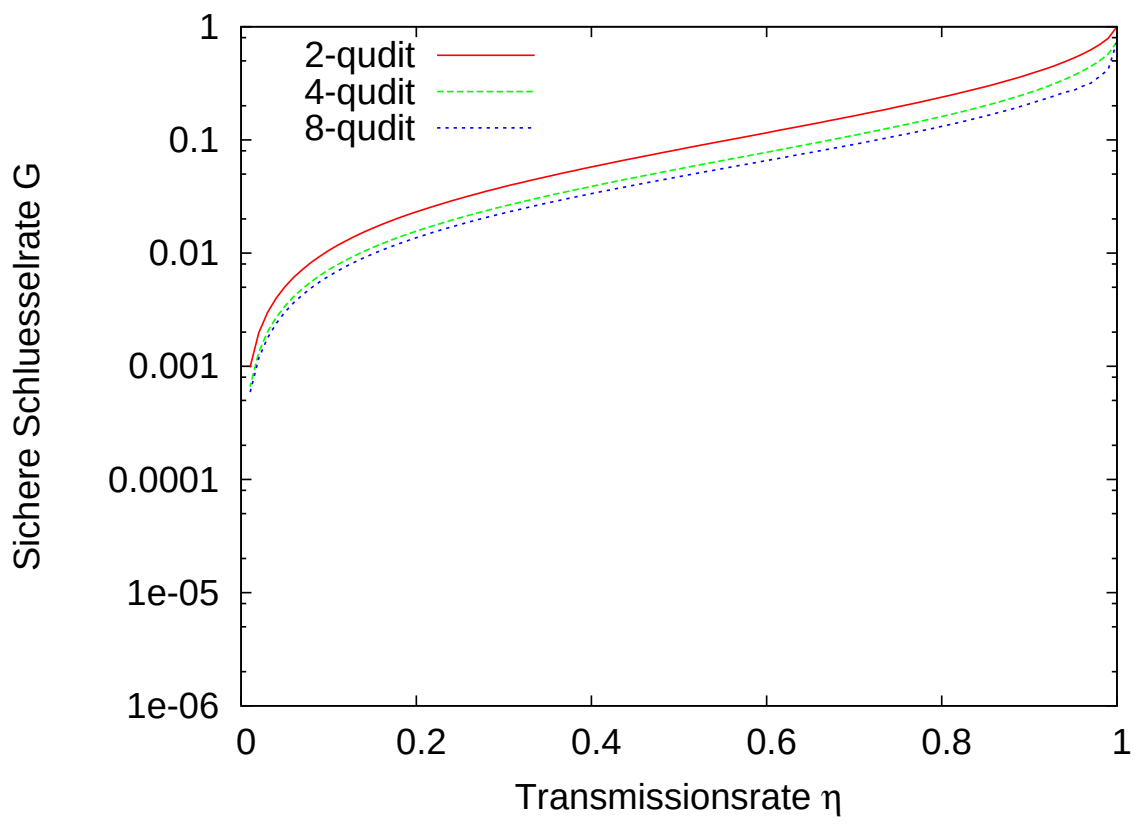

Abbildung 4.9.: Sichere Schlüsselrate (*reverse reconciliation*) für 2-, 4- und 8-Qudit-Zustände, normiert auf 1 für eine optimale Amplitude $|\alpha|$ und mit einer Messung auf der reellen Achse ($\chi = 0$).

**Satz 1.** *Seien die Matrizen* $A = (a_{ij}), C = (c_{ij}) \in \mathbb{C}^{n\times n}$ *mit* $c_{ij} = c_i\delta_{ij}$ *und* $|C| = \sqrt{C^\dagger C}$ *gegeben.*

*Dann gilt:*

$$\det(CAC^\dagger - \lambda E) = \det(|C|A|C| - \lambda E) \quad . \tag{4.62}$$

Der Beweis findet sich im Anhang A. Man benötigt nach diesem Satz also nicht die Phase der $c_l$, sondern nur deren Amplitude. D. h. man muss nur die Eigenwerte der Matrix

$$(\rho'^d_0)_{n,m} = |c_n||c_m| \frac{\sum\limits_{l=0}^{d-1} \mathrm{e}^{-i\frac{2\pi}{d}(n-m)(l+1)} p(\beta_x|\alpha_l^d)}{\sum\limits_{l=0}^{d-1} p(\beta_x|\alpha_l^d)} \tag{4.63}$$

berechnen, wozu man mithilfe numerischer Verfahren in der Lage ist. Dazu wird die numerische Bibliothek NAG (*Numerical Algorithms Group* aus Oxford) verwendet.

Der erste Term der Holevoinformation bleibt im Vergleich zur *direct reconciliation* unverändert und man kann die berechneten Resultate übernehmen.

### 4.5.2. Resultate

Für den Fall der *reverse reconciliation* lassen sich die sicheren Schlüsselraten analog dem Fall der *direct reconciliation* mit dem im Anhang B aufgeführen Program berechnen. Allerdings findet zur Diagonalisierung der Dichtematrix (4.60) die numerische Bibliothek NAG Anwendung.

Für eine Normierung auf 1, also die jeweilige Informationseinheit, sind die sicheren Schlüsselraten für Systeme höherer Ordnung nur unwesentlich schlechter, als das System mit zwei Zuständen (Abbildung 4.9). Interessant für die sichere Schlüsselrate sind ausschliesslich die Ergebnisse mit einer optimierten Amplitude. Renormiert man nun erneut die sichere Schlüsselrate auf die Betrachtung eines Bit, so erhält man bessere Ergebnisse für höhere Ordnungen unabhängig der Transmissionsrate $\eta$ für einen Messwinkel $\chi = 0$ (Abbildung 4.10(a)). In der zugehörigen Ausschnittvergrößerung (Abbildung 4.10(b)) sieht man ein analoges Verhalten des Systems mit 4 Zuständen. Die Transmissionsrate kann maximal zwei erreichen, allerdings ist die Messachse dafür ungünstig gewählt. Der Plot des Systems mit 8 Zuständen ist aufgrund der Matrixdiagonalisierung in 8 Dimensionen etwas unschön. Allerdings ist v. a. aus den Daten ersichtlich, dass für einen perfekten Kanal bereits ohne Optimierung der Amplitude die gleiche Rate erreicht wird wie im Falle der *direct reconciliation*.

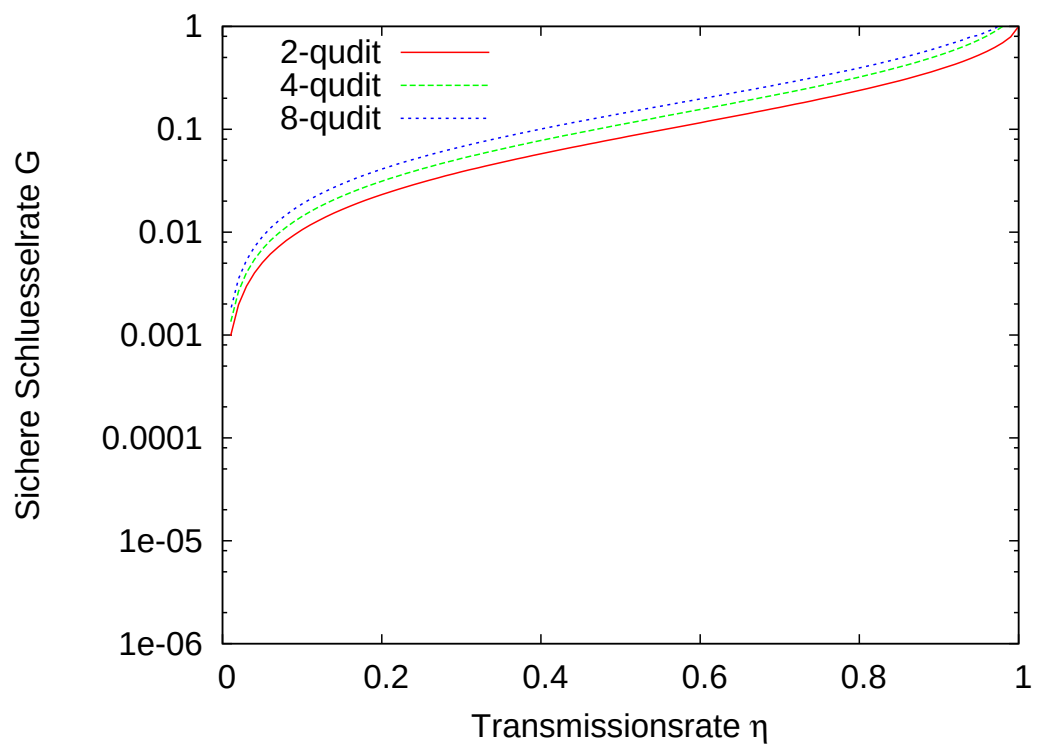

(a)

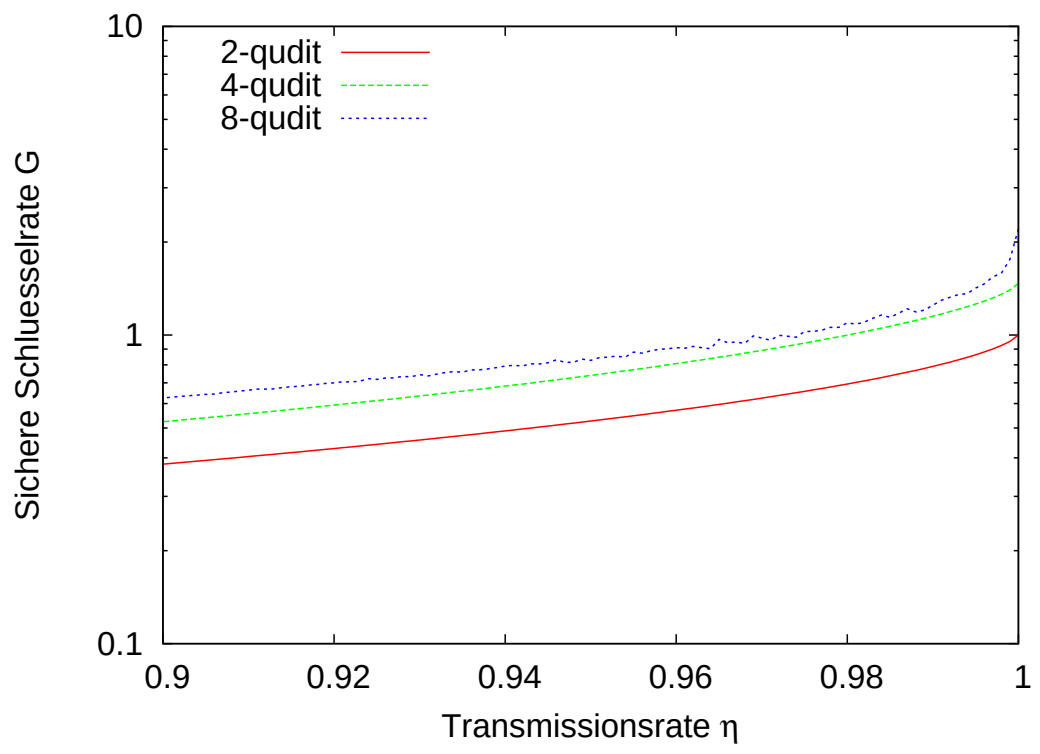

(b)

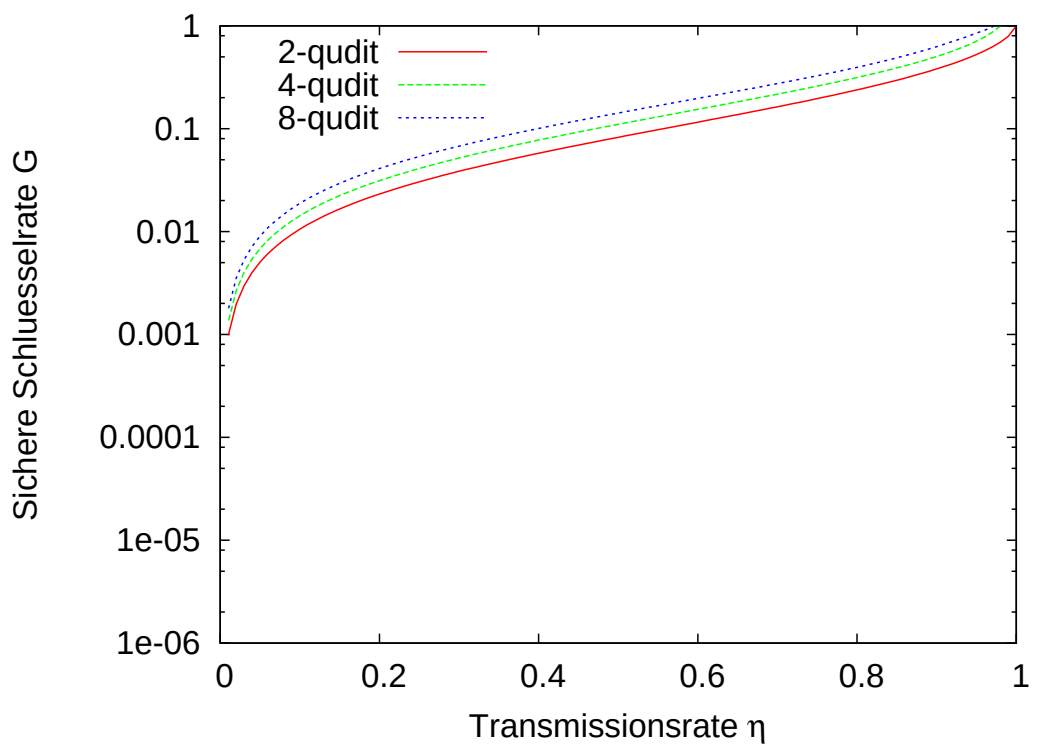

(c)

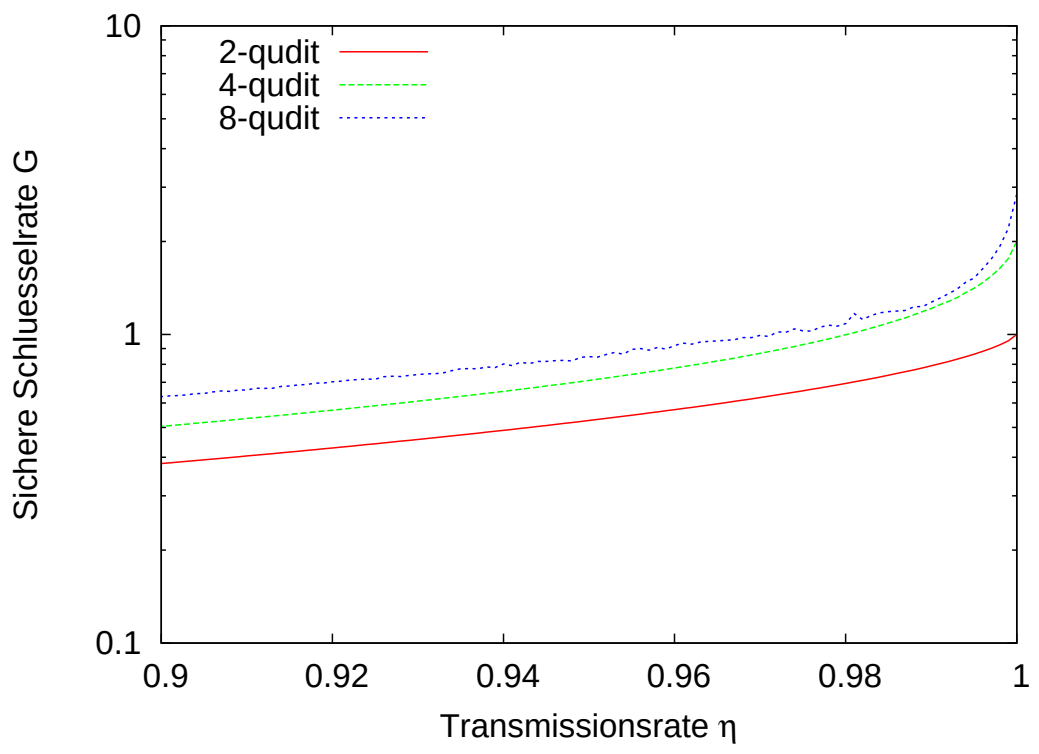

(d)

Abbildung 4.10.: Sichere Schlüsselrate im Falle der *reverse reconciliation* für 2-, 4- und 8-Qudit-Zustände, normiert auf ein Bit. Abbildung (a) zeigt den Fall für $\chi = 0$, Abbildung (b) eine Ausschnittsvergrößerung im Bereich perfekter Transmission. Für Abbildung (c) wurde die Meßachse optimiert mit $\chi = \arctan(1/3)$ für das 4-Qudit-System und $\chi = \arctan((1-\cos(\pi/4))/(3\sin(\pi/4)))$ für das 8-Qudit-System. Abbildung (d) zeigt wiederum die zugehörige Vergrößerung. Der unsaubere Verlauf resultiert aus Konvergenzproblemen der Diagonalisiermethoden der Bibliothek NAG.

Betrachtet man die zugehörigen Plots mit einer optimierten Messachse (Abbildungen 4.10(c) und 4.10(d)), so ändert sich der Verlauf des 4-Qudit-Systems im oberen Bereich wie erwartet. Allerdings ändert sich dort der Verlauf des 8-Qudit-Systems nicht. Ein Vergleich der Werte der Plots der optimierten und der nicht-optimierten Phase zeigt für das 4- und das 8-Qudit-System, eine völlige Übereinstimmung im Rahmen einer numerischen Ungenauigkeit, bis auf den höheren Wert des 4-Qubit-Systems für perfekte Transmission, der wiederum aus der größeren Symmetrie des 4-Zustands-Systems stammt.

## 4.6. Quetschen der Zustände

Eine weitere Verallgemeinerung des Protokolls lässt sich erreichen, indem man keine *kohärenten Zustände* betrachtet, sondern *gequetschte Zustände*. Als *gequetschte Zustände* werden in diesem Kapitel Zustände bezeichnet, welche entstehen, indem man einen *kohärenten Vakuumzustand* quetscht und anschließend verschiebt. Häufig bezeichnet man in der Literatur als *gequetschte Zustände* gequetschtes Vakuum. Neben der Idee der weiteren Verallgemeinerung steht die Idee, dass es ein Vorteil sein könnte, *gequetschte Zustände* auf einem Kreis anzuordnen, wenn man anschliessend nur auf einer Achse misst. Dies würde der Symmetrie dieser Achse entgegenkommen, wenn die Zustände in Richtung der Achse gequetscht werden. Die Verteilung wäre also in Meßrichtung schärfer.

### 4.6.1. Beschreibung

Da es sich um eine weitere Verallgemeinerung handelt, sollen die Parameter des Protokolls erhalten bleiben und nur ein weiterer Quetschparamter hinzugefügt werden. Weil Quetschung der Messung und damit der Projektion auf eine Achse dienen soll, ist eine Betrachtung unterschiedlicher Quetschwinkel nicht interessant, es wird $\phi = 0$ gewählt, um maximale Quetschung durch den Quetschradius $r$ zu erhalten, wenn für den Quetschparameter $\xi$ gilt $\xi = r \cdot \mathrm{e}^{i\phi}$.

Alice präpariert nun einen Zustand $|\alpha_k^d; \mu\rangle$, wobei mit $\mu$ der Quetschparameter unter der Bedingung $\phi = 0$ bezeichet wird, also der Quetschradius. Bob misst wieder mittels Homodynexperiment einen *kohärenten Zustand* $|\beta_x\rangle$.

### 4.6.2. Transinformation zwischen Alice und Bob

Um nun die Transinformation zwischen Alice und Bob zu bestimmen, benötigt man analog der Rechnung für die *kohärenten Zustände* die bedingte Wahrscheinlichkeit $p(\beta_x|\alpha_{k,\mu}^d)$ und die Wahrscheinlichkeit $p(\beta_x)$.

Allerdings muss man sich auch noch die Frage stellen, wie sich ein *gequetschter Zustand* in einem Strahlteiler verhält. Der betrachtete Zustand, den Alice präpariert, lässt sich darstellen durch $D(\alpha)S(\mu)|0\rangle$. Nimmt man nun diesem Zustand seine Quetschung vor dem Strahlteiler und quetscht ihn danach erneut, handelt es sich mathematisch um denselben Vorgang, da der Quetschoperator unitär ist. Man muss jedoch wissen, dass die zusätzliche Störung hinter dem Strahlteiler nun ebenfalls gequetscht ist. Diese Operation lässt sich schreiben als:

$$BS(\mu)D(\alpha)S(\mu) = S(\mu)BS(\eta)S^*(\mu)\underbrace{D(\alpha)S(\mu)}_{S(\mu)D(\alpha')}|0\rangle \tag{4.64}$$

$$= S(\mu)\underbrace{BS(\eta)D(\alpha')}_{D(\sqrt{\eta}\alpha')}|0\rangle \tag{4.65}$$

$$= D(\gamma)S(\mu)|0\rangle \quad . \tag{4.66}$$

Nun lassen sich $\alpha'$ und $\gamma$ bestimmen durch die *Bogoliubov-Transformation*:

$$\alpha' = \alpha\cosh r + \alpha^* \mathrm{e}^{2i\phi}\sinh r \tag{4.67}$$

$$\sqrt{\eta}\alpha' = \gamma\cosh r + \gamma^* \mathrm{e}^{2i\phi}\sinh r \quad , \tag{4.68}$$

also gilt $\gamma = \sqrt{\eta}\alpha$. Ein *gequetschter Zustand* verhält sich also beim Durchgang durch einen Strahlteiler genauso, wie ein *kohärenter Zustand*.

Für $p(\beta|\alpha^d_{k,\mu})$ gilt:

$$p(\beta|\alpha^d_{k,\mu}) = \frac{1}{\pi}|\langle\beta|\sqrt{\eta}\alpha^d_{k,\mu}\rangle|^2 \tag{4.69}$$

$$= \frac{1}{\pi}|\langle 0|D^*(\beta)D(\sqrt{\eta}\alpha^d_k)S(\mu)|0\rangle|^2 \tag{4.70}$$

$$= \frac{1}{\pi}|\langle 0|D^*(\beta - \sqrt{\eta}\alpha^d_k)S(\mu)|0\rangle|^2 \quad , \tag{4.71}$$

Die zusätzliche Phase, die man aufgrund der Gruppeneigenschaften des Verschiebeoperators erhält, wird durch den Betrag gleich 1. Nun gilt mit (2.33) und (2.57) und unter der Bedingung $\phi = 0$:

$$p(\beta|\alpha^d_{k,\mu}) = \frac{\mathrm{e}^{-|\beta-\sqrt{\eta}\alpha^d_k|^2}}{\pi\cosh r}\left|\sum_{n,m=0}^{\infty}\frac{(\beta-\sqrt{\eta}\alpha^d_k)^{*m}}{\sqrt{m!}}\frac{\sqrt{(2n)!}}{2^n n!}(-\tanh r)^n\langle m|2n\rangle\right|^2 \tag{4.72}$$

$$= \frac{\mathrm{e}^{-|\beta-\sqrt{\eta}\alpha^d_k|^2}}{\pi\cosh r}\left|\sum_{n=0}^{\infty}\frac{(\beta^*-\sqrt{\eta}(\alpha^d_k)^*)^{2n}}{2^n n!}(-\tanh r)^n\right|^2 \quad , \tag{4.73}$$

und weiterhin mit $\mathrm{e}^x = \sum_{n=0}^{\infty}\frac{x^n}{n!}$:

$$p(\beta|\alpha^d_{k,\mu}) = \frac{\mathrm{e}^{-|\beta-\sqrt{\eta}\alpha^d_k|^2}}{\pi\cosh r}\mathrm{e}^{-(\beta^*-\sqrt{\eta}(\alpha^d_k)^*)^2\frac{\tanh r}{2}-(\beta-\sqrt{\eta}\alpha^d_k)^2\frac{\tanh r}{2}} \tag{4.74}$$

$$= \frac{1}{\pi\cosh r}\mathrm{e}^{-\left(\beta_x^2(1+\tanh r)+\eta|\alpha|^2(1+\cos(\frac{4\pi}{d}k-2\chi)\tanh r)-2\sqrt{\eta}|\alpha|\beta_x\cos(\frac{2\pi}{d}k-\chi)(1+\tanh r)\right)}$$
$$\times\, \mathrm{e}^{-(1-\tanh r)\left(\beta_y^2-2\sqrt{\eta}|\alpha|\beta_y\sin(\frac{2\pi}{d}k-\chi)\right)} \tag{4.75}$$

Um nun $p(\beta_x|\alpha^d_{k,\mu})$ berechnen zu können, muss ein Integral über den letzten Faktor gebildet werden:

$$L = \int \mathrm{d}\beta_y \mathrm{e}^{-(1-\tanh r)\left(\beta_y^2 - 2\sqrt{\eta}|\alpha|\beta_y \sin(\frac{2\pi}{d}k-\chi)\right)} \tag{4.76}$$

$$= \mathrm{e}^{\eta|\alpha|^2 \sin^2(\frac{2\pi}{d}k-\chi)(1-\tanh r)} \cdot \int \mathrm{d}\beta_y \mathrm{e}^{-(1-\tanh r)\left(\beta_y - \sqrt{\eta}|\alpha|\sin^2(\frac{2\pi}{d}k-\chi)\right)^2} \tag{4.77}$$

$$= \mathrm{e}^{\eta|\alpha|^2 \sin^2(\frac{2\pi}{d}k-\chi)(1-\tanh r)} \cdot \int \mathrm{d}y \mathrm{e}^{-(1-\tanh r)y^2} \quad \text{mit } y = \beta_y - \sqrt{\eta}|\alpha|\sin^2\left(\frac{2\pi}{d}k-\chi\right) \tag{4.78}$$

Daraus folgt wegen $\int \mathrm{e}^{-ax^2}\mathrm{d}x = \sqrt{\pi/a}$ für $a > 0$:

$$L = \sqrt{\frac{\pi}{1-\tanh r}} \mathrm{e}^{\eta|\alpha|^2 \sin^2(\frac{2\pi}{d}k-\chi)(1-\tanh r)} \quad . \tag{4.79}$$

Somit gilt für $p(\beta_x|\alpha^d_{k,\mu})$:

$$p(\beta_x|\alpha^d_{k,\mu}) = \frac{\mathrm{e}^{-(\beta_x^2(1+\tanh r))}}{\sqrt{\pi(1-\tanh r)}\cosh r} \times \mathrm{e}^{-\eta|\alpha|^2\left(1+\cos(\frac{4\pi}{d}k-2\chi)\tanh r - \sin^2(\frac{2\pi}{d}k-\chi)(1-\tanh r)\right)} \times \mathrm{e}^{2\sqrt{\eta}|\alpha|\beta_x \cos(\frac{2\pi}{d}k-\chi)(1+\tanh r)} \tag{4.80}$$

Aus den Identitäten $\cos(2\phi) = 1 - 2\sin^2(\phi)$ und $\cos^2(\phi) + \sin^2(\phi) = 1$ folgt für den mittleren Faktor:

$$\mathrm{e}^{-\eta|\alpha|^2\left(1+\cos(\frac{4\pi}{d}k-2\chi)\tanh r - \sin^2(\frac{2\pi}{d}k-\chi)(1-\tanh r)\right)} \tag{4.81}$$

$$= \mathrm{e}^{-\eta|\alpha|^2 \cos^2(\frac{2\pi}{d}k-\chi)(1+\tanh r)} \tag{4.82}$$

Durch Anwenden der Binomischen Formel erhält man schliesslich

$$p(\beta_x|\alpha^d_{k,\mu}) = \frac{\mathrm{e}^{-(1+\tanh r)(\beta_x - \sqrt{\eta}|\alpha|\cos(\frac{2\pi}{d}k-\chi))^2}}{\sqrt{\pi(1-\tanh r)}\cosh r} \quad . \tag{4.83}$$

Wegen $p(\beta_x) = \sum_{k=0}^{d-1} p(\beta_x|\alpha^d_{k,\mu})p(\alpha^d_{k,\mu})$ und $p(\alpha^d_{k,\mu}) = 1/d \ \forall k$ lässt sich $p(\beta_x)$ nun ebenfalls leicht berechnen zu:

$$p(\beta_x) = \frac{\sum\limits_{l=0}^{d-1} \mathrm{e}^{-(1+\tanh r)(\beta_x - \sqrt{\eta}|\alpha|\cos(\frac{2\pi}{d}l-\chi))^2}}{d\sqrt{\pi(1-\tanh r)}\cosh r} \quad . \tag{4.84}$$

Es ist weiterhin leicht ersichtlich, dass man für $r = 0$ das Ergebnis für *kohärente Zustände* erhält (4.20), weil $\tanh(0) = 0$ und $\cos(0) = 1$.

### Detektorrauschen

Analog zu (3.43) und (3.44) in Kapitel 3.2.8 lassen sich die Wahrscheinlichkeiten der betrachteten Fälle mit Detektorrauschen berechnen. Da sich die Wahrscheinlichkeiten für *gequetschte Zustände* auf die für *kohärente Zustände* zurückführen lassen, wird hier nur die Erweiterung für *gequetschte Zustände* aus (4.84) angegeben. Es gilt:

$$p^{\text{NOISE}}(\beta_x) = \frac{\sum\limits_{l=0}^{d-1} \mathrm{e}^{-(1+\tanh r)(\beta_x - \sqrt{\eta}|\alpha|\cos(\frac{2\pi}{d}l-\chi))^2/(1+\delta)}}{d\sqrt{\pi(1-\tanh r)(1+\delta)}\cosh r} \quad , \tag{4.85}$$

und

$$p^{\text{NOISE}}(\beta_x|\alpha^d_{k,\mu}) = \frac{\mathrm{e}^{-(1+\tanh r)(\beta_x - \sqrt{\eta}|\alpha|\cos(\frac{2\pi}{d}k-\chi))^2/(1+\delta)}}{\sqrt{\pi(1-\tanh r)(1+\delta)}\cosh r} \quad . \tag{4.86}$$

### 4.6.3. Direct reconciliation

Als obere Schranke der Information, die Eve von Alice bekommt, gilt wiederum die Holevoinformation. Um diese berechnen zu können, benötigt man wiederum den gemittelten Dichteoperator der Zustände von Eve:

$$\bar{\rho} = \frac{1}{d}\sum_{k=0}^{d-1} |\epsilon^d_k;\mu\rangle\langle\epsilon^d_k;\mu| \tag{4.87}$$

Da auch die *gequetschten Zustände* keine Orthonormalbasis bilden, muss man auch hier eine geeignete Transformation finden. Betrachtet man die Transformation der *kohärenten Zustände* in die orthonormale Basis (4.39)

$$|\epsilon^d_k\rangle = \mathrm{e}^{-\frac{1}{2}|\epsilon|^2}\sum_{l=0}^{d-1}\mathrm{e}^{i\frac{2\pi}{d}kl}\mathfrak{e}^{|\epsilon|^2}_{d,l}|\Phi^d_l(|\epsilon|)\rangle = \sum_{l=0}^{d-1} c_l \mathrm{e}^{i\frac{2\pi}{d}kl}|\Phi^d_l(|\epsilon|)\rangle \quad ,$$

so beschreibt die Einheitswurzel dieser Gleichung die Kreissymmetrie der Anordnung der Zustände. Nur die Koeffizienten $c_l$, wie auch die Basisvektoren $|\Phi^d_l(|\epsilon|)\rangle$ sind von der eigentlichen Beschaffenheit der betrachteten Ausgangszustände, also der *kohärenten Zustände* abhängig. Dies wird unter anderem auch daran deutlich, dass die Koeffizienten $c_i$ nur durch die Normierung der *kohärenten Zustände* $\mathrm{e}^{-\frac{1}{2}|\epsilon|^2}$ und die Normierung der neuen Basis mittels $\mathfrak{e}^{|\epsilon|^2}_{d,l}$ zustande kommen.

Sind die betrachteten Zustände z. B. gequetsche Zustände, so bleibt die Einheitswurzel bei gleicher Anordnung der Zustände bestehen und die $c_l$ ändern sich. Es reicht aus, zu fordern, dass die betrachtete Basis orthonormal ist und es gilt dieselbe Zerlegung auch für *gequetschte Zustände*:

$$|\epsilon^d_k;\mu\rangle = \sum_{l=0}^{d-1} c'_l \mathrm{e}^{i\frac{2\pi}{d}kl}|\Phi^d_l(|\epsilon|)\rangle \quad .$$

Im weiteren Verlauf wird $c_l'$ wiederum als $c_l$ angegeben, ist allerdings nicht mit den Koeffizienten der *kohärenten Zustände* zu verwechseln.

Somit gilt analog:

$$|c_l|^2 = \frac{1}{d}\sum_{k=0}^{d-1} \mathrm{e}^{-i\frac{2\pi}{d}kl}\langle\epsilon_0^d;\mu|\epsilon_k^d;\mu\rangle \quad . \tag{4.88}$$

### Überlapp gequetschter Zustände

Der Überlapp zweier *gequetschter Zustände* berechnet sich wie folgt (im weiteren Verlauf wird auf die Angabe der Dimension $d$ der Zustände verzichtet):

$$\langle\epsilon_0;\mu|\epsilon_k;\mu\rangle = \langle 0|S^*(\mu)D^*(\epsilon_0)D(\epsilon_k)S(\mu)|0\rangle \tag{4.89}$$

$$= \mathrm{e}^{-\frac{1}{2}(\epsilon_k\epsilon_0^* - \epsilon_0\epsilon_k^*)}\langle 0|S^*(\mu)D(\epsilon_k-\epsilon_0)S(\mu)|0\rangle \tag{4.90}$$

$$= \mathrm{e}^{-\frac{1}{2}(\epsilon_k\epsilon_0^* - \epsilon_0\epsilon_k^*)}\langle 0|\underbrace{S^*(\mu)S(\mu)}_{1}D(\beta)|0\rangle \tag{4.91}$$

$$\text{mit } \beta = (\epsilon_k-\epsilon_0)\cosh r + \sinh r(\epsilon_k^* - \epsilon_0^*) \tag{4.92}$$

$$= \mathrm{e}^{-\frac{1}{2}(\epsilon_k\epsilon_0^* - \epsilon_0\epsilon_k^*)}\langle 0|\beta\rangle \tag{4.93}$$

$$= \mathrm{e}^{-i(1-\eta)|\alpha|^2\sin\left(\frac{2\pi}{d}k\right)}\mathrm{e}^{-\frac{1}{2}|\beta|^2} \tag{4.94}$$

Somit gilt für die Koeffizienten $c_l$:

$$|c_l|^2 = \frac{1}{d}\sum_{k=0}^{d-1}\mathrm{e}^{-i\frac{2\pi}{d}kl}\left(\mathrm{e}^{-i(1-\eta)|\alpha|^2\sin\left(\frac{2\pi}{d}k\right)}\mathrm{e}^{-\frac{1}{2}|(\epsilon_k-\epsilon_0)\cosh r+\sinh r(\epsilon_k^*-\epsilon_0^*)|^2}\right) \quad . \tag{4.95}$$

## 4.6.4. Reverse reconciliation

Der einzige Unterschied der *reverse reconciliation* zur *direct reconciliation* darin besteht, die Holevoinformation auch für die nicht-gemittelten Dichtematrizen zu berechnen, die Eve erhält. Da in diesem Ausdruck (4.56) nur Terme auftreten, die bereits für den Fall der *direct reconciliation* berechnet wurden und die Symmetrie zur numerischen Lösung aus (4.59) folgt, ist der Fall der *reverse reconciliation* für *gequetschte Zustände* bereits gelöst.

## 4.6.5. Resultate

Die *gequetschten Zustände* lassen sich analog zu Abb. 4.1 mit den zugehörigen Wahrscheinlichkeitsverteilungen $p(\beta_x)$ betrachten (Abb. 4.11(d)). Wie erwartet, werden die Verteilungen schmaler, die Symmetrie des Systems ändert sich nicht. Die Berechnung von $p(\beta_x)$ findet sich in Kapitel 4.6.2.

Entgegen der Erwartung, dass bei einer Quetschung in Richtung der Meßachse die sichere Schlüsselrate erhöht wird, sieht man in Abb. 4.12, dass sie sowohl im Falle der *direct reconciliation*, als auch im Falle der *reverse reconciliation* geringer wird. Ein negativer Quetschparameter ist damit gleichzusetzen, dass die Zustände so gequetscht werden, dass sie in Richtung der

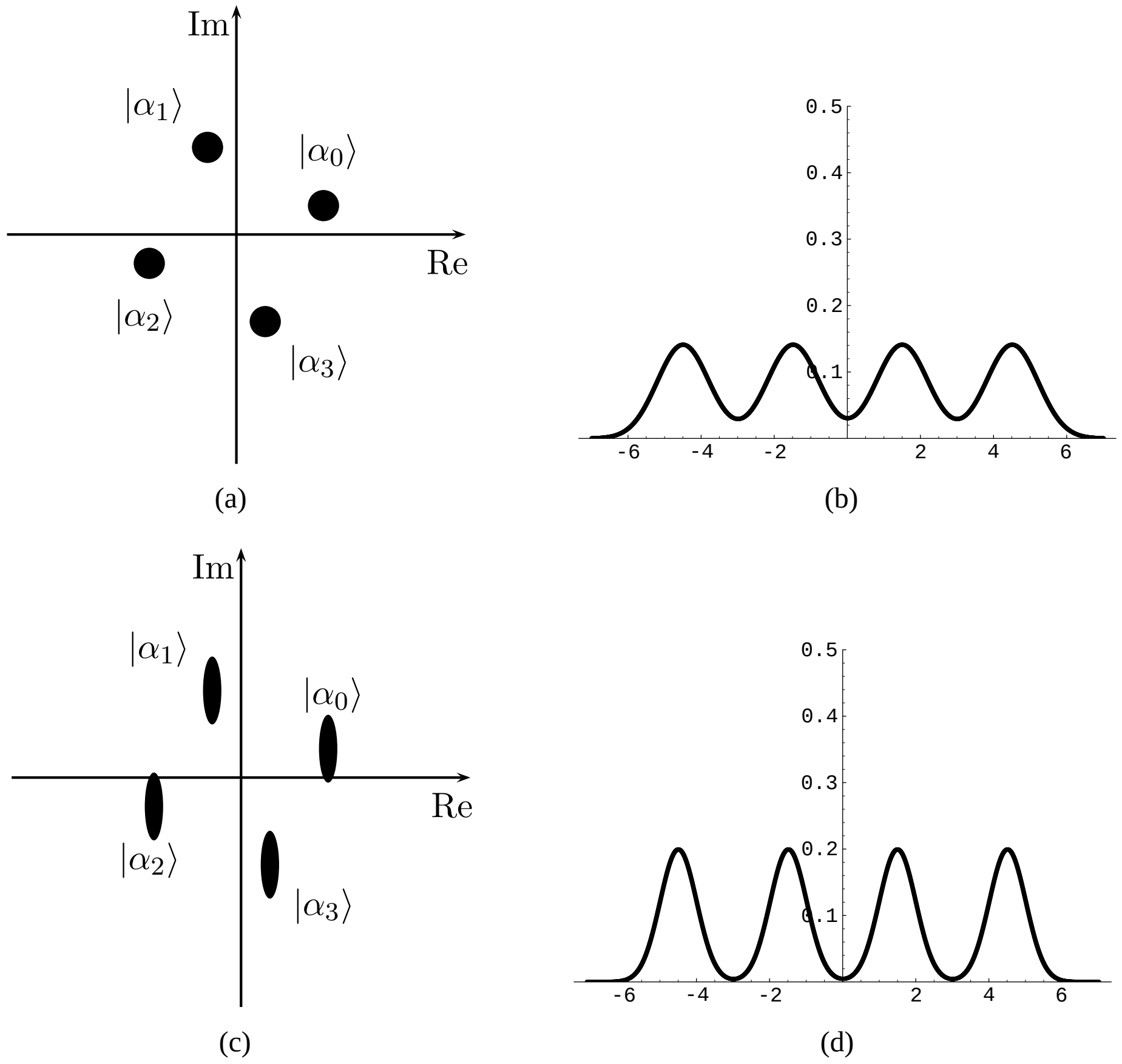

Abbildung 4.11.: Abb. (a) zeigt eine Skizze der vier *kohärenten Zustände* im Phasenraum in einem 4-Qudit-System, und Abb. (b) die zugehörige Wahrscheinlichkeitsverteilung $p(\beta_x)$ für eine Messung auf der reellen Achse mit $|\alpha| = 5$, $\eta = 0.9$ und $\chi = \arctan(1/3)$ (analog zu Abb. 4.1(c) und 4.1(f)).
Die Abbildungen (c) und (d) verdeutlichen den Effekt der Quetschung für eine extrem starke Quetschung von $r = 12$.

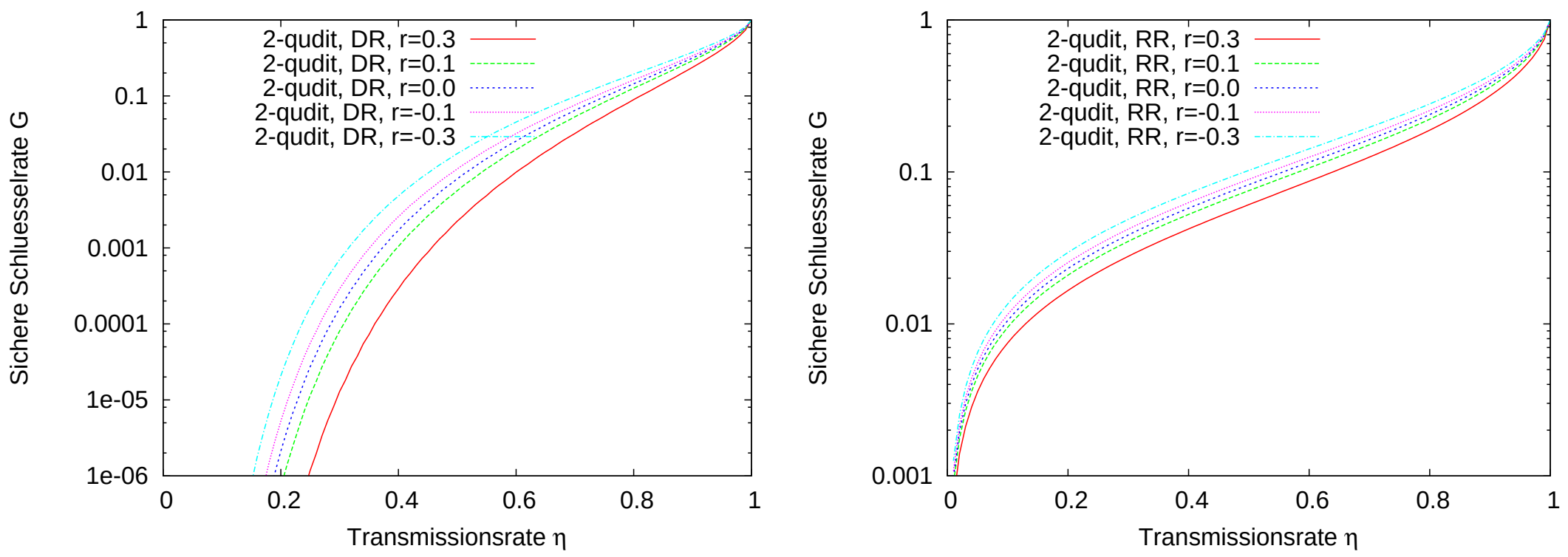

(a) (b)

Abbildung 4.12.: Die beiden Abbildungen vergleichen jeweils die sicheren Schlüsselraten über $\eta$ mit unterschiedlichen Quetschparametern von $r = \{0,3; 0,1; 0.0; -0,1; -0,3\}$ und der reellen Achse als Messache $\chi = 0$. Weiterhin sind sie über die Amplitude $|\alpha|$ optimiert und auf 1 normiert. Abb. (a) zeigt den Fall der *direct reconciliation* für ein 2-Qudit-System, Abb. (b) analog den Fall der *reverse reconciliation*.

Meßachse breiter werden (die Quetschung könnte auch mit einer Drehung der Messachse um $\phi = \pi/2$ erfolgen). Dieses doch erstaunliche Ergebnis bedeutet, dass Eve von der Quetschung mehr profitieren kann, als Alice und Bob. Es beruht also auf der Annahme, dass Eve eine bessere Meßmethode besitzt. Messungen in höherdimensionalen Systemen liefern ähnliche Ergebnisse.

Eine Erklärung für dieses Verhalten ist im Protokoll selbst zu finden. Eve profitiert von dieser Änderung, da sie in der Präparationsphase stattfindet, denn eine Quetschung der Zustände hilft, diese besser voneinander unterscheiden zu können. Jedoch ist die sichere Schlüsselrate bereits über die Amplitude $|\alpha|$ optimiert. Betrachtet man nun verschiedene Quetschungen bei gleicher Amplitude (Abb. 4.13(a), 4.13(c)) so zeigt sich ein völlig anderes Bild im Bereich nahezu perfekter Transmission (Abb. 4.13(b), 4.13(d)) sowohl im Falle der *direct* als auch der *reverse reconciliation*. Die sichere Schlüsselrate wird nun größer mit einer positiven Quetschung - wie eingangs erwartet.

Wie Abb. 4.14 zeigt, ist die optimale Amplitude $|\alpha|$ bei einer positiven Quetschung geringer. Vergleichbare Ergebnisse erhält man im *direct-reconciliated*-Fall. Man hat es also mit zwei konkurrierenden Parametern zu tun, um zwei Zustände besser voneinander unterscheiden zu können: auf der einen Seite einer Optimierung der Amplitude und auf der anderen Seite der Quetschung der Zustände. Im ersten Fall ändert sich bei steigender Amplitude die Verteilung $p(\beta_x)$ dahingehend, dass sich die Verteilungen zweier Zustände voneinander entfernen, bei einer positiven Quetschung allerdings werden die einzelnen Verteilungen schmaler. Da Alice und Bob *reconciliation*- und *postselektions*-Methoden verwenden, ist ihr Vorteil über Eve am grössten bei einer möglichst flachen Verteilung mit ausreichend weit entfernten Zuständen. Dies erreicht man mithilfe einer negativen Quetschung und einer größeren Amplitude $|\alpha|$.

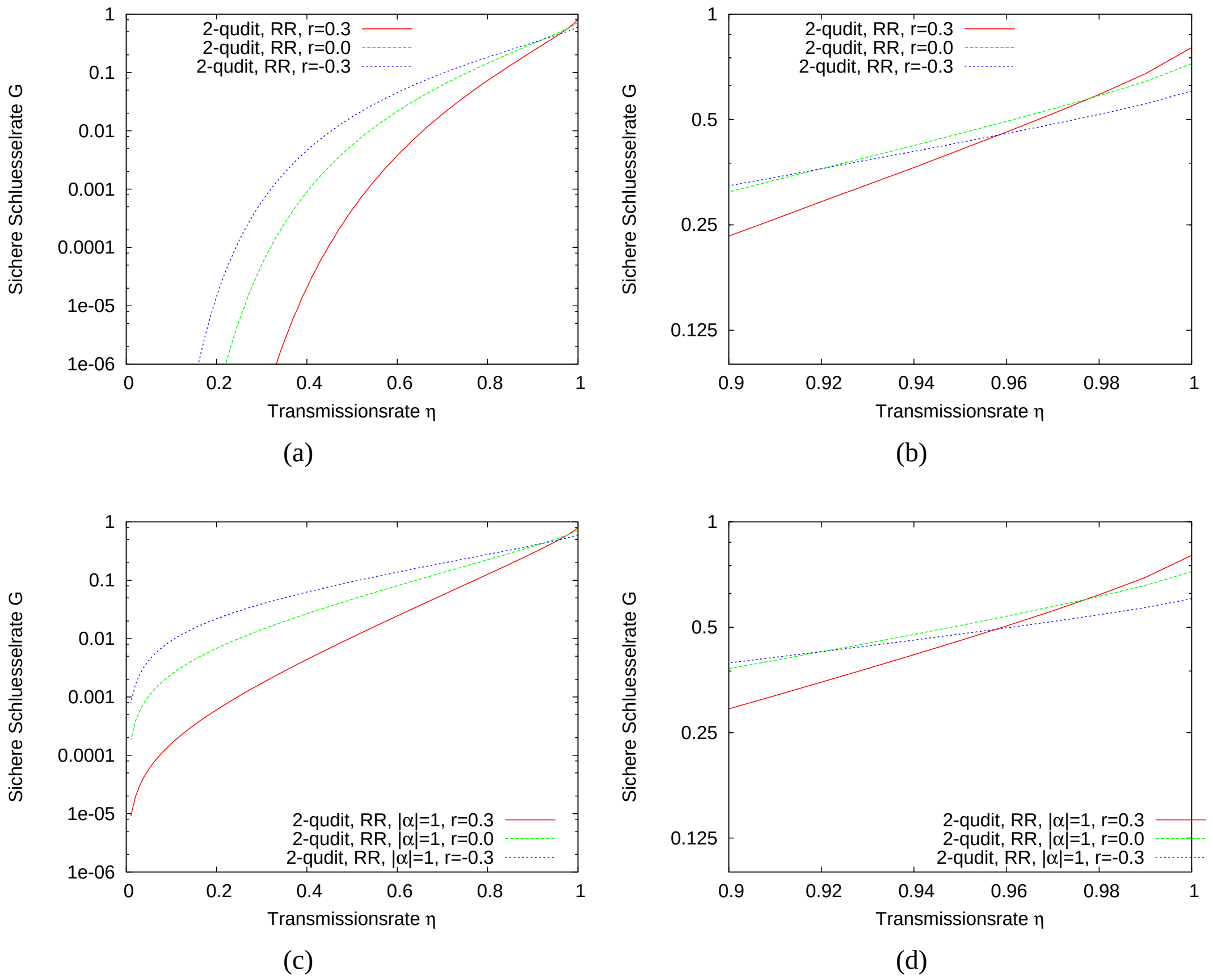

Abbildung 4.13.: Abb. (a) zeigt die sichere Schlüsselrate für ein 2-Qudit-System bei einer festen Amplitude von $|\alpha| = 1$ und *direct reconciliation* bei verschiedenen Quetschparametern $r = \{0,3; 0; -0,3\}$, Abb. (d) eine Vergrößerung im Bereich nahezu perfekter Transmission. Die Abbildungen (c) und (d) zeigen die entsprechenden Plots für den Fall der *reverse reconciliation*.

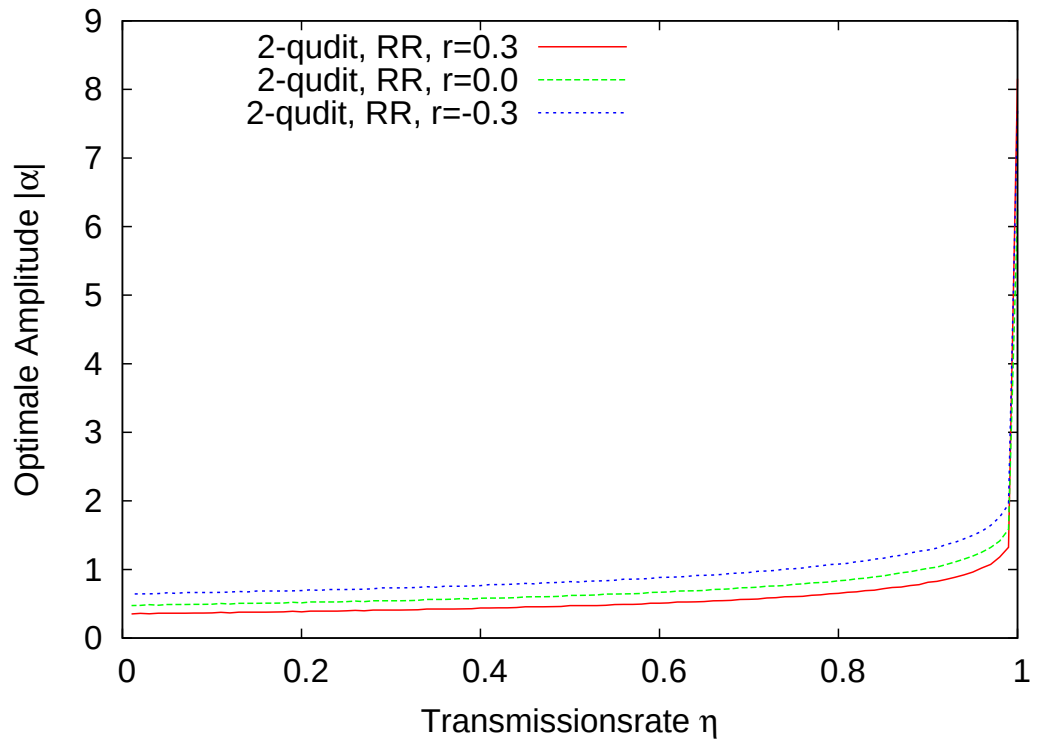

Abbildung 4.14.: zeigt die optimale Amplitude für ein 2-Qudit-System im Falle von *reverse reconciliation* bei verschiedenen Quetschparametern $r = \{0,3; 0; -0,3\}$.

## 4.7. Dual-Homodynmessung

Alternativ zu der Messung einer Achse lassen sich auch beide Achsen mit einem zusätzlichen Fehler messen. (Kapitel 2.5.5).

### 4.7.1. Information zwischen Alice und Bob

Da Bob in diesem Experiment den Zustand $\beta$ misst und nicht nur $\beta_x$, muss man dies bei der Berechnung der Entropien berücksichtigen. Die Transinformation zwischen Alice und Bob gilt weiterhin allgemein mit:

$$I(A:B) = H(A) - H(A|B) \quad , \tag{4.96}$$

die bedingte Entropie berechnet sich aber aus:

$$H(A|B) = -\int\int \mathrm{d}\beta_x \mathrm{d}\beta_y p(\beta) \sum_{k=0}^{d-1} p(\alpha_k|\beta) \log_d p(\alpha_k|\beta) \quad . \tag{4.97}$$

Die Entropie der von Alice präparierten Zustände ändert sich natürlich ebenfalls nicht, da nur die Messung eine andere ist. Analog zu (4.13) kann man wiederum die *a-posteriori*-Wahrscheinlichkeiten bestimmen:

$$p(\alpha_k|\beta) = \frac{p(\beta|\alpha_k)p(\alpha_k)}{\sum\limits_{l=0}^{d-1} p(\beta|\alpha_l)p(\alpha_l)} = \frac{p(\beta|\alpha_k)p(\alpha_k)}{p(\beta)} \quad . \tag{4.98}$$

Wegen des zweiten Strahlteilers geht der Zustand, den Bob an jeder seiner beiden Homodynexperimente erhält, über in

$$|\sqrt{\eta}\alpha\rangle \rightarrow |\sqrt{\frac{\eta}{2}}\alpha\rangle \quad . \tag{4.99}$$

Für $p(\beta|\alpha_k)$ gilt:

$$p(\beta|\alpha_k) = \frac{1}{\pi}\left|\langle\beta|\sqrt{\frac{\eta}{2}}\alpha_k\rangle\right|^2 \tag{4.100}$$

$$= \frac{1}{\pi}\mathrm{e}^{-|\sqrt{\frac{\eta}{2}}\alpha_k-\beta|^2} \quad . \tag{4.101}$$

Daraus folgt mit $|\beta\rangle = |\beta_x + i\beta_y\rangle$ und $|\alpha_k\rangle = |\alpha|| \mathrm{e}^{i\frac{2\pi}{d}k-\chi}\rangle$:

$$p(\beta|\alpha_k) = \frac{1}{\pi}\mathrm{e}^{-\left(\frac{\eta}{2}|\alpha|^2+\beta_x^2+\beta_y^2-\sqrt{2\eta}|\alpha|(\beta_x\cos(\frac{2\pi}{d}k-\chi)+\beta_y\sin(\frac{2\pi}{d}k-\chi)\right)} \quad . \tag{4.102}$$

Dieses Ergebnis ist für die eigentliche Messung interessant, da Bob $\beta_x$ und $\beta_y$ misst. Um allerdings die sichere Schlüsselrate berechnen zu können, ist es sinnvoll, diesen Ausdruck in Polarkoordinaten zu betrachten. Da die Messung in zwei Ebenen erfolgt und die möglichen Zustände auf einem Kreis angeordnet sind, ist dieses Problem kreissymmetrisch. Es genügt, nur über $1/d$ des Umfangs zu integrieren und das Ergebnis mit $d$ zu multiplizieren. Genau dies ist auch 1 Kanal. Man erhält in Polarkoordinaten mit $\beta_r = \sqrt{\beta_x^2 + \beta_y^2}$ und $\phi = \arccos(\beta_x/\beta_r)$:

$$p(\beta|\alpha_k) = \frac{\mathrm{e}^{-\frac{\eta}{2}|\alpha|^2 \sin^2\left(\frac{2\pi}{d}k-\chi+\phi\right)}}{\pi} \mathrm{e}^{-\left(\beta_r - \sqrt{\frac{\eta}{2}}|\alpha|\cos(\frac{2\pi}{k}d-\chi+\phi)\right)^2} \tag{4.103}$$

$$= \frac{\mathrm{e}^{-(\beta_r^2+\frac{\eta}{2}|\alpha|^2)}}{\pi} \mathrm{e}^{\sqrt{2\eta}|\alpha|\beta_r\cos(\frac{2\pi}{k}d-\chi+\phi)} \quad . \tag{4.104}$$

Damit gilt für $p(\beta)$ analog:

$$p(\beta) = \frac{\mathrm{e}^{-(\beta_r^2+\frac{\eta}{2}|\alpha|^2)}}{\pi} \sum_{l=0}^{d-1} \mathrm{e}^{\sqrt{2\eta}|\alpha|\beta_r\cos(\frac{2\pi}{l}d-\chi+\phi)} \quad . \tag{4.105}$$

Da die Dual-Homodynmessung auf zwei orthogonalen Achsen misst, sollte eine gemeinsame Drehung dieser Achsen invariant sein. Die Rechnung wurde mit dem Parameter der Meßachse $\chi$ durchgeführt und wie man sehen kann, ergibt sich so eine einfache Transformation des Phasenparameters $\phi$ zu $\phi' = -\chi + \phi$, das Ergebnis bleibt also unverändert und es gilt:

$$p(\beta) = \frac{\mathrm{e}^{-(\beta_r^2+\frac{\eta}{2}|\alpha|^2)}}{\pi} \sum_{l=0}^{d-1} \mathrm{e}^{\sqrt{2\eta}|\alpha|\beta_r\cos(\frac{2\pi}{l}d+\phi)} \quad . \tag{4.106}$$

### Darstellung als Bessel-Funktion

Der Ausdruck $\mathrm{e}^{\sqrt{2\eta}|\alpha|\beta_r\cos(\frac{2\pi}{k}d+\phi)}$ lässt sich auch durch eine *Laurent-Reihe* der *modifizierten Bessel-Funktion* $I_n(x)$ angeben. Mit der modifizierten Bessel-Funktion [Abr84]:

$$I_\nu(x) = \sum_{s=0}^{\infty} \frac{1}{s!(s+\nu)!} \left(\frac{x}{2}\right)^{2s+\nu} \tag{4.107}$$

als Laurent-Reihe in der Form

$$\mathrm{e}^{\left(\frac{x}{2}\right)\left(t+\frac{1}{t}\right)} = \sum_{-\infty}^{\infty} I_n(x) t^n \tag{4.108}$$

ergibt sich mit $t = \mathrm{e}^{i\left(\frac{2\pi}{d}k+\phi\right)}$ und $x = \sqrt{2\eta}|\alpha|\beta_r$:

$$\mathrm{e}^{\sqrt{2\eta}|\alpha|\beta_r\cos\left(\frac{2\pi}{d}k+\phi\right)} = \sum_{n=0}^{\infty} I_n(\sqrt{2\eta}|\alpha|\beta_r)\mathrm{e}^{in\left(\frac{2\pi}{d}k+\phi\right)} \quad . \tag{4.109}$$

## 4.7.2. Holevoinformation

Die Holevoinformation im *direct-reconciliated*-Fall ändert sich nicht, da nur Bob seine Messung verändert und Eve darauf nach Voraussetzung keinerlei Zugriff hat.

Der *direct-reconciliated*-Fall ist analog, wie in der *Homodynmessung* (Kapitel 4.4.1) aufgebaut. Es müssen nur die Wahrscheinlichkeiten $p(\beta|\alpha_k)$ und $p(\beta)$ durch (4.104) und (4.106) ersetzt werden.

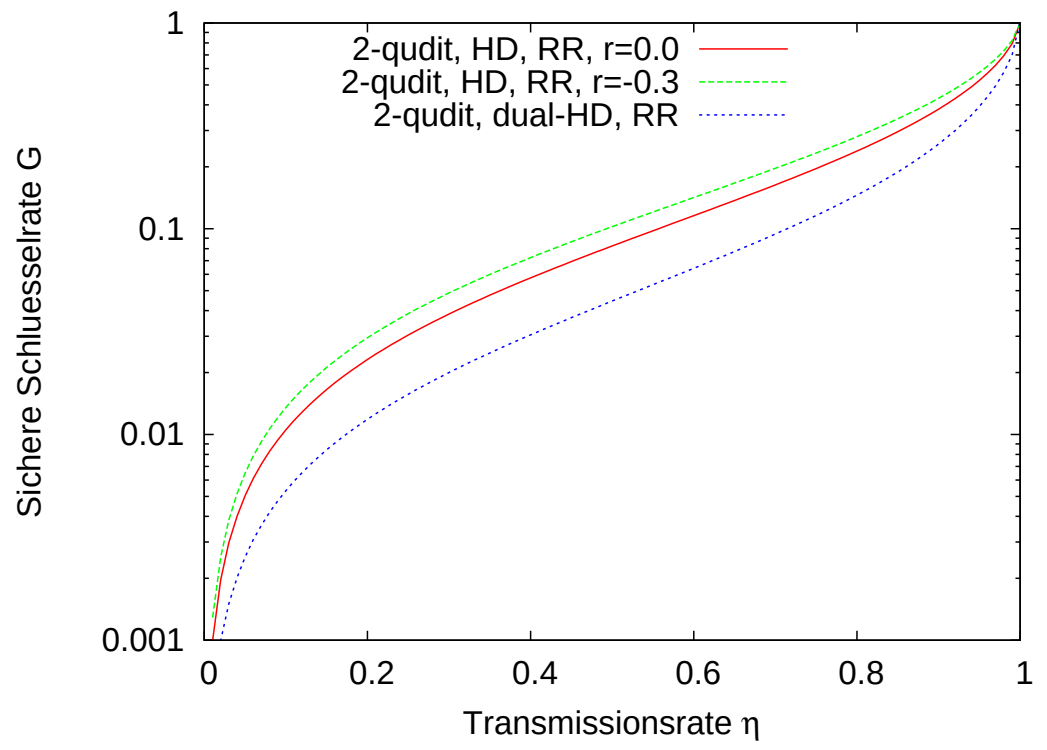

Abbildung 4.15.: Vergleich der sicheren Schlüsselrate im Falle der *reverse reconciliation* für 2-Qudit-Zustände, normiert auf 1 mit einer optimierten Amplitude. Verglichen wird das Ergebnis der Dual-Homodynmessung mit dem der Homodynmessung an einem *kohärenten* und einem *gequetschten Zustand* mit $r = -0.3$.

### 4.7.3. Resultate

Da bei einer Dual-Homodynmessung zwei kanonisch konjugierte Variablen gleichzeitig gemessen werden, opfert man eine scharfe Messung auf einer Achse zugunsten einer ungenaueren Messung auf zwei Achsen. Allerdings erhofft man sich hierdurch eine bessere Unterscheidung von Zuständen in Systemen höherer Ordnung, aufgrund der Symmetrie.

In einem 2-Qudit-System wäre also solch eine Messung nicht sinnvoll gegenüber einer normalen Homodynmessung, da der Wert von $\beta_y$ nicht interessant für die Unterscheidung ist. Wie man in Abb. 4.15 sehen kann, ist das Ergebnis der Homodynmessung deutlich besser.

Die Abbildungen 4.16(a) und 4.16(b) zeigen nun ein ähnliches Resultat für verschiedene Ordnungen wie bei der gewöhnlichen Homodynmessung. Allerdings sieht man einen sehr viel größeren Unterschied zwischen dem 2-Qudit und dem 4-Qudit-System. Auf der Vergrößerung in Abb. 4.16(b) ist zu erkennen, dass nun bei einer perfekten Transmission auch im 8-Qudit-System die volle Kapazität erreicht wird, also 3 Bit.

Genau dieser größere Unterschied führt zu dem erwarteten Ergebnis. Abb. 4.17 zeigt für den Fall eines 4-Qudit-Systems und den eine 8-Qudit-Systems jeweils, dass die erreichbare sichere Schlüsselrate auf bei einer Transmission größer $\eta = 0,5$ besser ist, als im Falle der Single-Homodynmessung. Dieser Effekt steigt mit größerer Transmission weiter an und erreicht dann die maximale Kapazität bei perfekter Transmission, wie bereits angesprochen.

#### Detektorrauschen

Analog zu (3.43) und (3.44) in Kapitel 3.2.8 lassen sich die Wahrscheinlichkeiten der betrachteten Fälle für die Dual-Homodynmessung mit Detektorrauschen berechnen. Es gilt:

$$p^{\mathrm{NOISE}}(\beta) = \frac{\mathrm{e}^{-(\beta_r^2 + \frac{\eta}{2}|\alpha|^2)/(1+\delta)}}{\pi} \sum_{l=0}^{d-1} \mathrm{e}^{\sqrt{2\eta}|\alpha|\beta_r \cos(\frac{2\pi}{l}d+\phi)/(1+\delta)} \quad . \tag{4.110}$$

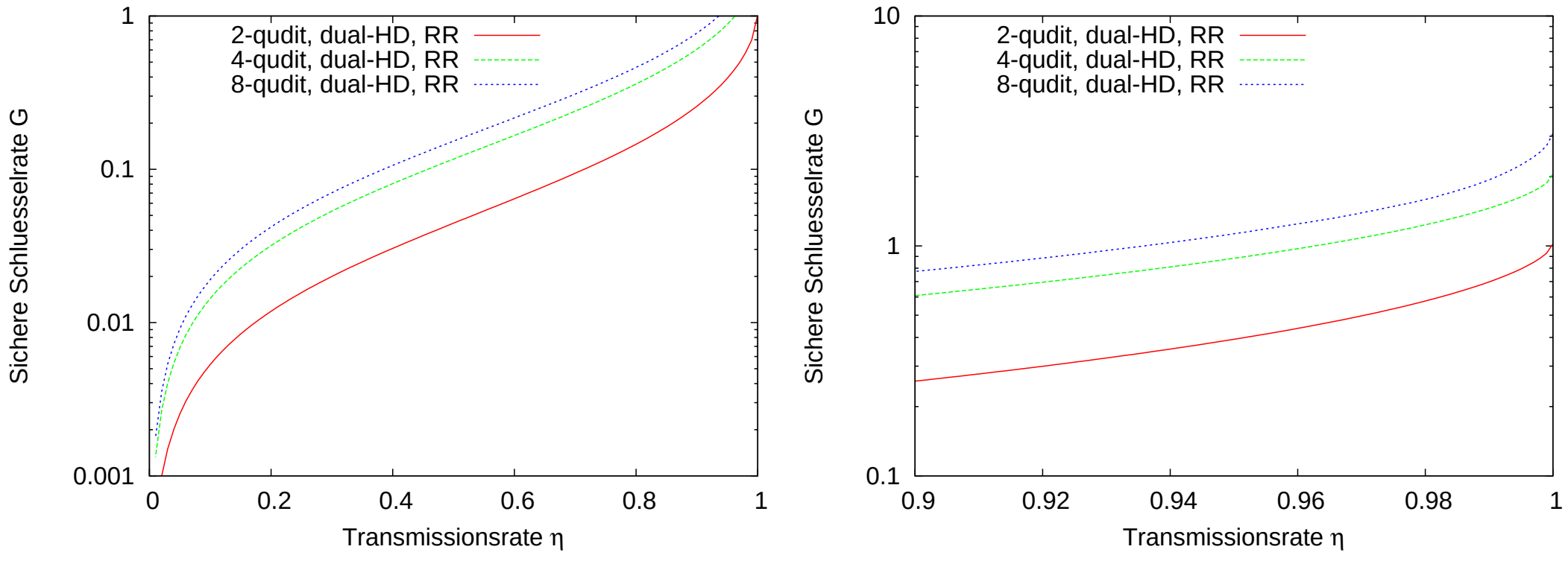

(a) (b)

Abbildung 4.16.: Abb. (a) zeigt die sichere Schlüsselrate im Vergleich zwischen einem 2-, 4-, und 8-Qudit-System für die Dual-Homodynmessung im *reverse-reconciliation*-Fall, normiert auf ein Bit. Auf der Vergrößerung in Abb. (b) lässt sich das Erreichen der maximalen Kanalkapazität erkennen.

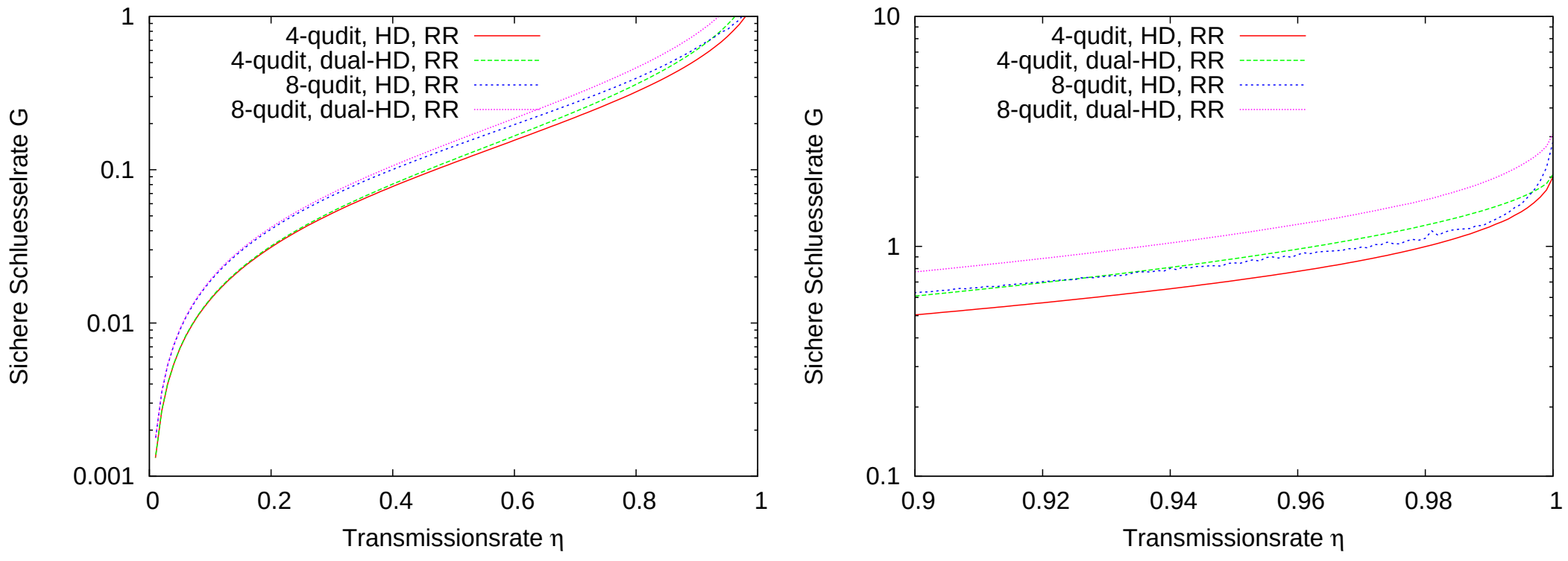

(a) (b)

Abbildung 4.17.: Abb. (a) zeigt die sichere Schlüsselrate im Vergleich jeweils zwischen der normalen Homodynmessung und der Dual-Homodynmessung für ein 4- und ein 8-Qudit-System im *reverse-reconciliation*-Fall, normiert auf ein Bit. Abb. (b) zeigt wiederum eine Vergrößerung im Bereich perfekter Transmission.

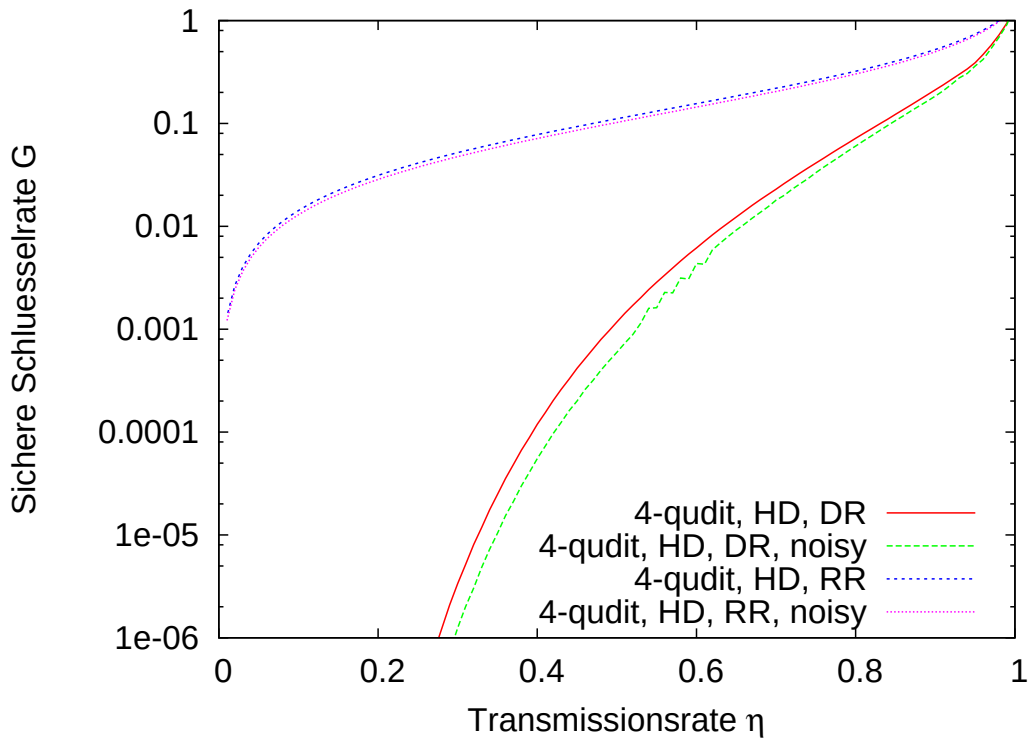

(a)

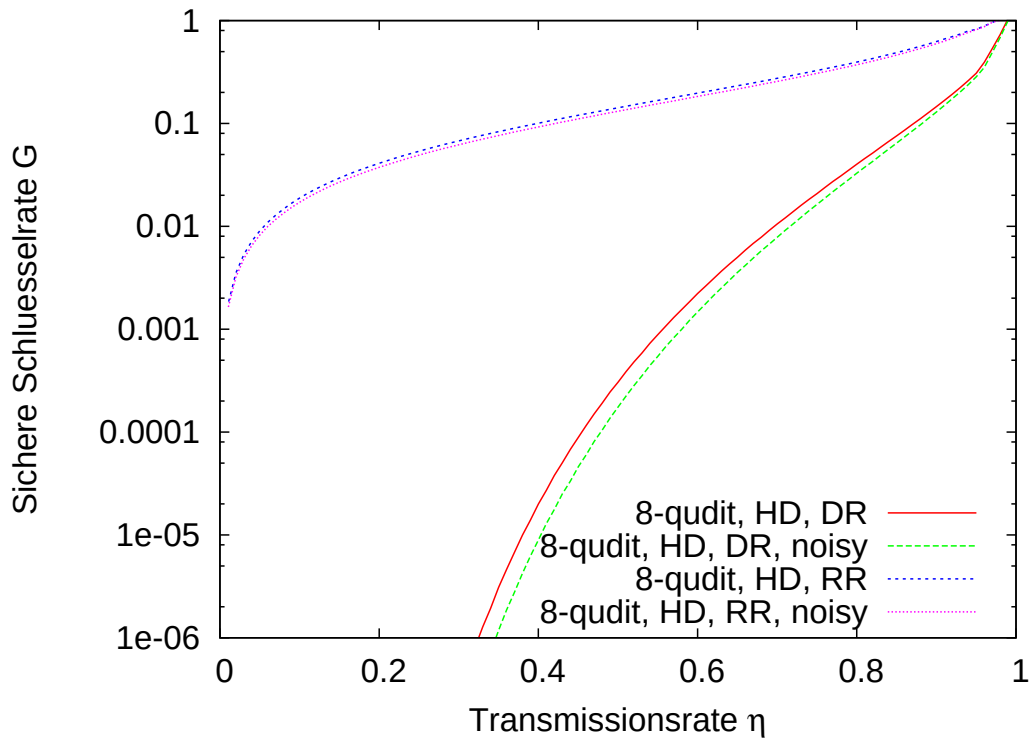

(b)

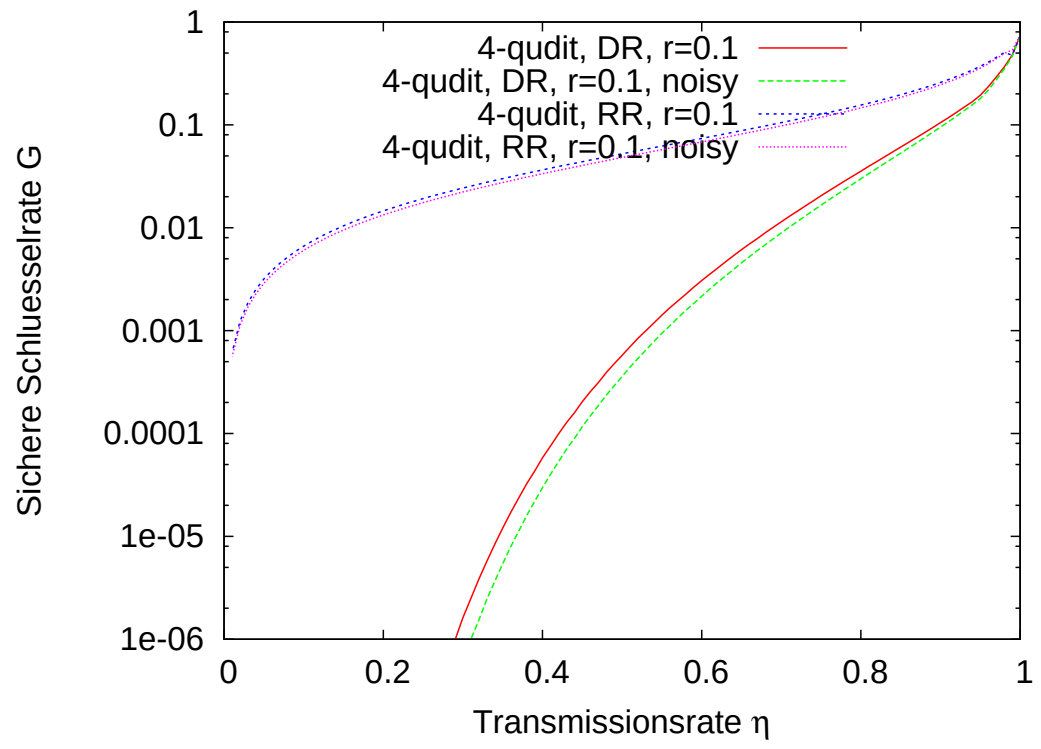

(c)

Abbildung 4.18.: Es werden jeweils die sichere Schlüsselrate eines Systems mit $\delta = 0.1$ Detektorrauschen verglichen mit der sicheren Schlüsselrate ohne Rauschen, jeweils mit *direct* und *reverse reconciliation*, normiert auf ein Bit. Abb. (a) zeigt die sichere Schlüsselrate für ein ungequetsches 4-Qubit-System, Abb. (c) für ein gequetschtes mit $r = 0.1$. Abb. (b) zeigt ein ungequetschtes 8-Qubit-System.

# 5. Zusammenfassung

Es wurde gezeigt, dass es mathematisch möglich ist, eine Erweiterung des Protokolls von Grosshans und Grangier zu formulieren. Diese erlaubt es, mit Qudits höherer Ordnung zu arbeiten und somit durch die Übertragung eines einzelnen *kohärenten Zustandes* über einen Quantenkanal mehr Informationen als ein Bit zu übertragen. Da in dieser Arbeit der Strahlteiler-Angriff diskutiert wurde, ist in den Betrachtungen die sichere Schlüsselrate meist über der Transmissionsrate des Kanals aufgetragen.

Zuerst wurde dieses Protokoll für gewöhnliche *kohärente Zustände* so erweitert, dass man von einer äquidistanten Verteilung auf einem Kreis im Phasenraum sprechen kann. Als wichtige Betrachtung gilt die Menge der übertragenen Information gemessen in Bit. Diese kann für Systeme höherer Ordnung natürlich auch größer als eins werden. Im Falle der *direct reconciliation* hat sich gezeigt, dass die Ergebnisse für Systeme höherer Ordnung schlechter werden. Das System mit 2 möglichen Zuständen zeigt die besten Resultate, ausser im Bereich nahezu perfekter Kanaltransmission. Im Falle der *reverse reconciliation* ergibt sich eine deutliche Erhöhung der sicheren Schlüsselrate mit einer Zunahme der betrachteten Ordnung. Man erkennt jedoch schon in einem System mit 8 Zuständen das Problem dieses Protokolls. Aufgrund der Asymmetrie zwischen einer Präparation auf einem Kreis und einer Messung auf einer Achse und einer damit verbundenen Projektion sind nicht mehr alle Zustände gleich gut voneinander unterscheidbar.

Durch Symmetriebetrachtungen wurde eine Basis gefunden, welche es erlaubt, auch andere, als *kohärente Zustände* in einem solchen Protokoll betrachten zu können, solange diese Zustände auf der angesprochenen Kreissymmetrie verteilt sind.

Mithilfe dieser Basis war es nun auch möglich, verallgemeinerte Zustände der *kohärenten Zustände* zu betrachten, und zwar *gequetschte Zustände*. Verwendet man also *gequetschte Zustände* für dieses Protokoll, die in Richtung der Messachse gequetscht sind, so erwartet man eine bessere Messung aufgrund der geringeren Unschärfe. Es zeigt sich jedoch, dass Eve aus dieser Quetschung mehr Profit schlagen kann als Alice und Bob, da man über Eve's Messung nichts weiss und somit davon ausgeht, dass sie mithilfe besserer Technologie die maximal zugängliche Information aus dem Kanal verwenden kann. Allerdings kann es sinnvoll sein, die Zustände negativ zu quetschen, also in Richtung der Meßachse zu verbreitern, um Eve's Messung bewusst zu stören. Es sei auch anzumerken, dass für eine stärkere Quetschung die optimale Amplitude kleiner wird.

Durch die Quetschung wurde das Protokoll in der Präparationsphase variiert. Eine Änderung der Messung lässt sich alternativ ebenfalls realisieren. Die letzte Betrachtung hat sich mit dem Fall beschäftigt, dass Bob nicht nur entlang einer Achse im Phasenraum misst, sondern entlang beider Achsen. Da es sich jedoch um zwei kanonisch konjugierte Variablen handelt, kann die Messung nicht mehr wie zuvor scharf in einer Variable stattfinden, sondern ist gleichverteilt mit einer Unschärfe auf beide Variablen. Dies erreicht Bob durch eine Dual-Homodynmessung. Das Eingangssignal von Alice wird mithilfe eines Strahlteilers in zwei unabhängige Homodyn-messungen beider konjugierter Parameter geführt. Der zweite Eingang des Strahlteilers sorgt für eben diese zusätzliche Unschärfe. Die Erwartung dieser Änderung ist die bessere Unterscheidbarkeit der einzelnen Zustände aufgrund der Symmetrie. Eine Messung in zwei Ebenen kann Zustände auf einem Kreis besser „symmetrisch“ voneinander unterscheiden. Die Resulate bestätigen genau diese Erwartung. Kann sich der Verlust an einer scharfen Messung im 2-dimensionalen Fall noch nicht auszahlen, da dies noch keine wirkliche Kreissymmetrie ist, so doch schon im 4-Dimensionalen. Eine Verbesserung ist vor allem im Bereich von Transmissionsraten größer $50\%$ deutlich zu erkennen. Im Fall perfekter Transmission wird sogar das Kanalmaximum erreicht.

## 5.1. Ausblick

Man kann auf der Grundlage dieser Erkenntnisse die betrachteten Protokolle mithilfe weiterer Parameter verändern und diskutieren. Würde bespielsweise Alice die Zustände so präparieren, dass sie nach einer Messung von Bob auf einer Achse äquidistant verteilt lägen, so könnte dies eine alternative Lösung des Symmetrieproblems zur Dual-Homodynmessung darstellen. Dank der berechneten Basis ist es auch möglich, allgemeinere Zustände zu betrachten, beispielsweise kohärent verschobene *Fock-Zustände* oder *Gauss-Zustände*.

Bisher existieren noch keine Sicherheitsbeweise für Schlüsselaustauschverfahren mit kontinuierlichen Variablen. Deshalb wäre es sinnvoll, die betrachteten Protokolle auch auf weitere möglichen Angriffe hin zu untersuchen.

Im Falle der Systeme mit *gequetschten Zuständen* wäre noch das Problem der experimentellen Realisierbarkeit zu klären und die Frage, ob mithilfe zusätzlicher Parameter diese Probleme in die theoretische Betrachtung mit einfließen könnten.

Diese Fragen sollten bezüglich weiterer Probleme gestellt werden, wie beispielsweise dem Betrachten von Einflüssen durch Rauschen jeglicher Art oder der Effizienz und Existenz von Protokollen zur Fehlerkorrektur.

# A. Methoden

**Satz 1.** *Seien die Matrizen* $A = (a_{ij}), C = (c_{ij}) \in \mathbb{C}^{n\times n}$ *mit* $c_{ij} = c_i\delta_{ij}$ *und* $|C| = \sqrt{C^\dagger C}$*.*
*Dann gilt:*

$$\det(CAC^\dagger - \lambda E) = \det(|C|A|C| - \lambda E) \tag{A.1}$$

**Lemma 1.** *Sei* $A = (a_1, a_2, \dots, a_n)$ *Matrix und* $a_k$ *Spaltenvektoren.*
*Dann gilt:*

$$\begin{aligned}\det(A - \lambda E) &= \det(a_1 - \lambda e_1, a_2 - \lambda e_2, \dots, a_n - \lambda e_n) \qquad \text{(A.2)}\\ &= \det(a_1, a_2, \dots, a_n)\\ &\quad + (-\lambda)(\sum_i \det(a_1, \dots, e_i, \dots, a_n))\\ &\quad + (-\lambda)^2(\sum_{i,j} \det(a_1, \dots, e_i, \dots, e_j, \dots, a_n))\\ &\quad \vdots\\ &\quad + (-\lambda)^n \qquad \text{(A.3)}\end{aligned}$$

*Beweis von Lemma 1.* Für die Determinante gilt die Multilinearität:

$$\det(\dots, \alpha a_j + \beta b_j, \dots) = \alpha \det(\dots, a_j, \dots) + \beta \det(\dots, b_j, \dots) \quad . \tag{A.4}$$

Daraus folgt mit $\det(A - \lambda E) = \det(a_1 - \lambda e_1, a_2 - \lambda e_2, \dots, a_n - \lambda e_n)$ das gewünschte Ergebnis. □

*Beweis von Satz 1.* Sei $b_m = (c_1 c_m^* a_{1m}, c_2 c_m^* a_{2m}, \dots, c_n c_m^* a_{nm})$ Spaltenvektor und $B = (b_1, b_2, \dots, b_n)$.
Dann gilt nach Lemma (1)

$$\begin{aligned}\det(B - \lambda E) &= \det(b_1 - \lambda e_1, b_2 - \lambda e_2, \dots, b_n - \lambda e_n) \qquad \text{(A.5)}\\ &= \det(b_1, b_2, \dots, b_n)\\ &\quad + (-\lambda)(\sum_i \det(b_1, \dots, e_i, \dots, b_n))\\ &\quad + (-\lambda)^2(\sum_{i,j} \det(b_1, \dots, e_i, \dots, e_j, \dots, b_n))\\ &\quad \vdots\\ &\quad + (-\lambda)^n \qquad \text{(A.6)}\end{aligned}$$

Für $\det(b_1, \ldots, b_n)$ gilt nach der Formel von Leibnitz:

$$\det(b_1, \ldots, b_n) = \sum_{\sigma \in S_n} \left( \operatorname{sgn}(\sigma) \prod_{i=1}^{n} c_i c^*_{\sigma(i)} a_{i\sigma(i)} \right) \tag{A.7}$$

$$= \sum_{\sigma \in S_n} \left( \operatorname{sgn}(\sigma) \prod_{i=1}^{n} |c_i|^2 a_{i\sigma(i)} \right) \tag{A.8}$$

$$= \sum_{\sigma \in S_n} \left( \operatorname{sgn}(\sigma) \prod_{i=1}^{n} |c_i||c_{\sigma(i)}| a_{i\sigma(i)} \right) \quad . \tag{A.9}$$

Also gilt:
$\det(b_1, \ldots, b_n) = \det(b'_1, \ldots, b'_n)$ mit $b'_m = (|c_1||c^*_m| a_{1m}, \ldots, |c_n||c^*_m| a_{nm})$.
Dies gilt auch für alle Minoren.
Daraus folgt durch erneutes Anwenden des Lemmas (1):

$$\det(b_1 - \lambda e_1, \ldots, b_n - \lambda e_n) = \det(b'_1 - \lambda e_1, \ldots, b'_n - \lambda e_n) \quad , \tag{A.10}$$

und schliesslich

$$\det(CAC^\dagger - \lambda E) = \det(|C|A|C| - \lambda E) \tag{A.11}$$

□

# B. C++-Code

Listing B.1: Code

```cpp
#include <iostream>
#include <cmath>
#include <complex>

using namespace std;

// compile with g++ -Wall -ggdb filename.cpp -o filename -lnag -lg2c

extern "C" void f02haf_( char &job_, char &uplo_, int &n_, complex<double> a_[], int &lda_,
        double w_[], double rwork_[], complex<double> work_[], int &lwork_, int &ifail_ );

double const PI=3.1415926535;
complex<double> II(0,1);

double min_ar   = 0.2;  // 0.2
double max_ar   = 12.0; // 16
double step_ar  = 0.1;  // 0.1
double min_bx   = 0.1;  // 0.1

double p_ak(double degree) {
        return 1.0/degree;
};

double p_bx_ak(double ar, int k, double bx, double n, int degree, double squeeze, double chi, double noise) {
        return exp(-(1.+tanh(squeeze))*
                (sqrt(n)*ar*cos(2.*PI*k/degree - chi) -bx)*
                (sqrt(n)*ar*cos(2.*PI*k/degree - chi) -bx)/(1.+noise)
        )/(cosh(squeeze)*sqrt(PI*(1.-tanh(squeeze))*(1.+noise)));
}

double p_b_ak(double ar, int k, double br, double bp, double n, int degree, double squeeze, double chi, double noise
      ) {
        return exp(-(br*br+n*ar*ar/2.)/(1.+noise))*
                exp(sqrt(2.*n)*ar*br*cos(2.*PI*(double)k/(double)degree -chi+bp))/PI;
}

double p_bx(double ar, double bx, double n, int degree, double squeeze, double chi, double noise) {
        double p = 0.;
        for (int k=0; k<degree; k++) {
                p += p_bx_ak(ar,k,bx,n, degree, squeeze, chi, noise);
        }
        return p/=degree;
}

double p_b(double ar, double br, double bp, double n, int degree, double squeeze, double chi, double noise) {
        double p = 0.;
        for (int k=0; k<degree; k++) {
                p += p_b_ak(ar,k,br,bp,n, degree, squeeze, chi, noise);
        }
        return p/=degree;
}

double H_AB(double ar, double bx, double n, bool RR, int degree, double squeeze, double chi, double noise) {
        double result = 0.;
        for (int k=0; k<degree; k++) {
                result -=(p_bx_ak(ar,k,bx,n, degree, squeeze, chi, noise)*p_ak(degree)/p_bx(ar,bx,n, degree, squeeze
                      , chi, noise))*
                   log(p_bx_ak(ar,k,bx,n, degree, squeeze, chi, noise)*p_ak(degree)/p_bx(ar,bx,n, degree,
                         squeeze, chi, noise))/log(degree);
        }
        return result;
}

double H_AB_DHD(double ar, double br, double bp, double n, bool RR, int degree, double squeeze, double chi, double
      noise) {
        double result = 0.;
        for (int k=0; k<degree; k++) {
                result -=(p_b_ak(ar,k,br,bp,n, degree, squeeze, chi, noise)*p_ak(degree)/p_b(ar,br,bp,n, degree,
                      squeeze, chi, noise))*
```

```
			log(p_b_ak(ar,k,br,bp,n, degree, squeeze, chi, noise)*p_ak(degree)/p_b(ar,br,bp,n, degree, squeeze, chi, noise))/log(degree);
	}
	return result;
}

complex<double> overlap (double ar, int k, double n, double squeeze, int d) {
	complex<double> b2(0,0);

	b2=2.0*(1.0 - n)*ar*ar*(1.0-cos(2.0*PI*k/d))*(cosh(squeeze)*cosh(squeeze)+sinh(squeeze)*sinh(squeeze));
	b2+=(1.0-n)*ar*ar*(2.0+2.0*cos(4.0*PI*k/d)-4.0*cos(2.0*PI*k/d))*sinh(squeeze)*cosh(squeeze);

	complex<double> ret(1,1);
	ret = exp(-II*(1.0-n)*ar*ar*sin(2.0*PI*k/d))*exp(-b2/2.0);
	return ret;
}

complex<double> fac (double d, int k, int l) {
	return exp(-II*(complex<double>)(2.0*PI*(double)(l*k))/d);
}

complex<double> calcCoeff(int l, double ar, double n, int degree, double squeeze) {
	complex<double> result(0,0);
	for (int k=0; k<degree; k++) {
		result += fac(degree,k,l)*overlap(ar,k,n,squeeze,degree);
	}
	result /= (double)degree;
	return result;
}


double reverseEntropyHD(double ar, double bx, double n, double d, double squeeze, double chi, double noise) {
	//cout << "general" << endl;
	int dim = (int)d;

	complex<double> matrix_[dim*dim];
	complex<double> sum(0.0,0.0);


	// calculate diagonal elements
	for (int l=0; l<dim; l++) {
		matrix_[l+l*dim] = real(calcCoeff(l, ar, n, d, squeeze));
	}

	complex<double> norm(0.0,0.0);
	for (int k=0; k<dim; k++) {
		norm += p_bx_ak(ar, k, bx, n, dim, squeeze, chi, noise);
	}

	// calculate lower part
	for (int i=1; i<dim; i++) {
		for (int j=0; j<i; j++) {

			// calculate first and second coefficient
			matrix_[i+j*dim] = sqrt(matrix_[i+i*dim]*matrix_[j+j*dim]);

			// calculate permutted probability
			sum = 0.0;
			for (int k=0; k<dim; k++) {
				sum += fac(d,i-j,k+1)*p_bx_ak(ar, k, bx, n, dim, squeeze, chi, noise);
			}
			matrix_[i+j*dim] *= sum/norm;
			matrix_[j+i*dim] = matrix_[i+j*dim];
		}
	}



	// NAG part

	int n_ = dim;

	double  z_[n_];

	double rwork_[3*n_];
	int lwork_ = 64*n_;
	complex<double> work_[lwork_];
	int ifail_ = 0;


	char job_ = 'N';
	char uplo_ = 'L';

	f02haf_(job_, uplo_, n_, matrix_, n_, z_, rwork_, work_, lwork_, ifail_ );
```

```


        double entropy=0.0;

        for (int i=0; i<n_; i++) {
                entropy+=(1.0/log((double)dim))*z_[i]*log(z_[i]);
        }

        if (isnan(entropy)) return 0.0;


        return -entropy;
}

double reverseEntropyDHD(double ar, double br, double bp, double n, double d, double squeeze, double chi, double noise) {
        //cout << "general" << endl;
        int dim = (int)d;

        complex<double> matrix_[dim*dim];
        complex<double> sum(0.0,0.0);


        // calculate diagonal elements
        for (int l=0; l<dim; l++) {
                matrix_[l+l*dim] = real(calcCoeff(l, ar, n, d, squeeze));
        }

        complex<double> norm(0.0,0.0);
        for (int k=0; k<dim; k++) {
                norm += p_b_ak(ar, k, br, bp, n, dim, squeeze, chi, noise);
        }

        // calculate lower part
        for (int i=1; i<dim; i++) {
                for (int j=0; j<i; j++) {

                        // calculate first and second coefficient
                        matrix_[i+j*dim] = sqrt(matrix_[i+i*dim]*matrix_[j+j*dim]);

                        // calculate permutted probability
                        sum = 0.0;
                        for (int k=0; k<dim; k++) {
                                sum += fac(d,i-j,k+1)*p_b_ak(ar, k, br, bp, n, dim, squeeze, chi, noise);
                        }
                        matrix_[i+j*dim] *= sum/norm;
                        matrix_[j+i*dim] = matrix_[i+j*dim];
                }
        }



        // NAG part

        int n_ = dim;

        double z_[n_];

        double rwork_[3*n_];
        int lwork_ = 64*n_;
        complex<double> work_[lwork_];
        int ifail_ = 0;

        char job_ = 'N';
        char uplo_ = 'L';

        f02haf_(job_, uplo_, n_, matrix_, n_, z_, rwork_, work_, lwork_, ifail_ );


        double entropy=0.0;

        for (int i=0; i<n_; i++) {
                entropy+=(1.0/log((double)dim))*z_[i]*log(z_[i]);
        }

        if (isnan(entropy)) return 0.0;


        return -entropy;
}

double ChiHD (double ar, double bx, double n, bool RR, int degree, double squeeze, double chi, double noise) {
        complex<double> ret(0,0);
        complex<double> sum(0,0);
```

```
	for (int l=0; l<degree; l++) {
		sum = calcCoeff(l, ar,n,degree,squeeze);
		ret -= sum*log(sum)/log((double)degree);
	}

	if (RR) ret -= reverseEntropyHD(ar, bx, n, degree, squeeze, chi, noise);

	return real(ret);
}

double ChiDHD (double ar, double br, double bp, double n, bool RR, int degree, double squeeze, double chi, double noise) {
	complex<double> ret(0,0);
	complex<double> sum(0,0);

	for (int l=0; l<degree; l++) {
		sum = calcCoeff(l, ar,n,degree,squeeze);
		ret -= sum*log(sum)/log((double)degree);
	}

	if (RR) ret -= reverseEntropyDHD(ar, br, bp, n, degree, squeeze, chi, noise);

	return real(ret);
}

double IdealHD(double ar, double bx, double n, bool RR, int degree, double squeeze, double chi, double noise) {
	return 1.-H_AB(ar, bx, n, RR, degree, squeeze, chi, noise)- ChiHD(ar, bx, n, RR, degree, squeeze, chi, noise);
}

double IdealDHD(double ar, double br, double bp, double n, bool RR, int degree, double squeeze, double chi, double noise) {
	return 1.-H_AB_DHD(ar, br,bp, n, RR, degree, squeeze, chi, noise)- ChiDHD(ar, br, bp, n, RR, degree, squeeze, chi, noise);
}

double AllInt(double ar, double n, bool DHD, bool RR, int degree, double squeeze, double precision, double limit, double chi, double noise) {
	double HDnew= 0.;
	double tmp = 0.;

	if (DHD) {
		for (double bp= 0.; bp < 2.*PI/(double)degree; bp+=precision*0.01) {
			for (double br=min_bx; br<=limit; br+=precision) {
				tmp = p_b(ar,br,bp, n, degree, squeeze, chi, noise)*IdealDHD(ar,br,bp, n, RR, degree, squeeze, chi, noise);
				if (tmp>0.) HDnew +=tmp*precision*precision*br*0.01;
			}
		}
	} else {

		for (double bx=min_bx; bx<=limit; bx+=precision) {
			tmp = p_bx(ar,bx, n, degree, squeeze, chi, noise)*IdealHD(ar,bx,n, RR, degree, squeeze, chi, noise);
			if (tmp>0.) HDnew +=precision*tmp;
		}
	}
	return HDnew;
}

double Integrate(double n, bool DHD, bool RR, int degree, double squeeze, double precision, double limit, double chi, double noise, double ar_fix) {
	double result = 0.;
	double bestAr = 0.;
	double HDnew = 0.;
	double stepping; // = 1.0;

	if (ar_fix!=0.) {
		result = AllInt(ar_fix, n, DHD, RR, degree, squeeze, precision, limit, chi, noise);
		bestAr = ar_fix;
	}
	else
	{
		//if (((max_ar+1.)/stepping)<5.) {
			stepping = (max_ar)/4.;
		//}
		for (double ar= 0.; ar <= max_ar+1.0; ar+=stepping) {
			HDnew = AllInt(ar, n, DHD, RR, degree, squeeze, precision, limit, chi, noise);

			if (HDnew > result) {
				result = HDnew;
				bestAr = ar;
			}
```

```
			}
			stepping /=2.;

			double bestArTmp = bestAr;
			double ar;

			do {

				ar=bestArTmp-stepping;
				HDnew = AllInt(ar, n, DHD, RR, degree, squeeze, precision, limit, chi, noise);

				if (HDnew > result) {
					result = HDnew;
					bestAr = ar;
				}

				ar=bestArTmp+stepping;
				HDnew = AllInt(ar, n, DHD, RR, degree, squeeze, precision, limit, chi, noise);

				if (HDnew > result) {
					result = HDnew;
					bestAr = ar;
				}
				bestArTmp = bestAr;
				stepping /=2.;
			} while (stepping>0.01);
	}

	if (bestAr>0) max_ar = bestAr+1.;

	if (DHD) result *=degree;
		else result *=2;

	cout << n << "\t" << result << "\t" << bestAr << endl;
	return result;
}

int main (int argc, char **argv) {

	int switcher=0;

	// paramter
	bool DHD=false;
	bool RR = false;
	int degree;
	double squeezing;
	double stepsize;
	double precision;
	double limit;
	double chi;
	double noise;
	double ar;

	if (argc>2) {
		if ( (argv[1][0]=='D') && (argv[1][1]=='H') && (argv[1][2]=='D')) DHD=true;
			else if ( (argv[1][0]=='D') && (argv[1][1]=='H')) DHD=false;

		if ( (argv[2][0]=='D') && (argv[2][1]=='R')) RR=false;
			else if ( (argv[2][0]=='R') && (argv[2][1]=='R')) RR=true;
	}

	if (argc>7) {
		degree   = (int)atoi(argv[3]);
		squeezing= (double)atof(argv[4]);
		stepsize        = (double)atof(argv[5]);
		precision       = (double)atof(argv[6]);
		limit           = (double)atof(argv[7]);
		if (!DHD) switcher = 1; else switcher = 2;
	}

	if (argc>8) {
		double denom= (int)atoi(argv[8]);
		if (denom==0) chi=0.;
		else chi = PI/denom;
	} else chi = 0.;

	if (argc>9) noise = (double)atof(argv[9]); else noise=0.;
	if (argc==11) ar= (double)atof(argv[10]); else ar = 0.;


	switch (switcher) {
		case 1:
		case 2:
		{
```

```
                for (double n = 1; n>=0; n-=stepsize) {
                    Integrate(n, DHD, RR, degree, squeezing, precision, limit, chi, noise, ar);
                }

                break;
            }

            default:
                cerr << "Usage: " << argv[0] << " HD/DHD DR/RR degree squeezing stepsize precision limit [denom] [noise] [ar]\n";
                return 1;
        }

}
```

# Literaturverzeichnis

Angaben zum Verfasser
---------------------
Name            : Ulrich Seyfarth
Geburtsdatum    : 29. Juni 1982
Geburtsort      : Hadamar

Deutscher Titel
---------------
Quantenkryptographie mit kontinuierlichen Variablen
Diskussion verschiedener Realisierungen mit Qudits

Englischer Titel
----------------
Continuous Variable Quantum Cryptography
Discussion of Various Realisations with Qudits

Angaben zur Diplomarbeit
------------------------
Technische Universität Darmstadt
Fachbereich Physik
Institut für Angewandte Physik
Arbeitsgruppe "Theoretische Quantenphysik"

Betreuer         : Prof. Dr. Gernot Alber
Zweitgutachter   : Dipl.-Phys. Kedar S. Ranade

Erklärung gemäß Paragraph 19, Absatz 6 DPO/AT
---------------------------------------------
Hiermit versichere ich, die vorliegende Diplomarbeit ohne Hilfe Dritter nur mit den angegebenen Quellen und Hilfsmitteln angefertigt zu haben. Alle Stellen, die aus den Quellen entnommen wurden, sind als solche kenntlich gemacht worden. Diese Arbeit hat in gleicher Form noch keiner Prüfungsbehörde vorgelegen.

Darmstadt, den 05. Mai 2008

_________________________

Ulrich Seyfarth